\documentclass[aps,twocolumn,nofootinbib]{revtex4}
\usepackage[T1]{fontenc}
\usepackage[latin9]{inputenc}
\usepackage{amsthm}
\usepackage{amsmath}
\usepackage{graphicx}
\begin{document}

\title{MC-EKRT: Monte Carlo event generator with saturated minijet production for initializing 3+1 D fluid dynamics in high energy nuclear collisions}

\author{Mikko Kuha, Jussi Auvinen, Kari J. Eskola, Henry Hirvonen, Yuuka Kanakubo, Harri Niemi}
\affiliation{University of Jyv\"askyl\"a, Department of Physics, P.O. Box 35, FI-40014 University of Jyv\"askyl\"a, Finland,}
\affiliation{Helsinki Institute of Physics, P.O.Box 64, FI-00014 University of
Helsinki, Finland}

\begin{abstract}
    We present a novel Monte-Carlo implementation of the EKRT model, MC-EKRT, for computing partonic initial states in high-energy nuclear collisions. Our new MC-EKRT event generator is based on collinearly factorized, dynamically fluctuating pQCD minijet production, supplemented with a saturation conjecture that controls the low-$p_T$ particle production. Previously, the EKRT model has been very successful in describing low-$p_T$ observables at mid-rapidity in heavy-ion collisions at the LHC and RHIC energies. As novel features, our new MC implementation gives a full 3-dimensional initial state event-by-event, includes dynamical minijet-multiplicity fluctuations in the saturation and particle production, introduces a new type of spatially dependent nuclear parton distribution functions, and accounts for the conservation of energy/momentum and valence-quark number. In this proof-of-principle study, we average a large set of event-by-event MC-EKRT initial conditions and compute the rapidity and centrality dependence of the charged hadron multiplicities and elliptic flow for the LHC Pb+Pb and RHIC Au+Au collisions using 3+1 D viscous fluid-dynamical evolution. Also event-by-event fluctuations and decorrelations of initial eccentricities are studied. The good agreement with the rapidity-dependent data suggests that the same saturation mechanism that has been very successful in explaining the mid-rapidity observables, works well also at larger rapidities.
\end{abstract}

\pacs{25.75.-q, 25.75.Nq, 25.75.Ld, 12.38.Mh, 12.38.Bx, 24.10.Nz, 24.85.+p}

\maketitle

%%%%%%%%%%%%%%%%%%%%%% SECTION %%%%%%%%%%%%%%%%%%%%%
\section{Introduction}
The theory of the strong interaction, Quantum Chromodynamics (QCD), predicts that at very high energy densities, at temperatures $T\gtrsim 150-160$~MeV  and at a vanishing baryochemical potential, strongly interacting matter is in the form of a quark-gluon plasma (QGP) \cite{Bazavov:2014pvz, Bazavov:2017dsy, Borsanyi:2013bia, Borsanyi:2010cj}. Such extreme conditions can be momentarily created and the properties of the QGP experimentally studied in laboratory by colliding heavy ions at ultrarelativistic energies at the CERN Large Hadron Collider (LHC) and the Brookhaven National Laboratory (BNL) Relativistic Heavy Ion Collider (RHIC). In these collisions, the "heating" of the matter necessary for the QGP formation is obtained from the kinetic energy of the colliding nuclei, through copious primary production of QCD quanta, quarks and gluons~\cite{ALICE:2022wpn}.

The QCD system formed in ultrarelativistic heavy-ion collisions is expected to experience various spacetime evolution stages: initial formation of a nearly-thermalized QGP, expansion and cooling of the QGP, transition of the QGP into a hadron resonance gas (HRG), expansion and cooling of the HRG, and finally decoupling of the HRG into non-interacting hadrons, out of which the resonances still decay before they can be detected. The dynamical expansion stages of QCD matter can be described with relativistic dissipative fluid dynamics \cite{Romatschke:2007mq, Luzum:2008cw, Bozek:2009dw, Schenke:2010rr, Schenke:2010nt, Schenke:2011bn, Song:2011qa, Niemi:2011ix, Gale:2012rq, Niemi:2012ry, Noronha-Hostler:2013gga, Niemi:2015qia, Ryu:2015vwa, Karpenko:2015xea, Giacalone:2017dud, Nijs:2020roc} which nowadays is a cornerstone in the event-by-event  analysis of heavy-ion observables.

The heavy-ion programs at the LHC and RHIC aim especially at the determination of the QCD matter properties, such as the temperature dependencies of the specific shear and bulk viscosities and other transport coefficients, from the experimental data. In practice, this can be achieved only by performing a fluid-dynamics based "global analysis", a simultaneous study of various different (low-transverse-momentum) observables from as many types of collision systems as possible. These analyses have evolved from pioneering works \cite{Niemi:2015qia, Gale:2012rq, Song:2011qa} (see also \cite{Hirvonen:2022xfv}) to those with a proper Bayesian statistical analysis and well defined uncertainty estimates  \cite{Novak:2013bqa, Bernhard:2016tnd, Bass:2017zyn, Bernhard:2019bmu, JETSCAPE:2020mzn, Nijs:2020roc, Auvinen:2020mpc, Parkkila:2021tqq, Parkkila:2021yha, Nijs:2023yab}. So far, the analyses have mainly focused on studies at mid-rapidity, where one assumes a longitudinally boost symmetric (but 3-dimensionally expanding) system described by the 2+1 D fluid dynamical equations of motion. The studies of rapidity-dependent observables requires a full 3+1 D implementation of viscous fluid dynamics~\cite{Shen:2022oyg, Bozek:2017qir, Bozek:2012qs, Pang:2018zzo, Pang:2015zrq, Pang:2012he, Denicol:2015nhu, Schenke:2011bn, Schenke:2010nt, Schenke:2010rr}. Recently, global analyses have been also extended into this direction~\cite{Auvinen:2017fjw, JETSCAPE:2023nuf, Soeder:2023vdn}. Moreover, neural networks have been developed for studying rare observables~\cite{Hirvonen:2023lqy, Hirvonen:2024ycx}.

In such global analyses, the results obtained for the QCD matter properties are strongly correlated with the the assumed fluid-dynamical initial conditions. Then, if the initial states are obtained from an ad hoc parametrization that is blind to QCD dynamics -- as is typically the case, see e.g. \cite{Bernhard:2016tnd,Bass:2017zyn, Bernhard:2019bmu, JETSCAPE:2020mzn, Nijs:2020roc, Parkkila:2021tqq, Parkkila:2021yha,Nijs:2023yab} -- it is not at all clear whether the initial densities such as the ones extracted from the global analysis could actually be realized in the studied nuclear collisions. It is therefore of paramount importance to try to study and model the QCD collision dynamics responsible for the QCD matter initial conditions. Works into this direction include the developments of the IP-Sat+MUSIC (Impact parameter dependent saturation + MUScl for Ion Collisions) model \cite{Schenke:2012wb,Gale:2012rq,Schenke:2010nt}, the EKRT (Eskola-Kajantie-Ruuskanen-Tuominen) model \cite{Eskola:1999fc,Eskola:2000xq,Paatelainen:2013eea,Niemi:2015qia}, the EPOS (Energy conservation + Parallel scattering + factOrization + Saturation) model \cite{Pierog:2009zt,Pierog:2013ria,Werner:2013tya, Werner:2023zvo,Werner:2023jps,Werner:2023fne}, the AMPT (A Multi-Phase Transport) model \cite{Pang:2014pxa, Pang:2015zrq}, and the Dynamical Core-Corona Initialization model \cite{Kanakubo:2019ogh, Kanakubo:2021qcw} with initial state generated by Pythia Angantyr \cite{Bierlich:2018xfw}, as well as initial state models such as in Refs.~\cite{Carzon:2019qja, Garcia-Montero:2023gex}.

In this work, we adopt, and significantly further develop, the pQCD and saturation -based EKRT model for computing event-by-event initial conditions of the QCD matter produced in nucleus-nucleus collisions at the LHC and at the highest RHIC energies. The leading idea in the EKRT model \cite{Eskola:1999fc,Eskola:2000xq,Paatelainen:2013eea, Niemi:2015qia} is that at sufficiently high collision energies the nucleus-nucleus collisions can be described as collisions of parton clouds rather than a collection of Glauber-model like nucleon-nucleon collisions. Then, gluons and quarks that are produced with transverse momenta ($p_T$) of the order of a few GeV, minijets, become so copiously produced \cite{Kajantie:1987pd,Eskola:1988yh} that their production processes start to overlap in the transverse coordinate space \cite{Eskola:1996ce}, which dynamically generates a saturation scale ($p_{\text{sat}}$) that suppresses softer particle production \cite{Paatelainen:2013eea}.

The original versions of the EKRT model \cite{Eskola:1999fc,Eskola:2000xq}, combined with longitudinally boost invariant 1+1 D ideal fluid dynamics, predicted successfully the LHC and RHIC hadron multiplicities and $p_T$ distributions at mid-rapidity in central collisions \cite{Eskola:1999fc,Eskola:2005ue}, and, with 2+1 D fluid dynamics, also the centrality dependence of these and of the elliptic flow coefficients ($v_2$) of the azimuth-angle asymmetries \cite{Kolb:2001qz,Niemi:2008ta}. Based on a well-defined (collinear- and infrared-safe)  pQCD calculation of minijet transverse energy production \cite{Eskola:2000my,Eskola:2000ji,Paatelainen:2013eea}, the model was extended to next-to-leading order (NLO) in Ref.~\cite{Paatelainen:2013eea}. Combined then with shear-viscous fluid dynamics, the NLO-improved EKRT model described well the centrality dependent hadron multiplicities, $p_T$ distributions and $v_2$ at mid-rapidity both at RHIC and LHC, systematically indicating a relatively low value for the QCD matter shear-viscosity-to-entropy ($\eta/s$) ratio \cite{Paatelainen:2013eea}.

An event-by-event version of the EKRT model (EbyE-EKRT) was developed in Ref.~\cite{Niemi:2015qia}. The pioneering global analysis of a multitude of LHC and RHIC bulk (low-$p_T$) observables presented in Ref.~\cite{Niemi:2015qia} demonstrated a very good overall agreement with the measurements, and resulted in improved constraints for the temperature dependence of $\eta/s$. Very interestingly (but not unexpectedly), also the Bayesian global analysis of LHC bulk observables of Ref.~\cite{Bernhard:2019bmu}, which used QCD-blind parametrized  initial states, confirmed that the initial density profiles predicted by the EbyE-EKRT \cite{Niemi:2015qia} and the IP-Sat models \cite{Schenke:2012wb} gave the best match with those obtained from the Bayesian inference.

The first attempt to perform a Bayesian global analysis of LHC and RHIC bulk observables using directly the EKRT initial states as input for the fluid-dynamics, for studying the effects of the EoS and for obtaining statistically controlled uncertainty estimates on the temperature dependence of $\eta/s$, can be found in Ref.~\cite{Auvinen:2020mpc}. The latest developments in the EKRT-initiated 2+1 D fluid-dynamics framework are a dynamically determined decoupling, which improves the description of peripheral collisions, and the inclusion of bulk viscosity. These developments are presented in Ref.~\cite{Hirvonen:2022xfv} together with a demonstration of a very good simultaneous global fit to bulk observables from various collision systems at the LHC and RHIC, and the corresponding extracted specific shear and bulk viscosities of QCD matter. Finally, the first study of how deep convolutional neural networks can be trained to predict hydrodynamical bulk observables from the EbyE-EKRT-generated energy density profiles, and how they can significantly speed up the statistics-expensive EbyE analysis of rare flow correlators especially, can be found in Ref.~\cite{Hirvonen:2023lqy}.

The predictive power of the EbyE-EKRT model originates from the underlying collinearly factorized NLO pQCD calculation. The model has been remarkably successful, especially in genuinely predicting bulk observables at mid-rapidity also for higher LHC energies, 5.02 TeV Pb+Pb collisions \cite{Niemi:2015voa}, as well as for collisions of deformed nuclei, 5.44 TeV Xe+Xe collisions at \cite{Eskola:2017bup} -- see the data comparisons e.g. Refs.~\cite{Niemi:2018ijm,Hirvonen:2022xfv}. However, there still is a number of shortcomings with the EKRT-model that need to be addressed.

First, for addressing also rapidity-dependent observables, the EbyE-EKRT initial state model should be extended to off-central rapidities and then coupled to 3+1 D viscous fluid dynamics.

Second, the average number of (or the average $E_T$ from) the parton-parton collisions is thus far in the EKRT saturation model computed as a product of a nuclear overlap function and ($p_T$ weighted) collinearly factorized integrated minijet cross section. This assumes essentially independent partonic collisions, and as discussed in Ref.~\cite{Eskola:1996ce}, especially towards larger rapidities at the LHC one easily violates the conservation of energy and baryon number. This problem clearly needs to be addressed together with the rapidity dependence.

Third, thus far in the EbyE-EKRT \cite{Niemi:2015qia}, the local fluctuations of the saturation scale, and thus of the computed energy densities, in the transverse coordinate plane are only of a geometrical origin, i.e.~they follow only from the sampled fluctuating positions of the nucleons inside the colliding nuclei. Dynamical, local EbyE fluctuations in the minijet multiplicity, inducing then further local EbyE fluctuations to the saturation scale and hence to the energy densities, should clearly be accounted for. Only by including these fluctuations can the EKRT model be relevantly applied to the studies of smaller collision systems, i.e.\ proton-nucleus and perhaps even proton-proton collisions.

Fourth, in an EbyE analysis the factorized minijet cross sections must be computed using nuclear parton distribution functions (nPDFs) that depend on the transverse position ($\bar{s}$) in each of the colliding nuclei. The spatial dependence can be modeled in terms of a power series of the nuclear thickness function, $T_A(\bar{s})$, as was done e.g. in EPS09s nPDFs \cite{Helenius:2012wd} that are used in EbyE-EKRT. The EbyE fluctuating $T_A$'s, however, often reach so large values (up to more than 3 times the largest average $T_A(0)$) that the $T_A$-applicability range of EPS09s is significantly exceeded. In EbyE-EKRT this problem was solved by an ad-hoc extrapolation of the saturation scale towards the larger values of $T_A$. Clearly, this problem is not EKRT-specific but should be addressed for the benefit of any factorized EbyE study of centrality dependence of hard processes, where spatial dependence of nPDFs is needed.

In this paper, we address these shortcomings and the arising uncertainties in solving them, for the first time in the EKRT-model framework. In particular, we introduce a completely new Monte Carlo EKRT event-generator, which we name \textit{MC-EKRT} \cite{MCEKRT}, for computing EbyE fluctuating initial states for fluid dynamics in nuclear collisions. We couple the MC-EKRT minijets to 3+1 D shear-viscous fluid dynamics \cite{Molnar:2014zha}, and discuss the various uncertainties in doing this. In this proof-of-principle paper we do not, however, aim at a full EbyE global analysis, yet, but instead study the model systematics by computing averaged initial conditions for each centrality class by summing over a large set of event-by-event MC-EKRT initial states. Running then 3+1 D shear-viscous fluid dynamics with these, we can meaningfully compare the MC-EKRT results against the measured pseudorapidity distributions of charged hadrons in different centrality classes, and also elliptic flow coefficients in semi-central collisions in Pb+Pb collisions at the LHC and Au+Au collisions at RHIC. We also study the decorrelation of eccentricities in spacetime rapidity, which was to our knowledge discussed first in \cite{Pang:2012he, Pang:2014pxa}.

The paper is organized as follows: In Sec.~\ref{S:MCEKRT} we define the MC-EKRT model framework and discuss how the previous shortcomings are solved. Section \ref{S:fluidsetup} discusses our fluid-dynamics setup, and how the 3+1 D fluid dynamics is initialized with the computed MC-EKRT minijet states. Comparisons against LHC and RHIC data, and the results for the decorrelation of eccentricities, are shown in Sec.~\ref{S:results}. Finally, conclusions and outlook are given in Sec.~\ref{S:conclusions}.

\section{Monte Carlo EKRT model setup}\label{S:MCEKRT}

Let us first see how the geometric saturation criterion that we will employ in the MC-EKRT set-up below, arises using collinearly factorized lowest-order pQCD $2\rightarrow 2$ gluonic processes as the basis and imagining the colliding nuclei as parton (gluon) clouds \cite{Eskola:1999fc, Paatelainen:2012at}. In an inelastic nucleus-nucleus collision at an impact parameter $\bar{b}_{AA}$,  the average transverse density of the number of gluon-gluon collisions that are producing minijets with $p_T$ above a cut-off $p_0$ and at rapidities $y_{1,2}$, is
\begin{eqnarray}
\frac{dN_{AA}^{2\rightarrow 2}({\bar b_{AA}})}{d^2{\bar s} }
&=&
 T_A({\bar s}_1)T_A({\bar s}_2)\frac{1}{2}\int_{p_0} dp_T^2dy_1dy_2 \cr
&\times&x_1 g(x_1,Q^2) x_2g(x_2,Q^2) \times \frac{d\hat \sigma}{d\hat t}^{2\rightarrow 2}
\end{eqnarray}
where $T_A({\bar s})$ is the standard nuclear thickness function obtained as an integral of the nuclear density over the longitudinal coordinate, ${\bar s}_{1,2}={\bar s}\pm {\bar b_{AA}}/2$
are the transverse coordinates, $g(x,Q^2)$ are the gluon PDFs, $x_{1,2} \sim p_T/\sqrt{s}$ are the longitudinal momentum fractions of the colliding gluons and $Q\sim p_T$ is the factorization/renormalization scale, $\hat t$ is a Mandelstam variable for the partonic scattering and  ${d \hat \sigma}^{2\rightarrow 2}/d\hat t \sim \alpha_s^2/p_T^4$ is the ${2\rightarrow 2}$ LO pQCD gluonic cross section.

On dimensional grounds, and ignoring the rapidity dependence, we may write for a symmetric system in central collisions \cite{Niemi:2015qia}
\begin{equation}
\frac{dN_{AA}^{2\rightarrow 2}(0)}{d^2{\bar s} } \sim (T_A x g) \times (T_A x g) \times \left(\frac{\alpha_s^2}{p_0^2}\right),
\label{eq:22scaling}
\end{equation}
where $x \sim p_0/\sqrt{s}$. Correspondingly, for $3\rightarrow 2$ processes, which can be expected to become important at small $x$, where the initial gluon densities become large, we would on dimensional grounds write, assuming here the double-PDFs from the nucleus 1 (and similarly for the other nucleus),
\begin{equation}
\frac{dN_{AA}^{3\rightarrow 2}(0)}{d^2{\bf s}}
\sim \left(T_A x g \right)^2 \times T_A x g \times \frac{\alpha_s}{p_0^2} \left(\frac{\alpha_s^2}{p_0^2}\right),
\end{equation}
where we have accounted for the extra power of $\alpha_s$ in the numerator, and for the $p_0^2$ in the denominator canceling the dimension of the extra $T_A$ there in the double-PDF. Saturation effects are expected to become dominant, and softer parton production suppressed, when
$dN_{AA}^{3\rightarrow 2} \sim dN_{AA}^{2\rightarrow 2}$,
i.e.\ when
\begin{equation}
T_A x g \sim\frac{p_0^2}{\alpha_s}.
\label{ISsat}
\end{equation}
Substituting this back to Eq.~\eqref{eq:22scaling}, and integrating over an effective nuclear transverse area $\pi R_A^2$ ($R_A$ being the nuclear radius), gives the geometrical EKRT scaling law, introduced in Ref.~\cite{Eskola:1999fc}
\begin{equation}
N_{AA}^{2\rightarrow 2}(0)\frac{\pi}{p_0^2} \sim \pi R_A^2,
\label{E:EKRTsat}
\end{equation}
where $\pi/p_0^2$ can be interpreted as a transverse formation-area for a produced dijet \cite{Eskola:1996ce,Eskola:1999fc}. Thus, the minijet production saturates when the minijet production processes fill the available transverse area in the nuclear collision.

In the MC-EKRT set-up introduced below, we will take the above geometric interpretation of saturation as our starting point, when deciding on an event-by-event and on a parton-by-parton basis, whether the produced minijet system becomes locally saturated. With the above  discussion, we would also like to emphasize  that saturation in the EKRT model is \textit{not} fusion of produced final-state gluons, but saturation of the minijet production processes themselves.

Our MC-EKRT simulation of a nucleus-nucleus ($A$+$B$) collision proceeds through the following steps, each of which will be discussed in more detail in this and the following sections.

\noindent\textbf{1.}
Sample the positions of the nucleons in $a\in A$ and $b\in B$ from the Woods-Saxon distribution, keeping track of the proton/neutron identity of each nucleon (Sec.~\ref{SS:ABconfig}).

\noindent\textbf{2.}
Sample the impact parameter for the $A$+$B$ collision similarly as in the MC Glauber model (Sec.~\ref{SS:impactparam}), and check whether the chosen trigger condition for the $A$+$B$ collision is fulfilled. If it is not, start again from  item 1 (Sec.~\ref{SS:triggering}).

\noindent\textbf{3.}
If the $A$+$B$ collision is triggered, find all the binary $ab$ pairs of nucleons, $a\in A$ and $b\in B$. Then go through the generated list of the $ab$ pairs and regard each $ab$ pair as a possible independent source of multiple minijet production. Sample the number of produced minijet pairs, dijets, for each $ab$ pair from a Poissonian probability distribution (Sec.~\ref{SSS:mpledijets}).

\noindent\textbf{4.}
For each produced dijet, sample the parton flavors and momenta from collinearly factorized LO pQCD cross sections (Sec.~\ref{SSS:dijetchemi}), using nuclear PDFs that depend on the transverse positions of $a\in A$ and $b\in B$ (Sec.~\ref{SSS:snPDFs}). For quark-initiated processes, decide (sampling the LO pQCD cross sections) whether the colliding quarks are valence quarks or sea quarks (Sec.~\ref{SSS:dijetchemi}). Sample also the transverse production point for each dijet from a Gaussian overlap function for each nucleon pair $ab$ (Sec.~\ref{SSS:mpledijets}).

\noindent\textbf{5.}
Consider all the generated dijets as candidates for the final minijet-state of this $A$+$B$ event. For filtering away the excess (unphysical) dijets, order the dijet candidates according to the transverse momentum $p_T$ of the minijets forming the dijet (Sec.~\ref{SS:filtering}).

\noindent\textbf{6.}
Filter the excess dijets in the order of decreasing $p_T$, by imposing a local geometric EKRT saturation criterion (cf.\ Eq.~\eqref{E:EKRTsat}). If a dijet gets filtered, both final-state partons are removed (Sec.~\ref{SS:filtering}).

\noindent\textbf{7.}
Filter the surviving dijets further by imposing conservation of energy and valence quark number for each nucleon, doing the filtering again in the order of decreasing $p_T$. Optionally, this filtering step can be ignored, or chosen to be done simultaneously with the dijet filtering in step 6 (Sec.~\ref{SS:filtering}).

\noindent\textbf{8.}
Collect the MC-EKRT minijet output data for the surviving dijets: the ${\bf p_T}$ vector, the rapidity, and the flavour of each minijet, along with the transverse location of each dijet's formation point, to be used in Sec.~\ref{S:initialization}.  Order the $A$+$B$ events according to the total minijet $E_T$ (a scalar sum of minijet $p_T$'s) for the centrality selection (Sec.~\ref{SS:centrality}).

A separate interface is then developed to initialize fluid dynamics, with the following steps:

\noindent\textbf{9.}
Propagate the surviving minijets as free particles to the proper time surface $\tau_0 = 1/p_0$, assuming that minijets with momentum rapidity $y$ move along the corresponding spacetime rapidity $\eta_s=y$. The parameter $p_0$ here is the smallest partonic $p_T$ allowed in the pQCD cross sections for the dijet candidates (Sec.~\ref{SS:freestreaming}).

\noindent\textbf{10.}
Feed the minijets into 3+1 D fluid dynamics as initial conditions at $\tau_0$: At each $\eta_s$ and transverse-coordinate grid cell, using a Gaussian smearing, convert the minijet transverse energy $E_T$ into a local energy density (Sec.~\ref{SS:smearing}).

\noindent\textbf{11.}
Run 3+1 D viscous fluid dynamics with these minijet initial conditions, in principle event by event. Note, however, that in the present exploratory study we are testing the model setup using averaged initial states for each centrality class (Sec.~\ref{SS:iniaveragin}). We do not couple the fluid dynamics with a hadron cascade afterburner but run fluid dynamics until the freeze-out of the system. Resonance decays are accounted for, as usual (Sec.~\ref{SS:IS_fluid}).

\noindent\textbf{12.}
Form the observables for which statistics is collected (Sec.~\ref{S:results}).

Next, we look at the above steps in more detail, and also specify the few parameters that the MC-EKRT minijet event generator has.

\subsection{Nucleon configurations of $A$ and $B$}
\label{SS:ABconfig}

First, we construct the nucleon structure of the colliding nuclei. Here, we essentially follow the procedure nowadays standard in the Monte Carlo Glauber approach \cite{Loizides:2017ack}. The distributions of the positions of the nucleons are taken to follow the nuclear charge densities extracted from low energy electron scattering experiments~\cite{DeJager:1974liz,DeVries:1987atn}. The lead nucleus, $\textrm{Pb}^{208}$ (used at the LHC) is assumed perfectly spherical, and as the gold nucleus $\textrm{Au}^{197}$ (used at RHIC) is also nearly spherical, the current version of the MC-EKRT assumes spherically symmetric nuclei $A$ and $B$.  Thus, the azimuthal angle $\phi\in\left[0,2\pi\right]$ and the cosine of the polar angle $\cos\theta\in\left[-1,1\right]$ are sampled from a uniform distribution, while the radial coordinate $r$ is sampled from the two-parameter Fermi (2pF) distribution, the Woods-Saxon distribution \cite{Woods:1954zz},
    \begin{equation}\label{E:2pF}
        \rho_{\text{WS}}(r)=\frac{\rho_0}{1+\exp\left(\frac{r-R_A}{d}\right)},
    \end{equation}
where $R_A$ is the nuclear radius and $d$ is the diffusion parameter. For the lead and gold nuclei we study here, $(R_A,d) = (6.624,\,0.550)\, \text{fm}$ and $(6.380,\,0.535)\, \text{fm}$, correspondingly \cite{DeJager:1974liz}.  The normalization constant $\rho_0$ is fixed by requiring the volume integral of $\rho_{\text{WS}}(r)$ to give $A$, but in the simulation here $\rho_0$ has no effect. The nuclei which have nucleons with positions closer to each other than $d_{\text{min}}=0.4\;$fm, are discarded and sampled again. The introduction of an exclusion radius $d_{\text{min}}$ is known to slightly deform the radial density profile \cite{Loizides:2014vua,Loizides:2017ack}, but we neglect this small effect here.

\subsection{Impact parameter sampling}
\label{SS:impactparam}

\begin{figure}[htb!]
        \includegraphics[width = 0.45\textwidth]{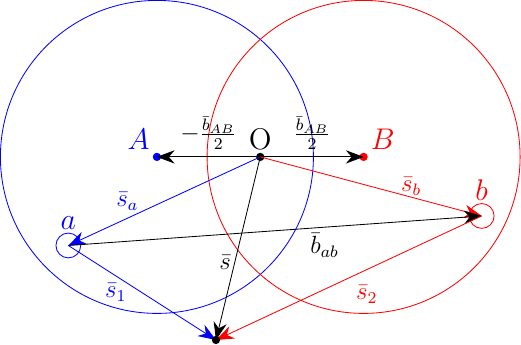}
\caption{Projection of the $A$+$B$ collision system into the transverse plane. The impact parameter vector $\bar{b}_{AB}$, extending from the center-of-mass of the nucleus $A$ to the center-of-mass of the nucleus $B$, is along the $x$-axis (axes are not shown), with the origin O in the middle. The location vectors of the nucleons $a\in A$ and $b\in B$ are $\bar{s}_a$ and $\bar{s}_b$, respectively. The impact parameter between the nucleons $a$ and $b$ is $\bar{b}_{ab}$.
}
        \label{F:Coordinates}
    \end{figure}

Next, the squared impact parameter, $\bar b_{AB}^2$, for the $A$+$B$ collision is sampled from a uniform distribution. As long as the colliding nuclei are spherically symmetric on the average, we do not need to randomly rotate the nuclei. We can fix the impact parameter vector, as a vector in the transverse $(x,y)$ plane, to be on the $x$-axis, pointing from the nucleus $A$ to the nucleus $B$ -- see Fig.~\ref{F:Coordinates}.

Once the positions of the nucleons in each nucleus -- $\{\bar s_a\}$ in $A$ and $\{\bar s_b\}$ in $B$ -- have been determined, the center of the mass of the projectile nucleus $A$ is shifted to $(-b_{AB}/2,0)$ and that of the target nucleus $B$ to $(b_{AB}/2,0)$, thus fixing the origin O of the collision frame. Finally, $Z_A\, (Z_B)$ of the nucleons in $A\, (B)$ are randomly labeled as protons and the rest as neutrons, i.e. we neglect possible effects arising from the differences of proton and neutron density distributions (such as a neutron skin), in this study.

\subsection{Trigger condition for the $A$+$B$ collision}
\label{SS:triggering}

Next, our simulation checks whether an inelastic collision between the generated nucleon configurations $A$ and $B$ takes place. We devise the trigger condition for the $A$+$B$ collision as follows: Assuming a hard-sphere scattering of two nucleons, $a\in A$ and $b\in B$, with a cross section $\sigma_{\text{trig}}^{ab}(s_{NN})$ at a nucleon-nucleon center-of-momentum system (CMS) energy $\sqrt{s_{NN}}$, an $A$+$B$ collision takes place if for at least one of the $ab$ pairs the squared transverse distance between $a$ and $b$ does not exceed ${\sigma_{\text{trig}}^{ab}(s_{NN})/\pi}$. In terms of the transverse-coordinate vectors introduced in Fig.~\ref{F:Coordinates}, with nucleons $a\in A$ and $b\in B$, their transverse positions at $\bar{s}_a$ and $\bar{s}_b$, and impact parameters $\bar {b}_{ab}$, the triggering condition for the $A$+$B$ collision is fulfilled if at least for one $ab$ pair
\begin{equation}
|\bar {b}_{ab}| =  |\bar{s}_b - \bar{s}_a| \leq  \sqrt{\sigma_{\text{trig}}^{ab}(s_{NN})/\pi}.
\end{equation}
If the above condition is not met, new nucleon configurations $A$ and $B$, and a new impact parameter $b_{AB}^2$  are generated. For the triggering cross section $\sigma_{\text{trig}}^{ab}(s_{NN})$ we use the inelastic nucleon-nucleon cross section $\sigma^{NN}_{\text{inel}} (s_{NN})$, calculated as
\begin{equation}
\sigma^{NN}_{\text{inel}} (s_{NN}) = \sigma^{NN}_{\text{tot}} (s_{NN}) - \sigma^{NN}_{\text{el}} (s_{NN}),
\end{equation}
where the total cross section $\sigma^{NN}_{\text{tot}}$ is obtained from a fit by COMPETE  \cite{COMPETE:2002jcr},
\begin{align}
\sigma^{NN}_{\text{tot}}(s_{NN})/\text{mb} &= 42.6s^{-0.46} -33.4s^{-0.545} \nonumber \\
				                                      & +0.307\log^2(s/29.1) +35.5,
\end{align}
and the elastic cross section from a fit by TOTEM \cite{TOTEM:2017asr},
\begin{equation}
        \sigma^{NN}_{\text{el}} (s_{NN})/\text{mb} = -1.617\log(s) +0.1359\log(s)^2 +11.84,
\end{equation}
with $s=s_{NN}/\text{GeV}^2$. For the CMS energies $\sqrt{s_{NN}}=$ 5020, 2700, 200 GeV, which we study here, this gives $\sigma^{NN}_{\text{inel}} (s_{NN}) =$ 69.14, 62.96, 41.78~mb, correspondingly.

We emphasize that $\sigma^{NN}_{\text{inel}}$ is here used only for the triggering of the nuclear collision, i.e.\ for determining the inelastic $A$+$B$ cross-section. It does not play any other role in what follows.

\subsection{Multiple dijet production}
\label{SS:nn}

\subsubsection{Probability distribution and nucleon thickness function}
\label{SSS:mpledijets}

If the trigger condition is fulfilled, the collision between $A$ and $B$ takes place.  The $A$+$B$ collision here is assumed to be a very high-energy one, and furthermore a collision of two large parton clouds, which are originating from the sampled nucleons and extending around the Lorentz contracted nuclei. In this case, the multiple minijets originating from each $ab$ pair are produced practically instantaneously around $z\sim 0$, and simultaneously everywhere in the transverse plane.

At this stage of our setup, all the $ab$ pairs can be considered to be fully independent from each other, they just divide the interaction of the two large nuclear parton-clouds into $ab$ contributions. Saturation and energy conservation, which will here be imposed in the order of decreasing minijet $p_T$, do not depend on the ordering of the $ab$ pairs, either. Thus, in our setup the ordering of the $ab$ pairs becomes irrelevant \footnote{Note, however, that if one models nuclear collisions as subsequent energy-conserving NN subcollisions (like e.g.\ in HIJING \cite{Wang:1991hta}), then the ordering (randomization) of the $ab$ pairs would be important.}.

Next, all the $ab$ nucleon pairs will be considered as potential sources for multiple minijet (dijet) production. In each $ab$ contribution, the candidate dijets are supposed to be produced independently from each other, hence Poissonian statistics is used in sampling the number of produced dijets. Then, the probability of producing $n\ge 0$ independent dijets from the pair $ab$, where the locations of $a$ and $b$, in the fixed nucleon configurations of this event, are $\bar s_a$ and $\bar s_b$, correspondingly, and whose impact parameter thus is $\bar{b}_{ab}=\bar s_b-\bar s_a$, is
\begin{equation}
        P_n(\{\bar s_a\},\{\bar s_b\},p_0,\sqrt{s_{NN}}) = \frac{\left(\bar N_\text{jets}^{ab}\right)^n}{n!} \text{e}^{-\bar N_\text{jets}^{ab}}, \label{E:Poisson}
\end{equation}
where the average number of dijets produced from the pair $ab$ is
\begin{eqnarray}
\bar N_\text{jets}^{ab}&\equiv& \bar N_\text{jets}^{ab}(p_0,\sqrt{s_{NN}},\{\bar s_a\},\{\bar s_b\})\label{N_a}\\
&=& T_{NN}(\bar{b}_{ab})\, \sigma^{ab}_{\text{jet}}(p_0,\sqrt{s_{NN}},\{\bar{s}_a\},\{\bar{s}_b\}) \label{N_b}
\label{E:barN}
\end{eqnarray}
where $\sigma_{\text{jet}}^{ab}$ is an integrated inclusive pQCD cross section for producing a pair of minijets with transverse momenta $p_T\ge p_0$ and any rapidities (details of obtaining $\sigma_{\text{jet}}^{ab}$ will be explained in Sec.~\ref{SSS:dijetchemi}), and with the notation $\{\bar{s}_a\}$ ($\{\bar{s}_b\}$) we underline that the computed pQCD cross section here depends both on the location $\bar{s}_a$ ($\bar{s}_b$) of the nucleon $a\in A$ ($b\in B$) and on the positions of all other nucleons in the nucleon configuration forming the nucleus $A$ ($B$) in each event. Above, $T_{NN}$ is the nucleon-nucleon overlap function,
\begin{eqnarray}
T_{NN}(\bar{b}_{ab})
&=& \int \!\text{d}^2 s\; T_{N}(\bar s- \bar{s}_a) T_{N}(\bar{s}-\bar{s}_{b}) \label{TNN_a}\\
&=& \int \!\text{d}^2 s_1\; T_{N}(\bar{s}_1) T_{N}(\bar{s}_1-\bar{b}_{ab}) \label{TNN_b}
\label{E:T_NN}
\end{eqnarray}
where the transverse vectors $\bar s - \bar{s}_a \equiv \bar{s}_1$  and $\bar{s}-\bar{s}_{b} = \bar{s}_1-\bar{b}_{ab}\equiv \bar{s}_2$ measure the transverse distance from the centers of the nucleons $a\in A$ and $b\in B$, correspondingly, see Fig.~\ref{F:Coordinates}. Here, $T_{N}$ is the nucleon thickness function, which is obtained from the spatial density distribution  $\rho_N$ as
\begin{equation}
        T_{N}(\bar{s}) = \int_{-\infty}^{\infty} dz \rho_N(\textbf{x}),
\end{equation}
where $\textbf{x} = (\bar{s},z)$. Both $T_N$ and $T_{NN}$ are normalized to one through the transverse integrals,
\begin{eqnarray}
   \int \!\text{d}^2 s\,  T_{N}(\bar{s}) &=& \int d^3 x\, \rho_N(\textbf{x}) = 1,\\
	\int \!\text{d}^2 s\,  T_{NN}(\bar{s}) &=& 1.
\end{eqnarray}

It should also be emphasized that in writing Eq.~\eqref{N_a} into the form of Eq.~\eqref{N_b}, we are assuming that the PDFs carry spatial dependence in that they do (quite strongly) depend on the locations $\bar s_a$ of $a\in A$ and $\bar s_b$ of $b\in B$, as well as on the positions of all the other nucleons in $A$ and $B$ (which all are fixed for one $A$+$B$ collision event), but that for each nucleon $a\in A$  and $b\in B$ we have fixed PDFs that do not depend on the variable $\bar s$ appearing in Eq.~\eqref{TNN_a}. This allows us to factorize the nucleon-nucleon overlap function $T_{NN}$ from the minijet cross section $\sigma^{ab}_{\text{jet}}$ in Eq.~\eqref{N_b}.

Following Ref.~\cite{Niemi:2015qia}, we extract $\rho_N$, and thereby $T_{N}$, from exclusive $J/\psi$ photo-production cross sections that have been measured in $\gamma + p \rightarrow J/\psi + p$ collisions at HERA~\cite{ZEUS:2002wfj}. As discussed e.g. in Ref.~\cite{Eskola:2022vpi}, the amplitude of this process is proportional to generalized parton distribution functions (GPDs) and a two-parton form factor $F_N(t)$ that depends on the Mandelstam variable $t$ and is linked to $\rho_N$ via a 3D Fourier transform,
\begin{equation}
F_N(t) = \int d^3x \, e^{i\textbf{q}\cdot \textbf{x}} \rho_N(\textbf{x}),
\end{equation}
where $|\textbf{q}|^2 = |t|$, and $F_N(0)=1$. As the GPDs become ordinary PDFs at the forward limit, and as the $J/\Psi$ mass scale is of the same order of magnitude as the dominant minijet $p_T$ scale, the above $\rho_N$ should to a good approximation describe also the corresponding partonic spatial density related to the PDFs we use here. The measured HERA cross sections show a behavior $d\sigma/dt \propto e^{-b|t|}$, with a slope parameter $b$ that depends on the photon-proton system c.m.s. energy $W$ as
\begin{equation}
b/\text{GeV}^{-2} = b_0 + 4\alpha'_P\log\left(\frac{{W}}{W_0}\right), \label{E:b_param}
\end{equation}
where $b_0$, $\alpha'_P$ and $W_0$ are constants.  Here, identifying $W=\sqrt{s_{NN}}$, our default choice  is the parametrization from Ref.~\cite{Flett:2021xsl} (also used in \cite{Eskola:2022vpi}), with $b_0=4.9$, $\alpha'_P=0.06$ and $W_0=90\;\text{GeV}$. Then, an inverse Fourier transform of $F_N(t)=\text{exp}({-b|t|/2})= \text{exp}(-b |\textbf{q}|^2/2)$ results in a 3D Gaussian density,
\begin{eqnarray}
\rho_N(\textbf{x})
&=& \int\frac{d^3q}{(2\pi)^3}   e^{-i\textbf{q}\cdot \textbf{x}} F_N(t)\\
&=& \left(\frac{1}{2\pi\sigma_N^2}\right)^{3/2} \exp\left(-\frac{|\textbf{x}|^2}{2\sigma_N^2}\right),
\end{eqnarray}
and a 2D Gaussian thickness function,
\begin{equation}
T_{N}(\bar{s})=\frac{1}{2\pi\sigma_N^2} \exp\left(-\frac{|\bar s|^2}{2\sigma_N^2}\right), \label{E:TN}
\end{equation}
with a width parameter $\sigma_N^2 \equiv b$. With the parametrization (\ref{E:b_param}), we have $\sigma_N = 0.478$ ($0.472$)~fm,  at $\sqrt{s_{NN}}= 5.02\, (2.76)$ TeV at the LHC, and $\sigma_N = 0.445$~fm for $\sqrt{s_{NN}}=200$ GeV at RHIC.

Then, with the Gaussian forms for $T_{N}$, also the nucleon-nucleon overlap function in Eq.~(\ref{E:T_NN}) can be expressed in a closed form, which also becomes a Gaussian,
\begin{equation}
T_{NN}(\bar{b}_{ab})=\frac{1}{4\pi\sigma_N^2} \text{exp}\left({-\frac{|\bar b_{ab}|^2}{4\sigma_N^2}}\right). \label{E:T_NN_gauss}
\end{equation}

Once the number of the independent dijet candidates has been sampled, each dijet candidate is assigned a spatial production point $\bar s$ that is sampled from the product distribution $T_N(\bar s -\bar s_a)T_N(\bar s - \bar s_b)$.

The modeling here is inspired by the eikonal minijet models \cite{Wang:1990qp,Durand:1987yv} which are high-energy limits of potential scattering, but we emphasize the different roles of the parameter $p_0$ in these models. In MC-EKRT, the impact parameter integral of the eikonal $1 - P_0 = 1 - \exp(-T_{NN}(b) \sigma_{\text{jet}}(p_0))$ is \textit{not} normalized to an inelastic NN cross section $\sigma^{NN}_{\text{inel}} (s_{NN})$ but is allowed to obtain larger values. Instead, the parameter $p_0$ needs to be chosen so small, of the order 1 GeV, that minijets are produced so abundantly that they overfill the coordinate space, so that saturation can become effective in regulating the smallest-$p_T$ minijet production. For this reason, our results are also fairly \textit{insensitive} to the value of $p_0$, unlike typically in the eikonal minijet models. Notice also that as we extend the value of $p_0$ to unphysically low values (but still keeping it in the pQCD region, $p_0\gg\Lambda_\text{QCD}$), and since we are considering the earliest moments in the collision, $\tau\lesssim 1/p_0$, we do not include any soft particle production component, but consider only the (semi)hard (mini)jet production in what follows.

\subsubsection{Dijet kinematics and parton chemistry}
\label{SSS:dijetchemi}

A key element in our MC-EKRT framework is the differential LO pQCD cross section of hard parton production \cite{Eichten:1984eu,Sarcevic:1988tu}
\begin{widetext}
\begin{equation}\label{E:diffSigmaJet}
        \frac{\text{d}\sigma^{ab}_{\text{jet}}(\{\bar{s}_a\},\{\bar{s}_b\})}{\text{d}p^2_{T}\text{d}y_1\text{d}y_2} = K\sum\limits_{ij\langle kl\rangle}x_1f_i^{a/A}(\{\bar{s}_a\},x_1,Q^2)\, x_2f_j^{b/B}(\{\bar{s}_b\},x_2,Q^2)\,\frac{\text{d}\hat{\sigma}^{ij\rightarrow kl}}{\text{d}\hat{t}}\left(\hat{s},\hat{t},\hat{u}\right),
    \end{equation}
\end{widetext}
where  $y_1$ and $y_2$ are the rapidities of the two final-state partons, $p_T$ is the transverse momentum of each of them, $f_i^{a/A}(\{\bar{s}_a\},x_1,Q^2)$ ($f_i^{b/B}(\{\bar{s}_b\},x_2,Q^2)$) is the nucleon-configuration-specific PDF of a parton flavor $i$ ($j$) of the bound nucleon $a\in A$ ($b\in B$) which is centered at $\bar s_a$ ($\bar s_b$) in the nucleon configuration of each event, and $x_1$ ($x_2$) is the parton's longitudinal momentum fraction, $Q^2$ is the factorization/renormalization scale which we set equal to $p_T$, and $\frac{\text{d}\hat{\sigma}^{ij\rightarrow kl}}{\text{d}\hat{t}}$ are the differential LO pQCD cross sections, which depend on the parton-level Mandelstam variables $\hat{s}$, $\hat{t}$, and $\hat{u}$. The notation $\langle kl\rangle$ indicates a sum over pairs of final-state partons, so that, say, $u_1g_2\rightarrow ug$ and $u_1g_2\rightarrow gu$ are the same process and hence are not to be counted as two separate ones, whereas $u_1g_2\rightarrow ug$ and $g_1u_2\rightarrow ug$ naturally are two different processes as the initial-state partons originate from different nucleons. Notice also that since we aim to follow the partons' identities as well, we do not introduce any $t,u$-symmetrized cross sections which are often used when observable jet cross sections are studied. In the present exploratory study, in the interest of the simulation speed and as there anyways are various other uncertainties and scale dependence present, we do not (yet) attempt to perform an NLO calculation similar to that in \cite{Eskola:2000my, Eskola:2000ji} but account for the missing higher order terms simply by a $K$-factor that is a constant for a fixed $\sqrt{s_{NN}}$ and that will be fitted to the $A$+$A$ data separately at the LHC and at RHIC.  Then, in LO, the momentum fractions can be expressed in terms of the transverse momentum $p_T$ and rapidities of each minijet as
\begin{equation}
x_{1,2} = \frac{p_T}{\sqrt{s_{NN}}}\left(\text{e}^{\pm y_1}+\text{e}^{\pm y_2}\right),
\end{equation}
and the Mandelstam variables become
    \begin{align}
        \hat{s}&=2p_T^2\left(1+\cosh(y_1-y_2)\right),\\%[0.5em]
        \hat{t}&=-p_T^2\left(1+\text{e}^{-(y_1-y_2)}\right), \\%[0.5em]
        \hat{u}&=-p_T^2\left(1+\text{e}^{+(y_1-y_2)}\right).
    \end{align}

Once the spatially dependent nuclear PDFs (PDFs of nucleons $a$ and $b$) have been devised (see discussion below),  Eq.~(\ref{E:diffSigmaJet}) can be integrated over the momentum phase space, to give the minijet cross section $\sigma^{ab}_{\text{jet}}(p_0,\sqrt{s_{NN}},\{\bar{s}_a\},\{\bar{s}_b\})$ which is employed in Eq.~(\ref{E:Poisson}). Explicitly, accounting for the symmetry factors for the identical final-state partons, we have
\begin{widetext}
\begin{equation}\label{E:SigmaJet}
\sigma^{ab}_{\text{jet}}(p_0,\sqrt{s_{NN}},\{\bar{s}_a\},\{\bar{s}_b\})
=
K \int dp_T^2 dy_1dy_2 \sum\limits_{ij\langle kl\rangle} \frac{1}{1+\delta_{kl}} x_1f_i^{a/A}(\{\bar{s}_a\},x_1,Q^2)\, x_2f_j^{b/B}(\{\bar{s}_b\},x_2,Q^2)\,\frac{\text{d}\hat{\sigma}^{ij\rightarrow kl}}{\text{d}\hat{t}}\left(\hat{s},\hat{t},\hat{u}\right),
\end{equation}
\end{widetext}
where, assuming a fixed lower limit $p_0=1$~GeV for $p_T$, the integration limits become
\begin{eqnarray}
p_0\le p_T\le\sqrt{s_{NN}}/2,\quad
|y_1|\le \text{arcosh}(1/x_T),\\
-\log(2/x_T - e^{-y_1})\le y_2\le \log(2/x_T-e^{y_1}),
\end{eqnarray}
with $x_T=2p_T/\sqrt{s_{NN}}$.

With these elements, the dijet kinematics and parton chemistry can be straightforwardly  generated. Once the number of independent dijets from an interaction of nucleons $a\in A$ and $b\in B$ has been determined using the Poissonian probabilities of Eq.~\eqref{E:Poisson}, the transverse momentum $p_T$
and rapidities $y_1,\, y_2$ of each (mini)jet in the dijet are obtained with rejection sampling from the differential minijet cross section (integrand) of Eq.~\eqref{E:SigmaJet}. With the fixed kinematic variables, we then sample Eq.~\eqref{E:SigmaJet} again for the parton process type that fixes the flavors of the participating partons. If the parton process involves a quark from either $a$ or $b$,  we also identify each participating quark as a sea quark or as a valence quark again on the basis of Eq.~\eqref{E:SigmaJet} (i.e.\ the PDFs, in this case, $(f^{a/A}_q - f^{a/A}_{\bar{q}})/f^{a/A}_q$ being the probability for obtaining a valence quark). Finally, one minijet in each dijet is assigned an azimuth angle $\phi$ from a flat distribution and its partner is then an angle $\pi$ apart in the $2\rightarrow 2$ kinematics assumed here.

\subsubsection{EbyE fluctuating spatial nuclear PDFs}
\label{SSS:snPDFs}

Systematic global analyses of collinearly factorized nuclear PDFs (nPDFs) indicate that bound-nucleon PDFs clearly differ from the free-proton PDFs, see e.g. Refs.~\cite{Eskola:2009uj,Eskola:2016oht,Eskola:2021nhw,Helenius:2021tof,AbdulKhalek:2022fyi,Duwentaster:2022kpv}. The resulting nuclear modifications in the bound-proton PDFs $f_i^{p/A}$ can be quantified with
\begin{equation}
f_i^{p/A}(x,Q^2) = R_i^{p/A}(x,Q^2)f_i^{p}(x,Q^2),
\label{E:R_iA}
\end{equation}
where $i$ denotes the parton flavor, $f_i^{p}$ is the free-proton PDF and $R_i^{p/A}$ is the nuclear modification. The corresponding neutron PDFs are obtained using isospin symmetry. The above PDFs and their modifications are, however, \textit{spatial averages} of the nPDFs, they do not account for the dependence of the nuclear density and especially not its fluctuations, i.e.\ for the fact that in the lowest-density regions the nuclear effects should vanish whereas in the high-density regions they should be larger than in the average $R_i^{p/A}$. These spatial effects can become significant especially in the small-$x$ region relevant for lowest-$p_T$ minijet production of interest here, hence they are an important contributing factor in computing hydrodynamic initial density profiles that directly influence the centrality dependence of observables like multiplicities and flow coefficients. Therefore, in an EbyE simulation such as MC-EKRT here, we cannot use the spatially averaged nPDFs but need to introduce \textit{EbyE-fluctuating} spatially dependent nPDFs (snPDFs), where the nuclear modifications are sensitive to the nucleon-density fluctuations from event to event. As we will discuss below, this turns out to be a non-trivial problem in an EbyE simulation where there are large density fluctuations present.

\begin{figure}[b]
        \includegraphics[width = 0.48\textwidth]{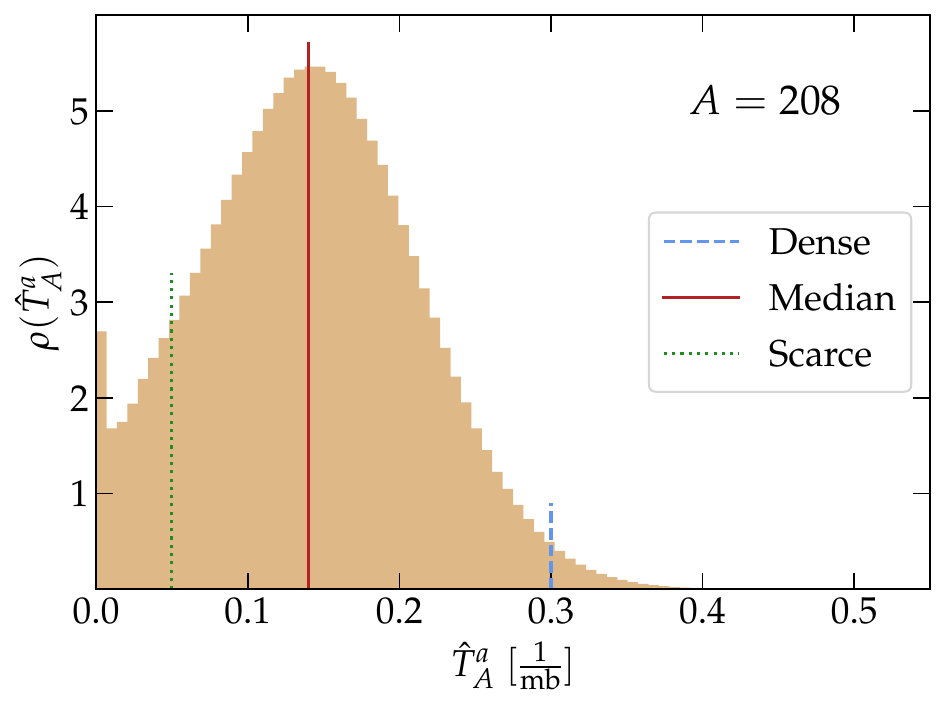}
        \caption{Normalized distribution of the average nuclear thickness function $\hat{T}_{A}^{a}$ experienced by a nucleon $a\in A$, defined in Eq.~\eqref{E:TAHat}, for the nucleus Pb$^{208}$, sampled from 10 000 nuclei. For comparison, optical Glauber $T_A^{\text{WS}}(0)\approx 0.212/\text{mb}$. The vertical lines indicate the example-density regions to which we refer as ``dense'' (dashed blue line), ``median'' (solid red line) and ``scarce'' (dotted green line). The peak at the smallest values of $\hat{T}_{A}^{a}$ arises at the edge of the nucleus where the inter-nucleon distance becomes larger than the nucleonic width $\sigma_N$.
				}
        \label{F:TAHistogram}
    \end{figure}

Originally, our idea was to directly utilize the available non-fluctuating snPDFs, such as EPS09s \cite{Helenius:2012wd}, where the nuclear modifications are encoded in as a power series of the average (optical Glauber) nuclear thickness function,  $T_A^{\text WS}(\bar s) = \int dz\, \rho_{\text{WS}}(\textbf{x})$, as follows:
\begin{equation}
f_i^{p/A}(\bar{s},x,Q^2)
= f_i^{p}(x,Q^2) r_i^{p/A}(\bar{s},x,Q^2),
\label{E:snPDF_ave}
\end{equation}
where $f_i^{p}$ again are the free-proton PDFs, and the nuclear modification part,
\begin{equation}
r_i^{p/A}(\bar{s},x,Q^2) = 1+ \sum_{n=1}^{4}c^i_n(x,Q^2)[T_A^{\text{WS}}(\bar{s})]^n,
\label{E:snPDF_r}
\end{equation}
where the coefficients $c^i_n$ are $A$-independent, is normalized to the known (EPS09 \cite{Eskola:2009uj}) average nuclear modifications,
\begin{equation}
R_i^{p/A}(x,Q^2) = \frac{1}{A}\int d^2s\, T_A^{\text{WS}}(\bar s) r_i^{p/A}(\bar{s},x,Q^2).
\label{E:snPDF_norm}
\end{equation}
Alternatively, as done e.g. in Refs.~\cite{Eskola:1991ec, Klein:2003dj,Vogt:2004hf}, one could in the interest of the simulation speed truncate the above power series at the second term, allow some residual $A$ dependence in the remaining single coefficient, and obtain
\begin{equation}
f_i^{p/A}(\bar{s},x,Q^2)
= f_i^{p}(x,Q^2) \left[ 1+ c^i_A(x,Q^2)T_A^{\text{WS}}(\bar{s}) \right],
\label{E:snPDF_truncated}
\end{equation}
where again the normalization to the average modifications $R_i^{p/A}(x,Q^2)$ would give
\begin{equation}
c^i_A(x,Q^2) = \frac{A}{T_{AA}(0)}\left( R_i^{p/A}(x,Q^2) - 1 \right),
\end{equation}
with $T_{AA}(0) = \int d^2s\, [T_A^{\text{WS}}(\bar s)]^2$. Then, with the nuclear density fluctuations present in an EbyE simulation, one could essentially just replace the average $T_A^{\text{WS}}$ by the fluctuating $T_A(\bar s) = \sum_a T_N(\bar s-\bar s_a)$, where $T_N$ is the Gaussian density from Eq.~\eqref{E:TN}. This procedure does \textit{not}, however, work, because in practice the maximal density at which the above approaches are applicable is the maximum of the average density \cite{Eskola:1988yh}, $T_A^{\text{WS}}(0)=2\rho_0d\log\left(1+\text{e}^{R_A/d}\right)$, and now with fluctuations we encounter densities that easily exceed this (see Fig.~\ref{F:TAHistogram} ahead), and can be even more than $3T_A^{\text{WS}}(0)$.

In particular with the latter approach above, in the small-$x$ nuclear shadowing region, where $R_i^{p/A}(x,Q^2)<1$ and thus $c^i_A(x,Q^2)<0$, when a negative $c^i_A$ is accompanied by a large enough $T_A(\bar s)$, the spatial PDFs become negative, which cannot be allowed in LO. A possible cure for this could be to introduce an exponentiated ansatz for the above power series (motivated by Ref.~\cite{Frankfurt:2011cs}),
\begin{equation}
1+ c^i_A(x,Q^2)T_A(\bar{s}) \rightarrow \text{exp}(c^i_A(x,Q^2)T_A(\bar{s})). \label{E:ansatz1}
\end{equation}
However, with density fluctuations, in the region where $T_A(\bar{s})\gg T_A^{\text{WS}}(0)$, also this form leads to too fast attenuating small-$x$ parton densities in that the density function $T_A(\bar{s})\text{exp}(c^i_A(x,Q^2)T_A(\bar{s}))$ (whose $\bar s$-integral is normalized to $R_i^{p/A}(x,Q^2)$), is not a monotonically rising function of $T_A(\bar{s})$ contrary to what it should be. This problem can be solved by using an another ansatz function, such as
\begin{equation}
1+ c^i_A(x,Q^2)T_A(\bar{s}) \rightarrow 1/(1 - c^i_A(x,Q^2)T_A(\bar{s})) \label{E:ansatz2}
\end{equation}
instead, which, when multiplied by $T_A(\bar{s})$, conveniently gives a positive-definite function that is monotonously rising with $T_A(\bar{s})$. In the antishadowing region where $c^i_A(x,Q^2)>0$, and where the $A$-dependence of the nuclear modification is modest in any case,  such a function would at large $T_A$'s lead to violation of the per-nucleon momentum sum rule that is assumed in the global PDF analyses. We have tested that this problem can be solved approximately (conserving momentum on a percent level) by choosing a more modestly increasing logarithmic function
\begin{equation}
1+ c^i_A(x,Q^2)T_A(\bar{s}) \rightarrow 1+\log(1+c^i_A(x,Q^2)T_A(\bar{s})). \label{E:ansatz3}
\end{equation}
Equations \eqref{E:ansatz2} and \eqref{E:ansatz3} above are therefore the functional choices we make in what follows.

Now, exploiting these preliminary observations, we can construct the needed snPDFs, $f_i^{a/A}(\{\bar{s}_a\},x_1,Q^2)$, which are sensitive to the location $\bar s_a$ of the nucleon $a$ in the nucleus $A$, and thereby also to the surrounding nucleon density in each event (indicated by $\{\bar{s}_a\}$), but which do \textit{not} depend on the intra-nucleon density $T_N(\bar{s})$ of the nucleon $a$ or its fluctuations. This is the approximation which we have used in writing Eq.~(\ref{E:barN}) in its form, where the minijet cross section depends spatially only on the locations of the nucleons $a$ and $b$ but does not contain any transverse-coordinate integrals.

First, for each fixed nucleon configuration in the nucleus $A$ (correspondingly for $B$), we define a nuclear thickness function $T_A^a$ from where the contribution from the nucleon $a$, whose center is at $\bar s_a$, has been excluded,
\begin{equation}
T_A^a(\bar{s})\equiv \sum\limits_{a'\neq a}^A T_N^{a'}(\bar s-\bar s_{a'}).
\label{E:T_Aa}
\end{equation}
Then the average nuclear thickness function experienced by the nucleon $a\in A$ can be defined as
\begin{align}
        \hat{T}_{A}^a(\{\bar s_a\})
				&\equiv \frac{\int \!\text{d}^2 \bar{s}\; T_{N}(\bar{s}-\bar{s}_a) T_A^a(\bar{s})}{\int \!\text{d}^2 \bar{s}\; T_{N}(\bar{s}-\bar{s}_a)} \\
								&= \sum\limits_{a'\neq a}^A \int \!\text{d}^2 \bar{s}\;  T_{N}(\bar{s}-\bar{s}_a)  T_{N}^{a'}(\bar{s}-\bar{s}_{a'}) \\
        &= \sum\limits_{a'\neq a}^A T_{NN}^{aa'}\left(\bar b_{aa'}\right),\label{E:TAHat}
\end{align}
where we have used the normalization of $T_N^a$ and Eqs.~(\ref{E:T_Aa}) and (\ref{E:T_NN}) with
$\bar b_{aa'} = \bar s_{a'}-\bar s_a$, and where the overlap functions $T_{NN}^{aa'}(\bar b_{aa'})$ are of the same Gaussian form as that in Eq.~(\ref{E:T_NN_gauss}). Two things are to be noted here: First, for a specific nucleon $a$  in a nucleus $A$ with a fixed (random) nucleon configuration, $\hat{T}_{A}^a(\{\bar s_a\})$ is a fixed number, whose value depends on the positions of the other nucleons ($a'$) relative to the nucleon $a$. Second, the effect of the above self-exclusion is that in the region of very low nucleon density, which is the case in an event where a single nucleon $a$ is far from other nucleons $a'$, the density $T_N(\bar s-\bar s_{a'})$ vanishes, bringing thus also $\hat{T}_{A}^a$ appropriately to zero. The distribution of $\hat{T}_{A}^a$ for a lead nucleus is shown in Fig.~\ref{F:TAHistogram}.

Now, essentially using $\hat{T}_{A}^a$ in place of $T_A^{\text{WS}}$, we define the EbyE fluctuating snPDFs for a nucleon $a$ analogously to the above discussion, as follows:
\begin{equation}
f_i^{a/A}(\{\bar{s}_a\},x,Q^2)  = f_i^{a}(x,Q^2) r_i^{a/A}(\{\bar{s}_a\},x,Q^2),
\label{E:snPDF_def}
\end{equation}
where $\bar s_a$ is the location of the nucleon $a\in A$, which is fixed for each nucleon configuration (i.e., in each event), and the nuclear modification is
\begin{widetext}
\begin{equation}
r_i^{a/A}(\{\bar{s}_a\},x,Q^2)
= \theta(c^i_A(x,Q^2))\left[1+\log(1+c^i_A(x,Q^2)\hat{T}_{A}^a(\{\bar s_a\}))\right]\\
+ \frac{\theta(-c^i_A(x,Q^2))}{1-c^i_A(x,Q^2)\hat{T}_{A}^a(\{\bar s_a\})},
\label{E:snPDFs_fluct}
\end{equation}
\end{widetext}
where $\theta$ is the Heaviside step function. Because of the reasons discussed above, we have chosen the above functional forms for  ensuring an appropriate behaviour of the modifications in $\hat{T}_{A}^a$, accurate enough momentum conservation, and a correct small-$\hat{T}_{A}^a$ limit. As explained above, at the limit of vanishing nucleon density, i.e. if $a$ is an isolated single nucleon far away from other nucleons, $\hat{T}_{A}^a\rightarrow 0$ and thus also $r_i^{a/A}(\{\bar{s}_a\},x,Q^2)\rightarrow 1$.

The coefficient function $c^i_A(x,Q^2)$ in Eq.~(\ref{E:snPDFs_fluct}) is determined for fixed $x$ and $Q^2$ by requiring that the average modification, which is obtained by averaging first over all the nucleons $a$ in each nucleus and then averaging over a large sample of nuclei $A$, becomes $R_i^{p/A}(x,Q^2)$ of Eq.~(\ref{E:R_iA}),
\begin{align}
R_i^{p/A}(x,Q^2) &= \left\langle
\frac{1}{A}\sum_a r_i^{a/A}(\{\bar{s}_a\},x,Q^2)
\right\rangle_{\!\!\{\!A\!\!\:\}} \\
&\equiv F\left(c^i_A(x,Q^2)\right),
\end{align}
where $\langle\dots\rangle_{\{\!A\!\!\:\}}$ denotes the latter average. Note that here for each parton flavor $i$ we are summing the modifications $r_i^{a/A}$ that are related to the bound proton's $R_i^{p/A}$ (e.g. related to $R_{u_V}^{p/A}$ we sum $r_{u_V}^{p/A}$ from $Z$ protons and $r_{d_V}^{n/A}=r_{u_V}^{p/A}$ from $A-Z$ neutrons). Since we assume isospin symmetry and as the locations of the protons and neutrons are sampled from the same Woods-Saxon distribution, we do not need to keep track of the nucleon identity here but can take all nucleons to be just protons. The function $F(c^i_A)$ is a monotonous function of $c^i_A$, so it can be inverted to yield the normalization function
    \begin{equation}
        c^i_A(x,Q^2) = F^{-1}\left(R_i^{p/A}(x,Q^2)\right).
    \label{E:snPDF_coef}
    \end{equation}
The function $F$ can be calculated numerically for any given $c^i_A$ by sampling a large number of nuclei $A$. The inverse can then be approximated by creating an interpolation function for a list of numerically calculated values of $F\left(c^i_A(x,Q^2)\right)$, and then inverting that interpolation function. In what follows, in computing the nucleon-configuration-specific PDFs
$f_i^{a/A}(\{\bar s_a\},x,Q^2)$ in Eq.~\eqref{E:snPDF_def}, we obtain the coefficients $c_A^i(x,Q^2)$ in Eq.~\eqref{E:snPDF_coef} using the EPS09LO average modifications \cite{Eskola:2009uj}, and the free-proton PDFs correspondingly from the CT14LO set \cite{Dulat:2015mca} using the LHAPDF library \cite{Buckley:2014ana}.

\begin{figure}[tbh]
\hspace{-0.4cm} \includegraphics[width = 0.5\textwidth]{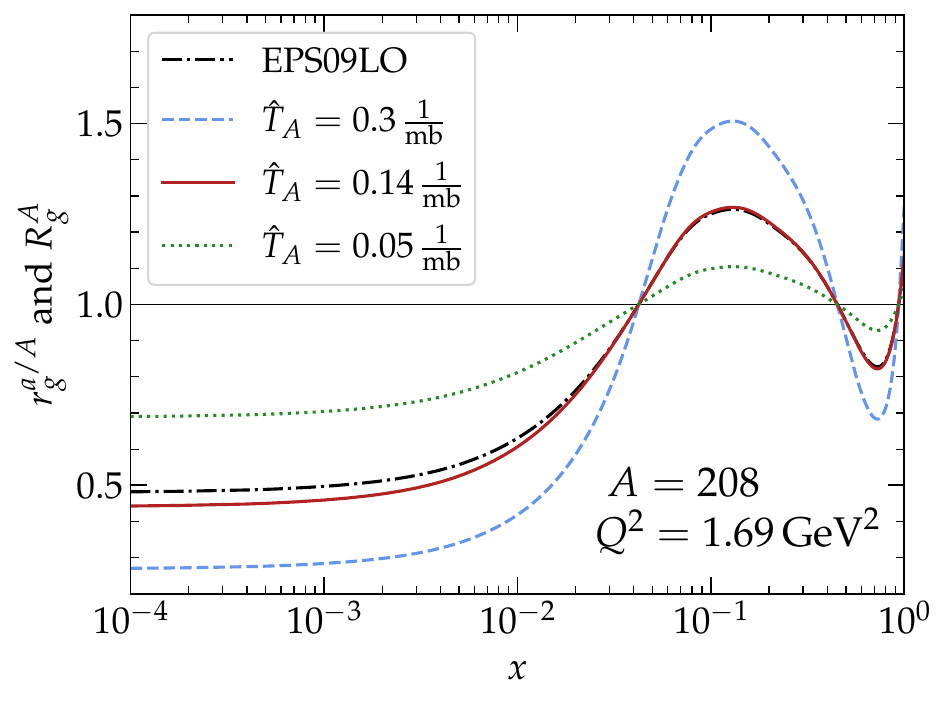}
\caption{Comparison of the snPDF gluon modification $r_g^{a/A}(\{\bar{s}_a\},x,Q^2=1.69\,\text{GeV}^2)$ of Eq.~\eqref{E:snPDFs_fluct} (dashed blue, solid red and dotted green curves) with	the average EPS09LO~\cite{Eskola:2009uj} gluon modification $R_g^A(x,Q^2=1.69\,\text{GeV}^2)$ (dashed-dotted black curve) for the nucleus Pb$^{208}$. The fixed values $\hat{T}_{A}^a(\{\bar s_a\})=$ 0.3, 0.14, and 0.05 1/mb, chosen here as input for $r_g(\{\bar{s}_a\},x,Q^2=1.69\,\text{GeV}^2)$, are representatives for a nucleon in the dense,  median, and scarce density regions, correspondingly, see Fig.~\ref{F:TAHistogram}.}
\label{F:pdfComparison}
\end{figure}

In Fig.~\ref{F:pdfComparison} we compare the spatially dependent, nucleon-configuration-specific gluon modifications $r_g^{a/A}(\{\bar{s}_a\},x, Q^2)$, computed from Eq.~(\ref{E:snPDFs_fluct}), with the average nuclear gluon modifications $R_g^A(x,Q^2)$, obtained from the EPS09LO nPDFs, for a lead nucleus at a scale $Q^2=1.69\,\mathrm{GeV}^2$. To illustrate how in the densest (scarcest) regions the nuclear effects become larger (smaller) than in the average modification $R_g^A(x,Q^2)$, we show the snPDF gluon modifications for three different fixed values of the average thickness function $\hat{T}_{A}^a(\bar s_a)$.

    \begin{figure}[t!]
        \includegraphics[width = 0.5\textwidth]{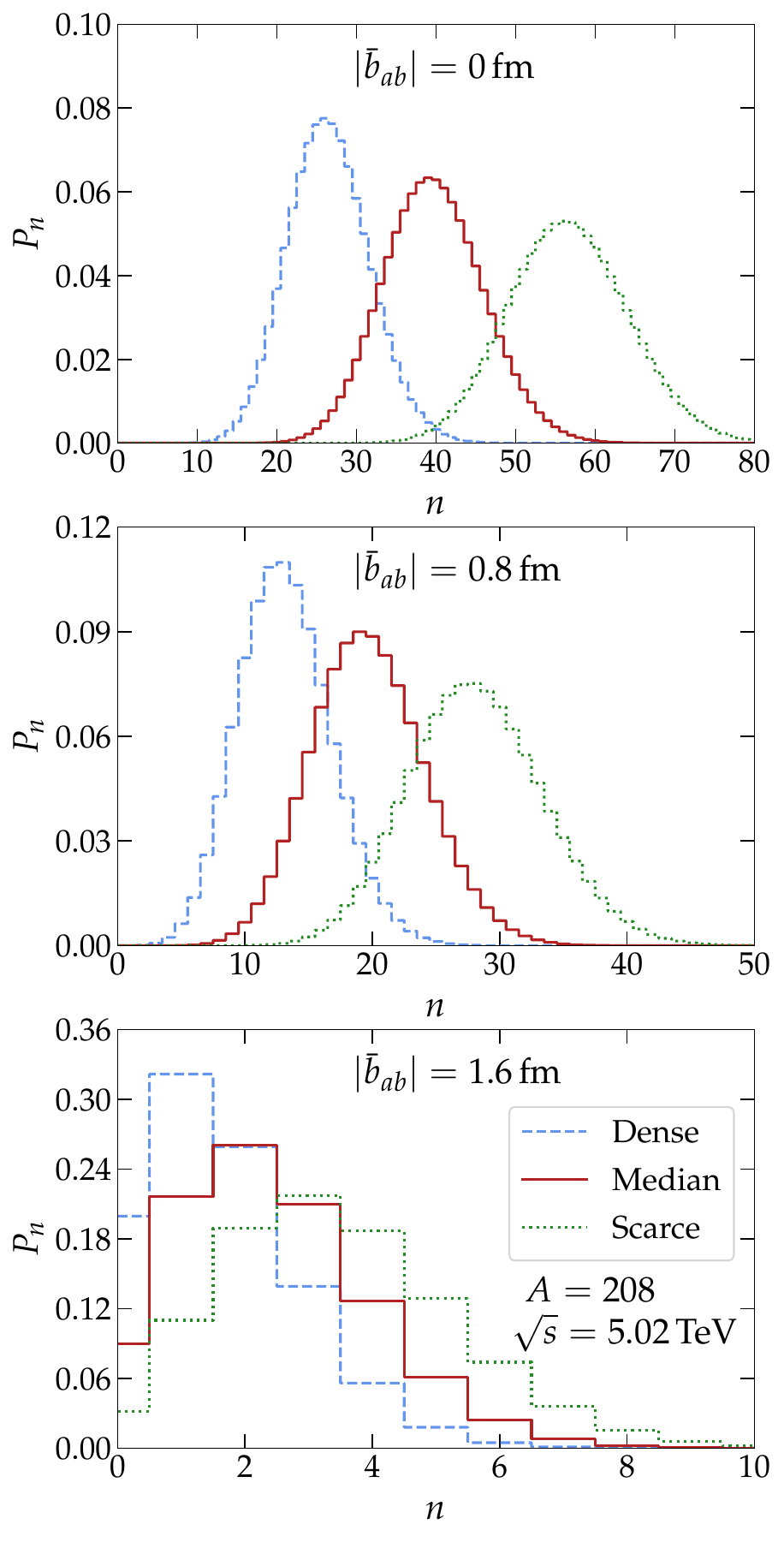}
        \caption{Examples of the Poissonian probability distributions $P_n$ of Eq.~(\ref{E:Poisson}) for multiple candidate-dijet production with partonic $p_T\ge 1\,\text{GeV}$ from a nucleon pair $ab$, $a \in A$ and $b \in B$, at some fixed nucleon-nucleon impact parameters $\bar b_{ab}$ in Pb+Pb collisions at $\sqrt{s_{NN}}=$ 5.02 TeV. The values used for $\sigma_{\text{jet}}^{ab}$ (with $K=2$) here are chosen so that they represent the cases where both of the nucleons originate from the same dense (dashed blue curves), median (solid red curves) and scarcest (dotted green curves) density regions as in Fig.~\ref{F:pdfComparison}.
}
        \label{F:MinijetProbability}
    \end{figure}

We have now discussed the elements necessary for obtaining the nucleon-nucleon overlap function $T_{NN}(\bar{b}_{ab})$ and the integrated minijet cross section $\sigma^{ab}_{\text{jet}}(\{\bar{s}_a\},\{\bar{s}_b\},p_0,\sqrt{s_{NN}})$ that go into the calculation of the probability distributions of multiple minijet production in nucleon-nucleon collisions in Eq.~(\ref{E:Poisson}). Figure \ref{F:MinijetProbability} shows examples of these distributions in Pb+Pb collisions at $\sqrt{s_{NN}}=$ 5.02 TeV and  $p_0=1$~GeV at three different nucleon-nucleon impact parameters $\bar b_{ab}$, and choosing both nucleons, $a\in A$ and $b\in B$, from the same densest, scarcest and median density regions of $A$ and $B$ as in Fig.~\ref{F:pdfComparison}, $\hat{T}_{A}^a(\bar s_a)= \hat{T}_{B}^b(\bar s_b) =$ 0.3, 0.05, and 0.14 1/mb. The figure nicely illustrates the large fluctuations of the minijet multiplicity due to various sources. The minijet multiplicity is heavily sensitive not only to the nucleon-nucleon impact parameter $\bar b_{ab}$ (the larger $\bar b_{ab}$ the smaller $\bar N_\text{jets}^{ab}$) but also to the spatial dependence of the nPDFs (large fluctuations at fixed $\bar b_{ab})$. We also see the role of shadowing and its spatial dependence, in that the colliding nucleons that come from the densest (scarcest) nuclear density regions produce clearly less (more) dijet multiplicity than those who originate from the median-density regions.

    \begin{figure}[t!]
        \includegraphics[width = 0.5\textwidth]{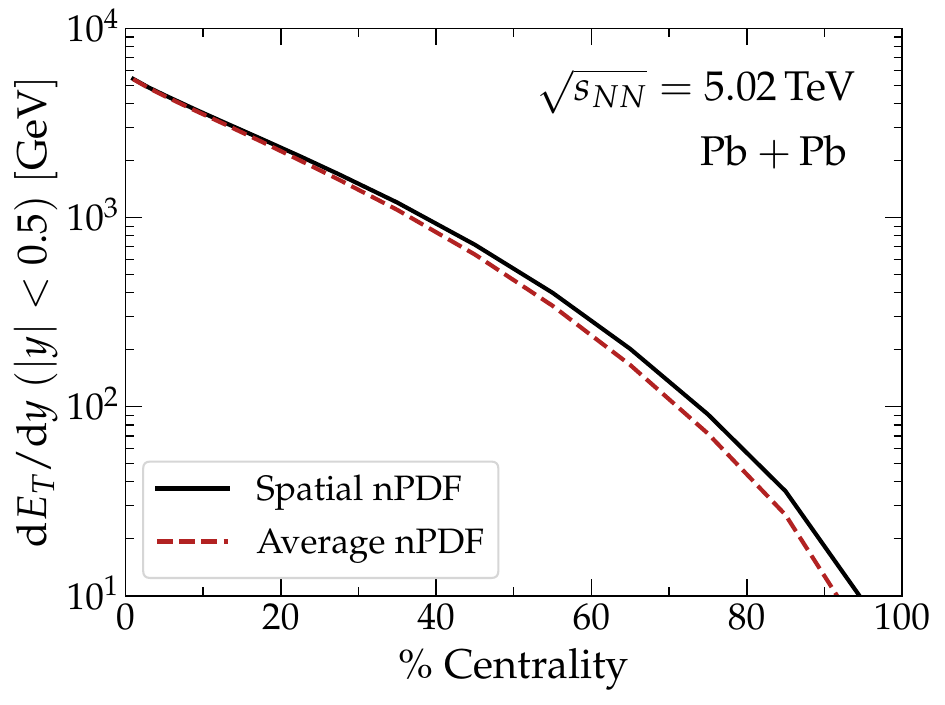}
        \caption{Minijet transverse energy in the mid-rapidity unit as a function of the collision centrality in $\sqrt{s_{NN}}=5.02\,$TeV Pb+Pb collisions, as predicted from the MC-EKRT model using snPDFs (solid lines) and spatially averaged nPDFs (dashed lines). Here $K=2$ and $\kappa_{\text{sat}}=2$, see Sec.~\ref{SS:filtering} for the details of minijet filtering and Sec.~\ref{SS:centrality} for the details of the centrality selection.}
        \label{F:snPDFeffects}
    \end{figure}

In Fig.~\ref{F:snPDFeffects}, we show the centrality dependence of the produced minijet transverse energy at mid-rapidity that is obtained from our MC-EKRT model with snPDFs and with average nPDFs. The figure very clearly demonstrates why it is important to account for the spatial dependence of the nPDFs (the details of the centrality selection and the imposed minijet filtering will be discussed below). As can be seen in the figure, in central collisions, where the minijet production on the average originates from the average nuclear-overlap regions (volume effect), the spatial nuclear effects due to the snPDFs average essentially to those obtained with spatially averaged nPDFs. Towards peripheral collisions, however, where scarcer regions of the nuclei are colliding and where the nuclear effects in the snPDFs become smaller, the difference to the average-nPDF results becomes increasingly larger. As the figure shows, we can expect easily over 20~\% changes relative to the average-nPDF results, which is a significant effect when we compare the MC-EKRT results (after hydrodynamic evolution) with experimental data (Sec.~\ref{S:results} ahead).

%%%
\subsection{Minijet filtering by saturation and  conservation of energy \& valence quark number}
\label{SS:filtering}

After the dijet candidates have been generated from all the nucleon--nucleon pairs as described in Sec.~\ref{SS:nn}, the next step in the MC-EKRT simulation is to filter away the excessive dijets on the basis of saturation, and conservation of energy/momentum and valence-quark numbers. Ideally of course the energy/momentum conservation should not be needed at all, as ideal multiparton distributions should conserve momentum, but as these are not available, and especially not to all orders as would be required here in the context of saturation, we have to impose energy/momentum conservation separately from the saturation. As we assume saturation to be the decisive dynamical mechanism that regulates minijet production at low transverse momenta, saturation-based filtering is done first, and conservation of momentum only after that. With such phenomenological details, experimental data is our guide as well: we have tested, averaging over the minijets falling into the mid-rapidity unit and feeding them into 2+1 D hydrodynamics event-by-event, that we reproduce systematically more correctly the measured ratio of the flow coefficients $v_2$ and $v_3$ \cite{Hirvonen:workinprogress} when the energy/momentum-conservation filtering is performed after the saturation-filtering and also when the latter filtering has as little effect as possible.

As is obvious, any kind of filtering breaks the factorization assumption of our pQCD calculation as the produced minijets are then not anymore independent of each other. The higher-twist effects (causing saturation here) die out in inverse powers of the virtuality $Q^2$, so that at the highest values of $p_T$, factorization is expected to hold. Also the global analysis of nPDFs \cite{Eskola:2021nhw, AbdulKhalek:2022fyi,Duwentaster:2022kpv} and jet production in minimum-bias proton-nucleus collisions \cite{CMS:2014qvs} indicate this to be the case. Thus, to maintain factorization at the highest values of $p_T$, the list of all candidate dijets in an $A$+$B$ collision is next ordered in decreasing $p_T$. Both filterings are then done, separately, in this order, starting from the jets with highest values of $p_T$, and rejecting all those dijets that fulfill the filtering conditions.

Guided by the geometric EKRT saturation criterion, Eq.~(\ref{E:EKRTsat}), each dijet is assumed to have a spatial uncertainty area of a radius $\propto 1/p_T$ in the transverse plane around the dijet production point. Consider a dijet candidate whose transverse momentum is $p_T^{\text{cand}}$, and transverse production point is $\bar{s}^{\text{cand}}$. All of the previously accepted dijets with corresponding parameters $p_T\ge p_T^{\text{cand}} $ and $\bar{s}$ are then inspected, and if for any of them
    \begin{equation}
        |\bar{s}-\bar{s}^{\text{cand}}| < \frac{1}{\kappa_{\mathrm{sat}}}\left(\frac{1}{p_T}+\frac{1}{p_T^{\text{cand}}}\right),\label{E:saturation}
    \end{equation}
the dijet candidate is rejected.   The parameter $\kappa_{\mathrm{sat}}$ introduced here is an external fit parameter, which acts as a ``packing factor'' in determining how close to each other the dijets can be produced. Notice that parametrically $\kappa_{\mathrm{sat}}^2\propto K_{\mathrm{sat}}$ of Ref.~\cite{Niemi:2015qia}, and that the smaller $\kappa_{\mathrm{sat}}$ the stronger the saturation, i.e. the more dijet candidates get rejected.

After the saturation filtering above, the remaining, still $p_T$-ordered, list of accepted dijets is then subjected to the filtering according to energy/momentum conservation. Again here it is not obvious, or even clear, whether the momentum should be conserved for each nucleon separately, or only for the whole nucleus as a parton cloud, or something in between. Here, to be consistent with what is typically done in the global analyses of the nPDFs, we require energy conservation at the nucleon level as a default. We do, however, test also the case where no separate energy/momentum conservation is required in addition to saturation.

To force the energy/momentum conservation (energy conservation, for short) per nucleon for a given dijet candidate with momentum fractions $x_1^\text{cand}$ in a projectile nucleon $a\in A$ and $x_2^\text{cand}$ in a target nucleon $b\in B$, we proceed as follows: Assume that we have a list of $n$ already accepted dijets that involve the same projectile nucleon $a$, and $m$ previously accepted dijets that involve the same target nucleon $b$. These dijets have momentum fractions $(x_1^{(1)},\dots,x_1^{(n)})_a$ and $(x_2^{(1)},\dots,x_2^{(m)})_b$ associated with $a$ and $b$, respectively. Now, if either
\begin{equation}
        x_1^\text{cand} + \sum\limits_{i=1}^{n} x_1^{(i)} > 1 \quad \text{or} \quad
        x_2^\text{cand} + \sum\limits_{j=1}^{m} x_2^{(j)} > 1,
\label{E:mom_consv}
\end{equation}
the dijet candidate is rejected due to the breaking of the per-nucleon energy budget.

The third filtering, performed simultaneously with the above energy conservation, is the forcing of the valence quark number conservation. As explained earlier in Sec.~\ref{SSS:dijetchemi}, we can keep track of whether each candidate dijet involves valence quarks from the nucleons $a\in A$ and/or $b\in B$. If a candidate dijet involves a valence quark of a specific flavor either from $a$ or from $b$, and if either $a$ or $b$ has already consumed all its valence quarks of that flavor in the prior parton scatterings at $p_T > p_T^\text{cand}$, then the candidate dijet is rejected. For the multiplicities and elliptic flow that we will study later in this paper, this filtering causes a negligible effect but we nevertheless build it in for interesting further studies in the future.

\begin{figure*}[t!]
        \includegraphics[width=0.49\textwidth]{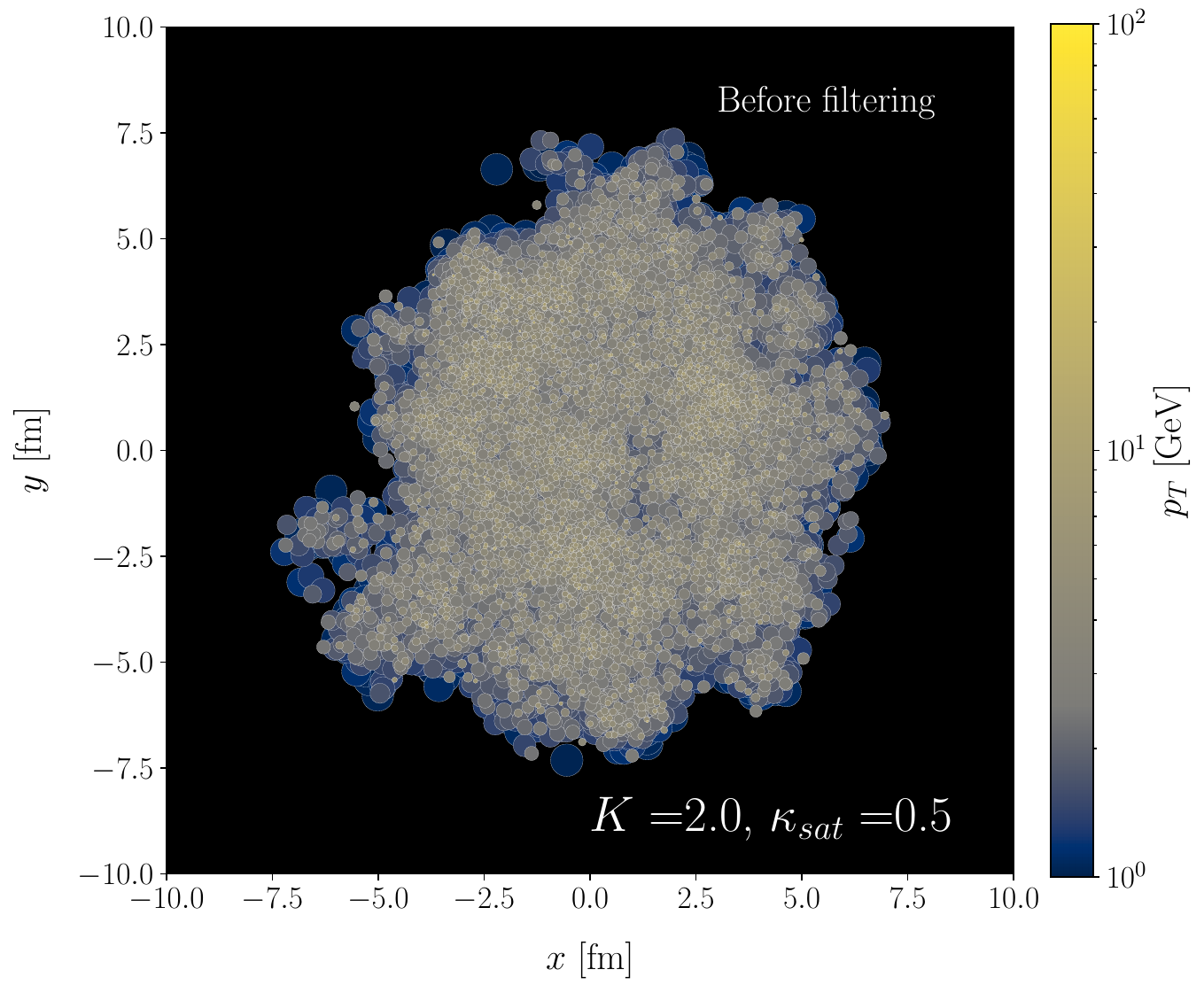}
        \hspace{0.1cm}
        \includegraphics[width=0.49\textwidth]{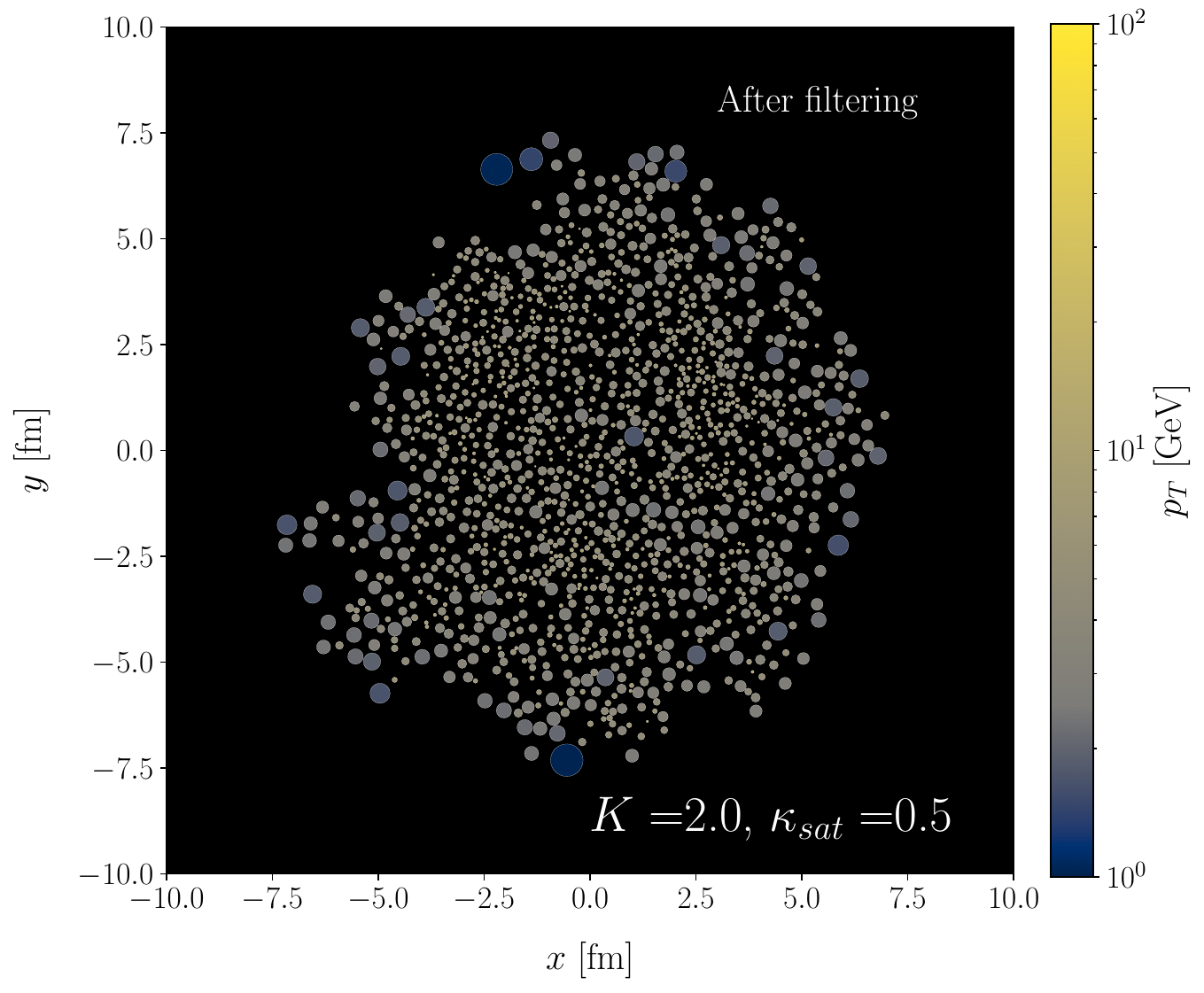}
        \caption{Illustration of the effects of saturation and energy-conservation in the transverse distribution of produced dijets in one central event. Left panel shows the production points of all the candidate dijets, and the right panel the case after the filterings. The radius of the disk surrounding each dijet production point is $1/(\kappa_{\rm sat} p_T)$. Here for the illustration, we use $K=2$, and $\kappa_{\rm sat} = 0.5$.}
        \label{F:Filters_transverse_plane}
\end{figure*}

As an illustration, in Fig.~\ref{F:Filters_transverse_plane} we show the transverse-plane distribution of dijet production points before and after the filterings in a single central event. The radius of each disk surrounding the production points is $1/(\kappa_{\rm sat} p_T)$. As seen in the left panel, the candidate dijets overoccupy the transverse plane. As a result of applying the saturation condition of Eq.~\eqref{E:saturation}, none of the disks overlap in the right panel.

\begin{figure*}[t!]
        \includegraphics[width=0.9\textwidth]{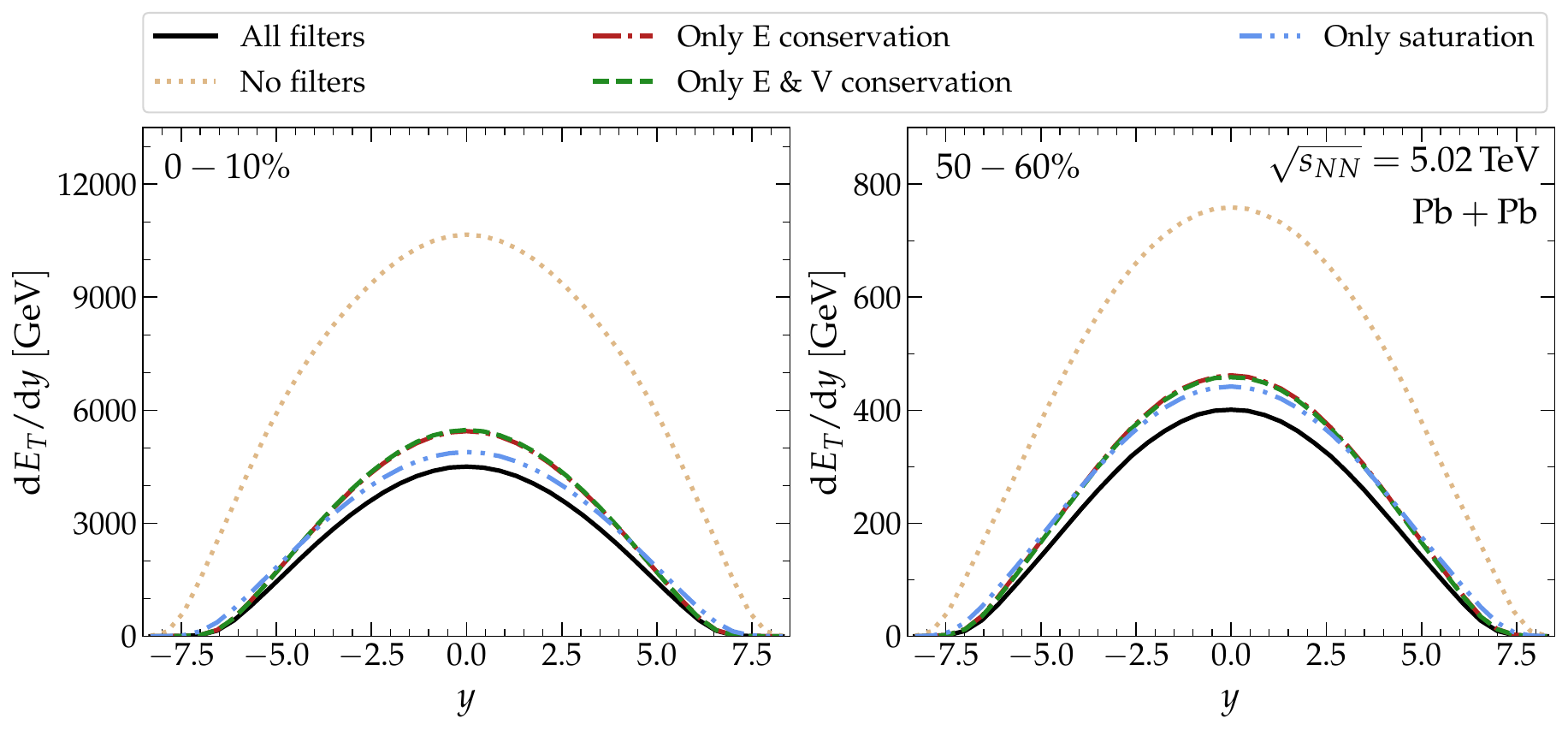}
        \caption{The effects of the EKRT saturation, energy-conservation and valence-quark number conservation filters to the production of minijet transverse energy $\text{d}E_T/\text{d}y$, as a function of rapidity $y$, in central (left panel) and peripheral Pb+Pb collisions (right panel) at $\sqrt{s_{NN}}=5.02\,$TeV. Here $K=2$ and $\kappa_{\mathrm{sat}}=2$.}
        \label{F:Filters}
    \end{figure*}

\subsection{Centrality selection}
\label{SS:centrality}

To determine which centrality percentile each $A$+$B$ collision belongs to, one needs to classify the events according to, e.g., the produced minijet transverse energy $E_T$ in a chosen rapidity window.  Alternatively, when running hydrodynamics with the minijet initial conditions, converting $E_T$ into initial state densities,  one can use either initial state entropy or final state multiplicity as the criterion. In this work, in the interest of simulation speed, we do the centrality selection according to the total minijet $E_T$ produced (after the filterings) anywhere in rapidity. We have checked that the results would be very similar if e.g. a central rapidity unit would be used. Concretely then, for a simulation of, say, 10 000 $A$+$B$ collisions, the 0-5 \% centrality class refers to the collection of 500 collisions with the highest total transverse energy.

\subsection{Systematics of minijet filtering}

\begin{figure*}[t!]
        \includegraphics[width=0.45\textwidth]{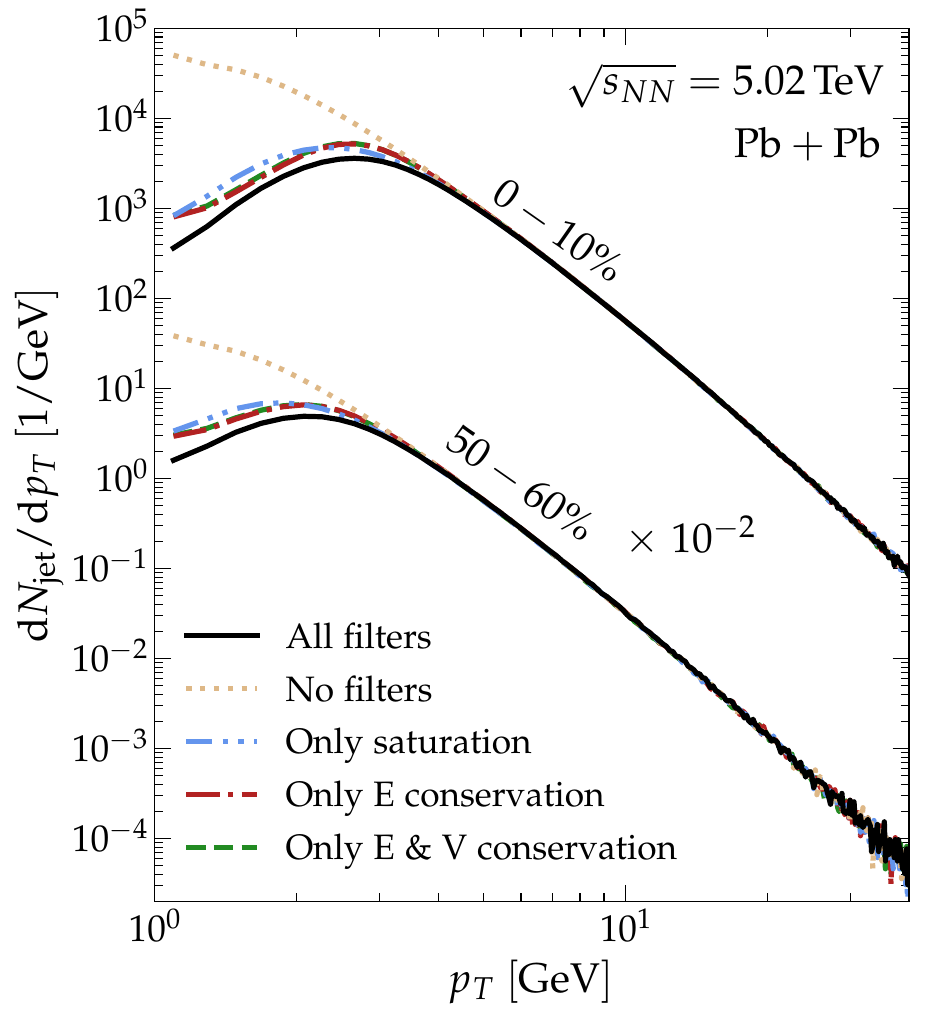}
        \includegraphics[width=0.45\textwidth]{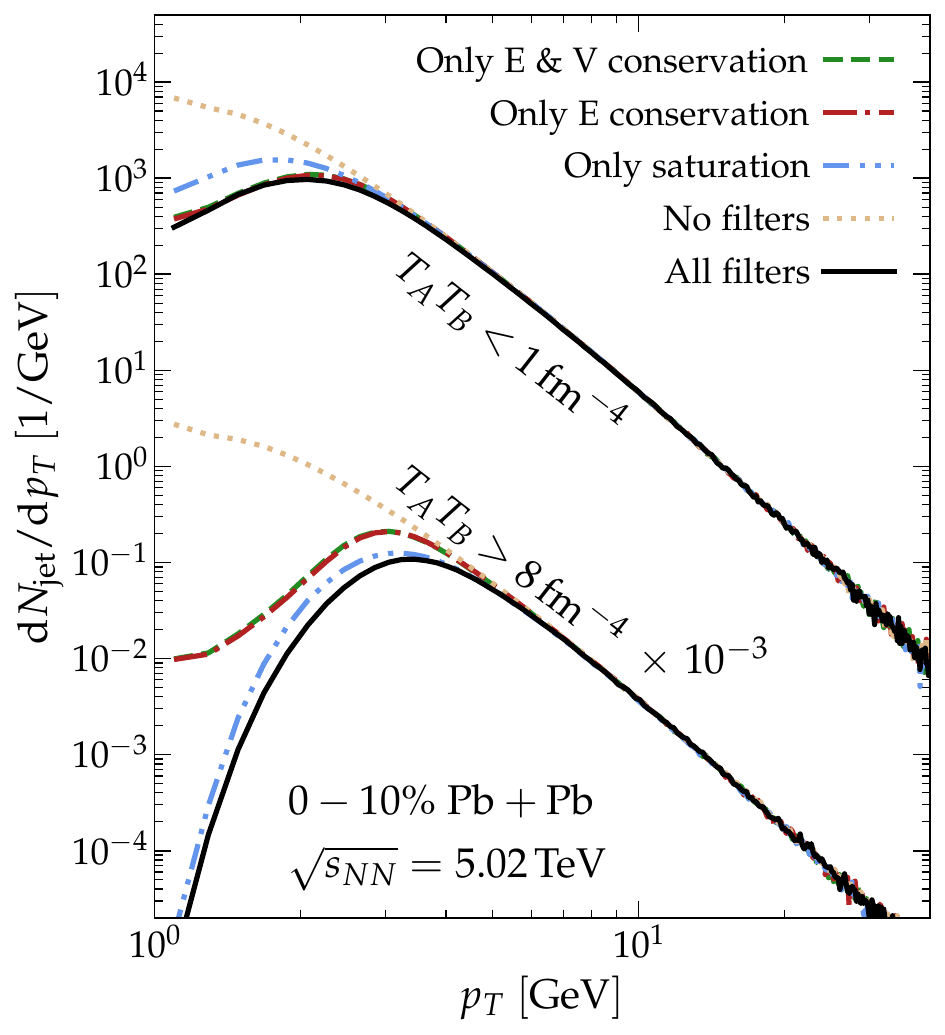}
        \caption{The effects of the EKRT saturation and the energy- and valence-quark conservation filters to the transverse momentum distribution of produced (mini)jets in Pb+Pb collisions at $\sqrt{s_{NN}}=5.02\,$TeV, in 0-10\% and 50-60\% centrality classes (left panel) and in the different overlap-density regions in the 0-10\% central collisions (right panel). Here $K=2$ and $\kappa_{\mathrm{sat}}=2$. For the lower set of curves in both panels, notice the downward scalings by the factors indicated in the panels.}
        \label{F:filter-pT}
    \end{figure*}

Figures \ref{F:Filters} and \ref{F:filter-pT} illustrate the effect of the three filters. Figure \ref{F:Filters} shows the rapidity distribution of the transverse energy originating from the dijets, obtained as a scalar sum of minijet $p_T$'s, plotted for 0-10 \% central (left panel) and 50-60 \% central (right panel) Pb+Pb collisions at  $\sqrt{s_{NN}}=5.02\,$TeV, computed with $K=2$, and $\kappa_{\rm sat} = 2$. The figure demonstrates first a considerable reduction of $E_T$ when going from all the candidate dijets (dotted brown curves) down to those who pass the saturation filter (dashed-double-dotted  blue curves), and then a clearly smaller reduction down to those who pass also the energy-conservation and valence quark filters in addition (solid black curves). As expected, for this quantity the effect of the valence quark filtering is very small (see the overlapping dashed-dotted red and dashed green curves). Interestingly, however, we notice that imposing only the energy-conservation filter without saturation (dashed-dotted red curves) would lead to a similar result in $E_T$ as the saturation filter alone, which essentially is a result of ordering the dijet candidates according to the minijet $p_T$. Here again, we note that although not visible in these plots, we have checked that the  $v_2/v_3$ ratio prefers a strongest possible saturation~\cite{Hirvonen:workinprogress}, and also that imposing only the energy-conservation filter (when realized as in here) typically leads to too narrow rapidity distributions.

Figure \ref{F:filter-pT} then, correspondingly, shows the $p_T$ distribution of (mini)jets at all rapidities, originating from the dijets which have not been filtered at all (dotted brown curves), from those dijets that survived first the saturation filter (dashed-double-dotted blue curves) and then also the energy-conservation and valence-quark filters (black solid curves). In the left panel, we see -- as is expected by construction -- how factorization in central collisions (upper set of curves) remains unbroken at $p_T\gtrsim 5$~GeV, while both filters start to have an effect at $p_T\lesssim 5$~GeV. In peripheral collisions (lower set of curves), where the minijet multiplicities are smaller and therefore saturation becomes effective at smaller $p_T$, factorization remains unbroken until slightly smaller values of $p_T$ than in central collisions. We again also see how saturation filter, the one imposed first, dominates here over that of energy conservation, and also that the saturation filter tends to remove dijets at slightly larger values of $p_T$ than the energy-conservation filter (see dashed-double-dotted blue and the dotted-dashed red curves). Also here the valence quark conservation causes a negligible effect.
The right panel of Fig. \ref{F:filter-pT} is to demonstrate the difference of (mini)jet production in different spatial regions of central collisions: In the dilute overlap regions (upper set of curves) the factorization-breaking saturation and energy-conservation effects set in at clearly smaller values of $p_T$ than in the regions of densest overlap (lower set of curves).

\begin{figure*}[htb!]
\includegraphics[width=0.95\textwidth]{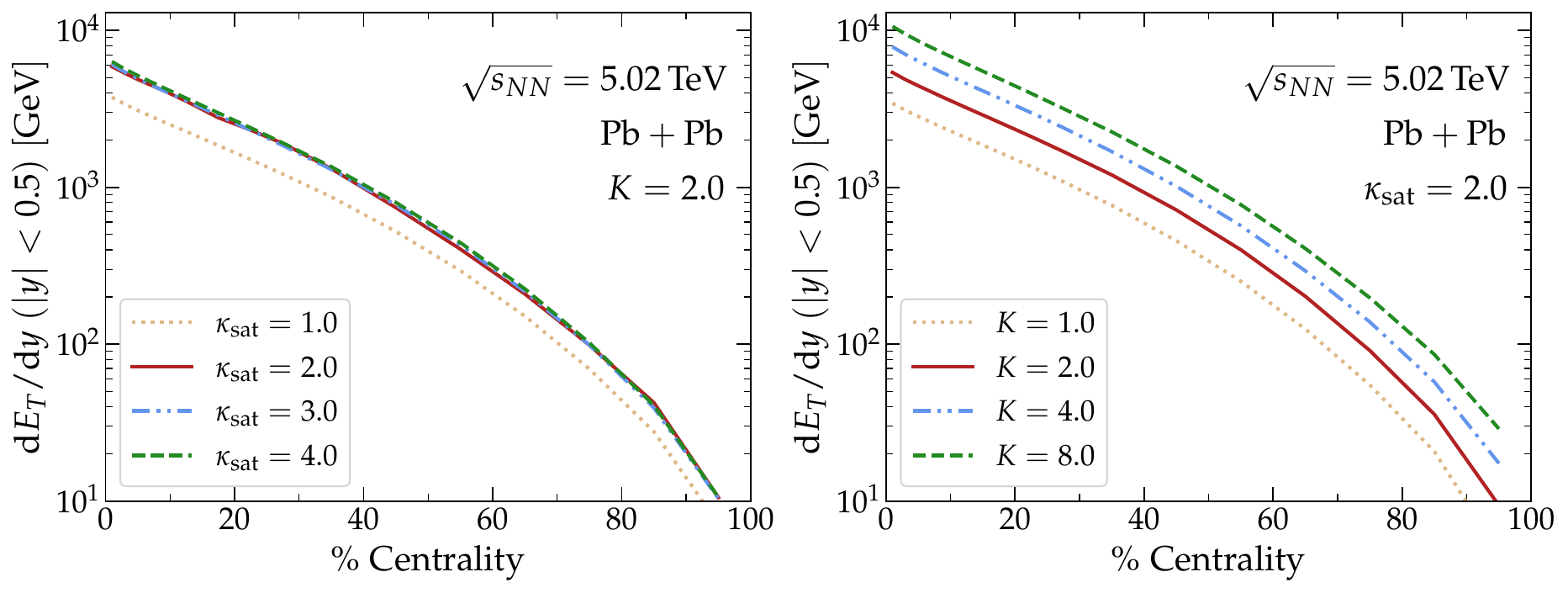}
\caption{Centrality dependence of minijet transverse energy in the mid-rapidity unit in Pb+Pb collisions at $\sqrt{s_{NN}}=5.02\,$TeV, corresponding to one fixed value of $K=2$ and various values of $\kappa_\text{sat}$ (left panel), and to one fixed value of $\kappa_\text{sat} = 2$ and various values of $K$ (right panel), with all filters imposed. The red solid curve is the same in both panels.}
        \label{F:parameter_effect}
    \end{figure*}

\begin{figure*}[htb!]
\includegraphics[width=0.95\textwidth]{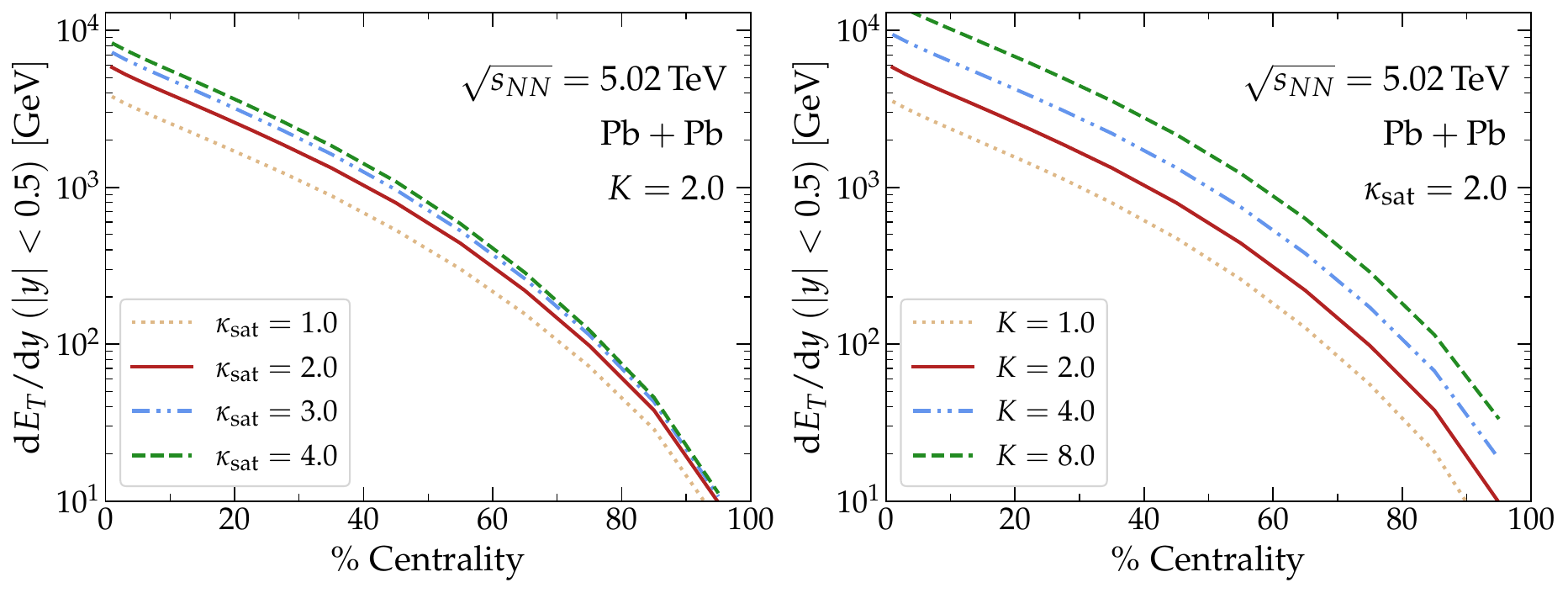}
\caption{The same as in Fig.~\ref{F:parameter_effect} but with only the saturation filter imposed.}.
        \label{F:parameter_effect_NoE}
    \end{figure*}

Figures \ref{F:parameter_effect} and \ref{F:parameter_effect_NoE} show the minijet transverse energy production in the central rapidity unit as a function of centrality in Pb+Pb collisions at $\sqrt{s_{NN}}=5.02\,$TeV, computed with various values of the fit parameters $K$ and $\kappa_\text{sat}$, with all filters imposed in Fig.~\ref{F:parameter_effect}, and with only the saturation filter imposed Fig.~\ref{F:parameter_effect_NoE}. As can be seen from the right panels, where
$\kappa_\text{sat} = 2$ is fixed, changing $K$ changes mainly the overall normalization but essentially not the centrality slope of the produced $E_T$ (and hence the final multiplicities as well). The energy-conservation filter weakens the $K$ dependence, because with a larger $K$-factor the energy-conservation filter removes more candidate dijets. The left panels in turn show how, for a fixed value of $K = 2$, changing $\kappa_\text{sat}$ changes both the normalization and especially the centrality slope. Here the energy-conservation filter in turn weakens the $\kappa_\text{sat}$ dependence, as with a larger $\kappa_\text{sat}$ there is less saturation and more minijet production and the energy conservation filter becomes more efficient in removing candidate dijets. In any case, as long as $\kappa_\text{sat}$ does not become too large, and especially if only the saturation-filter is imposed,  $\kappa_\text{sat}$ serves as a centrality-slope parameter for the mid-rapidity multiplicities, whereas the $K$-factor controls mainly their normalization. This observation is exploited in what follows (Sec.~\ref{S:results}), in finding the possible values for $\kappa_\text{sat}$ and $K$ with which we can reproduce the measured charged-hadron multiplicities.

\section{Fluid dynamical evolution and particle spectra}
\label{S:fluidsetup}

The MC-EKRT computation gives the initially produced parton state. In order to compare with the measured data, we need to first propagate the partons to a proper time $\tau_0$ for initializing the 3+1 D fluid dynamics, then compute the subsequent spacetime evolution of the matter, and eventually determine the experimentally measurable momentum spectra of hadrons.

\subsection{Fluid dynamical framework}
\label{SS:IS_fluid}

The spacetime evolution is computed using 3+1 D fluid dynamics, applying the code package developed in Ref.~\cite{Molnar:2014zha}. The fluid dynamical framework employed is the relativistic dissipative second-order transient fluid dynamics \cite{Denicol:2012cn}, originally formulated by Israel and Stewart~\cite{Israel:1979wp}.

The basic equations of motion governing the evolution of a fluid are the local conservation laws for energy, momentum and conserved charges, like the net-baryon number. In the following we, however, will neglect the conserved charges. In this case the state of the fluid is given by its energy-momentum tensor that can be decomposed with the help of the Landau-picture fluid 4-velocity $u^{\mu}$ as
\begin{equation}
T^{\mu\nu} = e u^\mu u^\nu - P \Delta^{\mu\nu} + \pi^{\mu\nu},
\label{eq:energymomentum}
\end{equation}
where $\Delta^{\mu\nu} = g^{\mu\nu} - u^{\mu}u^{\nu}$ is a projection operator,  $e = T^{\mu\nu} u_\mu u_\nu$ is the energy density in the local rest frame, $P = -\frac{1}{3}\Delta_{\mu\nu}T^{\mu\nu}$ is the isotropic pressure, and $\pi^{\mu\nu} = T^{\langle \mu\nu \rangle}$ is the shear-stress tensor. The angular brackets project the symmetric and traceless part of the energy-momentum tensor that is orthogonal to the fluid 4-velocity. We will also neglect the bulk viscous pressure, and the isotropic pressure is given by the equation of state (EoS) of the strongly interacting matter at zero net-baryon density, $P = P(e)$. In the Landau picture the fluid 4-velocity is a time-like, normalized eigenvector of the energy-momentum tensor, defined by $T^{\mu}_{\,\,\nu} u^\nu = e u^\mu$. The energy diffusion current $W^{\mu} = \Delta^{\mu\alpha}T_{\alpha\beta}u^\beta$ is then zero and does not contribute to the energy-momentum tensor.

In the formalism by Israel and Stewart~\cite{Israel:1979wp}, the equations of motion for the remaining dissipative quantity, shear-stress tensor, are given by \cite{Denicol:2012cn, Molnar:2013lta}
\begin{eqnarray}
 \tau_\pi \frac{d}{d\tau}\pi^{\langle \mu \nu \rangle} + \pi^{\mu\nu}  = 2\eta \sigma^{\mu\nu}  + 2 \tau_\pi \pi_\alpha^{\langle \mu}\omega^{\nu\rangle \alpha} \notag \\
  \qquad - \delta_{\pi\pi} \pi^{\mu\nu} \theta-\tau_{\pi\pi} \pi_{\alpha} ^{\langle \mu} \sigma^{\nu\rangle \alpha} + \varphi_7 \pi_\alpha^{\langle \mu} \pi^{\nu\rangle \alpha},
\end{eqnarray}
where $\sigma^{\mu\nu} = \nabla^{\langle \mu}u^{\nu\rangle}$ and $\omega^{\mu\nu} = \frac{1}{2}\left(\nabla^\mu u^\nu - \nabla^\nu u^\mu\right)$ are the strain-rate and vorticity tensors, respectively, $\theta = \nabla^\mu u_\mu$ is the volume expansion rate, and the gradient is defined as $\nabla^\mu = \Delta^\mu_{\;\alpha} \partial^\alpha$. The coefficient $\eta$ is the shear viscosity, $\tau_\pi = 5\eta/(e+P)$ is the shear relaxation time, and the remaining coefficients of the second-order terms are taken from the 14-moment approximation to massless gas \cite{Denicol:2010xn, Denicol:2012cn, Molnar:2013lta}, i.e.\ $\delta_{\pi\pi} = (4/3)\tau_{\pi}$, $\tau_{\pi\pi} = (10/7)\tau_{\pi}$, and $\varphi_7 = 9/(70P)$. The shear viscosity over entropy density $\eta/s$ is chosen such that it roughly reproduces the elliptic flow in semi-central collisions. For the EoS of strongly interacting matter we use the $s$95$p$-v1 \cite{Huovinen:2009yb} parametrization, which interpolates between the lattice QCD at high temperatures and the hadron resonance gas model at low temperatures. The partial chemical freeze-out at $T=150$ MeV is encoded into the hadronic part of the EoS as temperature-dependent chemical potentials for each hadron, $\mu_h = \mu_h(T)$ \cite{Huovinen:2007xh}.

The Israel-Stewart equations together with the conservation laws are solved numerically in 3+1 dimensions~\cite{Molnar:2014zha} using the SHASTA algorithm \cite{borisbook} in $(\tau, x, y, \eta_s)$--coordinates, where
\begin{equation}
\tau = \sqrt{t^2 - z^2}
\end{equation}
is the longitudinal proper time, and
\begin{equation}
\eta_s = \frac{1}{2}\ln\left(\frac{t+z}{t-z}\right)
\end{equation}
is the spacetime rapidity. The grid resolution is $\Delta \eta_s = 0.15$, $\Delta x = \Delta y  = 0.15$ fm, and $\Delta \tau = 0.05$ fm. For further details of the algorithm, see Refs.\ \cite{Molnar:2009tx, Molnar:2014zha}.

The final spectra of free hadrons are obtained by computing the Cooper-Frye integrals \cite{Cooper:1974mv} on a constant-temperature decoupling surface, with $T_{\rm dec} = 130$ MeV. The momentum distributions of hadrons on the decoupling surface are given by the 14-moment approximation, so that the single-particle momentum distribution function of a hadron $h$ is
\begin{equation}
f_h(p^\mu, x) = f_{0h}\left(1 + \left(1 \pm f_{0h}\right) \frac{p_\mu p_\nu \pi^{\mu\nu}}{2T^2(e+P)}\right),
\end{equation}
where $+$($-$) is for bosons (fermions), $p^{\mu}$ is the 4-momentum of a hadron $h$, and $f_{0h} = f_{0h}(T, \mu_h)$ is the corresponding Bose-Einstein or Fermi-Dirac equilibrium distribution function. The Cooper-Frye integral is computed for all the hadrons included into the hadron resonance gas part of the EoS. As explained in Ref.~\cite{Molnar:2014zha}, after computing the full spectra of hadrons, $dN_h/dydp_T^2d\phi$, the spectra are interpreted as probability densities and they are randomly sampled to obtain a set of hadrons with 4-momenta $p_i^{\mu}$. For the unstable hadrons the corresponding 2- and 3-particle strong and electromagnetic decays are then computed. The sampling procedure is then repeated several times in order to get smooth momentum distributions for the hadrons that are stable under strong decays.

\subsection{Initialization}
\label{S:initialization}

The equations of fluid dynamics take the energy-momentum tensor as an initial condition at a fixed initial proper time $\tau_0$. However, an MC-EKRT event consists of a set of partons, and we need to convert this set to the corresponding $T^{\mu\nu}(\tau_0, x, y, \eta_s)$ using the momenta of the produced particles. There are two essential ingredients in this. First, we need to propagate the particles to a fixed proper time $\tau_0 = 1/p_{0}$, and for the determination of densities from a finite set of particles, we need to define an averaging volume where the components of the energy-momentum tensor are computed.

Naively, the grid size, e.g.\ $\Delta x$ or $\Delta \eta_s$ in the numerical algorithm to solve the Israel-Stewart theory would provide such an averaging volume. However, the grid defines rather a discretization of the continuous fields in the hydrodynamic equations of motion, and in principle we should be able to take the limit to the continuum, i.e.\ $\Delta x, \Delta \eta_s \rightarrow 0$, and at this limit densities are no longer well defined smooth functions. Thus, we should distinguish between the averaging volume and the numerical resolution. The procedure with which we define the averaging volume through Gaussian smearing and obtain the corresponding densities is described below. We note that here we will eventually only construct the local energy density from the MC-EKRT computation, and neglect the initial velocity and shear-stress components. Moreover, we do not take into account the event-by-event fluctuations in the hydrodynamical phase, but compute the initial conditions as averages over a large sample of MC-EKRT events. However, the procedure below can be extended to the computation of all the components of $T^{\mu\nu}$. We will leave the studies that take into account the event-by-event fluctuations as well as a complete $T^{\mu\nu}$ initialization as a future work.

\subsubsection{Free streaming}
\label{SS:freestreaming}

Each parton $i$ in an MC-EKRT event has the following information: transverse coordinate $\mathbf{x}_{\perp,0i}$ of the production point, transverse momentum $\mathbf{p}_{T i}$, and rapidity $y_i$. All partons are massless in this work. We assume that each parton is produced at the location $\mathbf{x}_{\perp,0i}$ and $z_i=0$ at time $t=0$. The partons are assumed to travel as free particles along straight line trajectories. In this case, the spacetime rapidity $\eta_{s,i}$ of the parton $i$  becomes equivalent to its momentum rapidity $y_i$, and longitudinal coordinate of the propagating parton is given by $z_i(t) = t \tanh \eta_{s,i}$. The transverse position of the parton at Cartesian coordinate time $t$ is given by $\mathbf{x}_{\perp i} (t) = \mathbf{x}_{\perp,0i} + t \mathbf{p}_{T i} /E_i$, where $E_i = p_{T i} \cosh y_i$. However, we need to initialize fluid dynamics at a fixed proper time $\tau_0 = t/\cosh \eta_s$ in the $\tau$-$\eta_s$ coordinate system, in which case the parton's coordinates become $(\tau_0, \mathbf{x}_{\perp i} (\tau_0), \eta_{s,i})$, where $\mathbf{x}_{\perp i} (\tau_0) = \mathbf{x}_{\perp,0i} + \tau_0 \mathbf{p}_{T i}/p_{T i}$.

\subsubsection{Smearing}
\label{SS:smearing}

In general, the four-momentum $p^\alpha =(p^\tau, \mathbf{p}_T, p^\eta)$ of a particle at a spacetime location $x^\alpha = (\tau, \mathbf{x}_\perp, \eta_s)$ in the $\tau$-$\eta_s$ coordinates is obtained as
\begin{align}
\label{E:FourMomentum_hyperbolic}
    p^\alpha = \frac{\partial x^\alpha }{\partial x'^\mu } p'^\mu
             = \left(
            \begin{matrix}
                p_{T} \cosh(y - \eta_{s}) \\
                \mathbf{p}_T \\
                \tau^{-1} p_{T} \sinh(y -\eta_{s})
            \end{matrix}
            \right),
\end{align}
where $x'^{\mu}$ and $p'^{\mu}$ are the corresponding spacetime point and four-momentum in the Cartesian coordinates.

The total number of partons $N$ that flow through a surface, whose surface element 4-vector is $d\Sigma_\mu$, can be written as
\begin{equation}
 N = \int d\Sigma_\alpha N^{\alpha}(\tau, \mathbf{x}_\perp, \eta_s),
 \label{E:Nparticles_definition}
\end{equation}
where the particle 4-current $N^\alpha$ in the $\tau$-$\eta_s$ coordinates can be written using Eq.~\eqref{E:FourMomentum_hyperbolic} as
\begin{equation}
 N^{\alpha}(\tau, \mathbf{x}_\perp, \eta_s) = \int \frac{d^3 p}{p^\tau}\tau p^{\alpha} f(\tau, \mathbf{x}, \mathbf{p}),
\end{equation}
where we defined $d^3 p = d^2 \mathbf{p}_T dp^\eta$, and $f$ is a scalar momentum distribution function at a constant $\tau$. For a constant-$\tau$ surface, the surface element 4-vector has only the $\tau$ component, $d\Sigma_\tau =  d^2\mathbf{x}_\perp d\eta_s \tau$, and the total number of partons can be written as
\begin{equation}
 N = \int d^2\mathbf{x}_\perp d\eta_s \tau \int d^2 \mathbf{p}_T dp^\eta \tau f(\tau, \mathbf{x}, \mathbf{p}).
 \label{E:Nparticles}
\end{equation}
Now, following Ref.~\cite{vanLeeuwen}, the scalar momentum distribution function for a set of $N$ partons can be written in terms of delta functions in coordinate and momentum space as
\begin{equation}
f(\tau, \mathbf{x}, \mathbf{p}) = \sum_{i = 1}^{N} \delta^{(3)}(\mathbf{x}-\mathbf{x}_{i}) \delta^{(3)}(\mathbf{p}-\mathbf{p}_{i})/|{\rm det}(g)|,
\label{E:distribution}
\end{equation}
where $\mathbf{x}_{i} =(\mathbf{x}_{\perp i}, \eta_{s,i})$ is the three-location and $\mathbf{p}_{i} = (\mathbf{p}_{T i}, p^\eta_i)$ is the three-momentum of the particle $i$ at proper time $\tau$, and ${\rm det}(g) = -\tau^2$ is the determinant of the metric tensor $g_{\mu\nu} = {\rm diag}(1, -1, -1, -\tau^2)$. The summation is over all the particles. Substituting Eq.~\eqref{E:distribution} into Eq.~\eqref{E:Nparticles}, it is easy to verify that we consistently arrive at the correct number of particles, i.e.\ in our case the number of partons from an MC-EKRT event. Similarly, the components of the energy-momentum tensor can be expressed as
\begin{equation}
 T^{\alpha\beta}(\tau, \mathbf{x}_\perp, \eta_s) = \int \frac{d^2 \mathbf{p}_T dp^\eta}{p^\tau}\tau p^{\alpha}p^{\beta} f(\tau, \mathbf{x}, \mathbf{p}).
\end{equation}

In what follows, we will assume that $p^\eta_i = 0$, so that $y_i = \eta_{s,i}$. Changing the integration variable from $p^\eta$ to rapidity $y$ using Eq.~\eqref{E:FourMomentum_hyperbolic}, the integral can be then written as
\begin{align}
\label{eq:energymomentum_ebye}
 T^{\alpha\beta} & = \sum_{i}\int d^2\mathbf{p}_T dy \frac{p^{\alpha}p^{\beta}}{p^\tau}\frac{1}{\tau} \cosh(y - \eta_{s}) \\
 & \!\!\!\!\times \delta^{(2)}(\mathbf{x}_\perp - \mathbf{x}_{\perp i})\delta(\eta_s - \eta_{s, i}) \delta^{(2)}(\mathbf{p}_T - \mathbf{p}_{T i}) \delta(y - \eta_s). \notag
\end{align}
The resulting $\delta(y - \eta_s)$ ensures that $y = \eta_s$, i.e.\ initial longitudinal scaling flow holds even after we replace the spatial delta functions by Gaussian smearing functions below.

To obtain a smooth density profile for relativistic hydrodynamics from the partons, we replace the spatial delta functions with Gaussian distributions,
\begin{equation}
\delta^{(2)}(\mathbf{x}_\perp - \mathbf{x}_{\perp i})\delta(\eta_s - \eta_{s, i}) \rightarrow g_\perp (\mathbf{x}_\perp; \mathbf{x}_{\perp i})g_\parallel (\eta_s; \eta_{s,i}),
\end{equation}
with
\begin{align}
\label{E:SmearingInitial_trans}
    g_\perp (\mathbf{x}_\perp; \mathbf{x}_{\perp i}) = \frac{C_\perp}{2\pi \sigma_\perp^2}  \exp \left[ -\frac{(\mathbf{x}_\perp - \mathbf{x}_{\perp i})^2}{2\sigma_\perp^2 } \right], \\
\label{E:SmearingInitial_long}
    g_\parallel (\eta_s; \eta_{s,i}) = \frac{C_\parallel}{ \sqrt{2\pi \sigma_\parallel^2}}  \exp \left[ -\frac{(\eta_s - \eta_{s,i})^2}{2\sigma_\parallel^2 } \right],
\end{align}
where $\sigma_\perp$ and $\sigma_\parallel$ are the widths of the distributions in the transverse and longitudinal directions, respectively. Both $\sigma_\perp$ and $\sigma_\parallel$ are considered to be free parameters of our model. Equations \eqref{E:SmearingInitial_trans} and \eqref{E:SmearingInitial_long} are normalized as
\begin{align}
    \label{E:SmearingNormInitial}
    \int d^2\mathbf{x}_\perp d\eta_s g_\perp (\mathbf{x}_\perp; \mathbf{x}_{\perp i}) g_\parallel (\eta_s; \eta_{s,i})  = 1.
\end{align}
To reduce the computational costs, we impose a cut-off on the smearing range to $\pm 3\sigma$ in each direction from the centre of the Gaussian distribution. However, the cut-off on the integration range and the numerical error originating from the discretization of Gaussian functions violate the normalization condition in Eq.~\eqref{E:SmearingNormInitial}. Therefore, the constants $C_\perp$ and $C_\parallel$ in Eqs.~\eqref{E:SmearingInitial_trans} and \eqref{E:SmearingInitial_long} are adjusted in every $f_i(\tau, \mathbf{x}, \mathbf{p})$ so that the unit normalization is ensured. We checked, however, that $C_\perp$ and $C_\parallel$ are almost unity with the current parameters in the simulations.

With these choices, the initial value of the $T^{\tau \tau} (\tau_0, \mathbf{x}_\perp, \eta_s)$ component of the energy-momentum tensor in hydrodynamics is given as
\begin{align}
\label{E:EMtensor_initialization_pt}
   T^{\tau \tau} &(\tau_0,  \mathbf{x}_\perp, \eta_s) = \\
   & \frac{1}{\tau_0}\sum_i p_{T i} g_\perp (\mathbf{x}_\perp; \mathbf{x}_{\perp i})g_\parallel (\eta_s; \eta_{s,i}). \notag
\end{align}

In this exploratory study, as we do not yet consider a more detailed spacetime picture of parton production \cite{Eskola:1993cz}, pQCD showering and secondary collisions of partons, and especially as we consider only averaged initial conditions, we follow Ref.~\cite{Niemi:2015qia} and compute only the above initial $T^{\tau\tau}(\tau_0)$ component, and ignore the initial bulk pressure and shear-stress tensor, as well as set $T^{\tau i}(\tau_0) = 0$, or equivalently set the spatial components of the four-velocity $u^{\mu}(\tau_0) = \gamma(1, \mathbf{v}_T(\tau_0), v^\eta(\tau_0))$ initially to zero. Here $v^\eta(\tau_0) = 0$ follows from the condition $y = \eta_{s}$ that corresponds to $v_z = z/t$ in the collision frame. The remaining diagonal components of the energy-momentum tensor are then given by the EoS as $T^{ij}(\tau_0) = P(e(\tau_0))\delta^{ij}$, where now in the absence of initial transverse flow, $e(\tau_0) = T^{\tau\tau}(\tau_0)$.

\begin{figure}[htb]
 \centering
\includegraphics[width=0.46\textwidth]{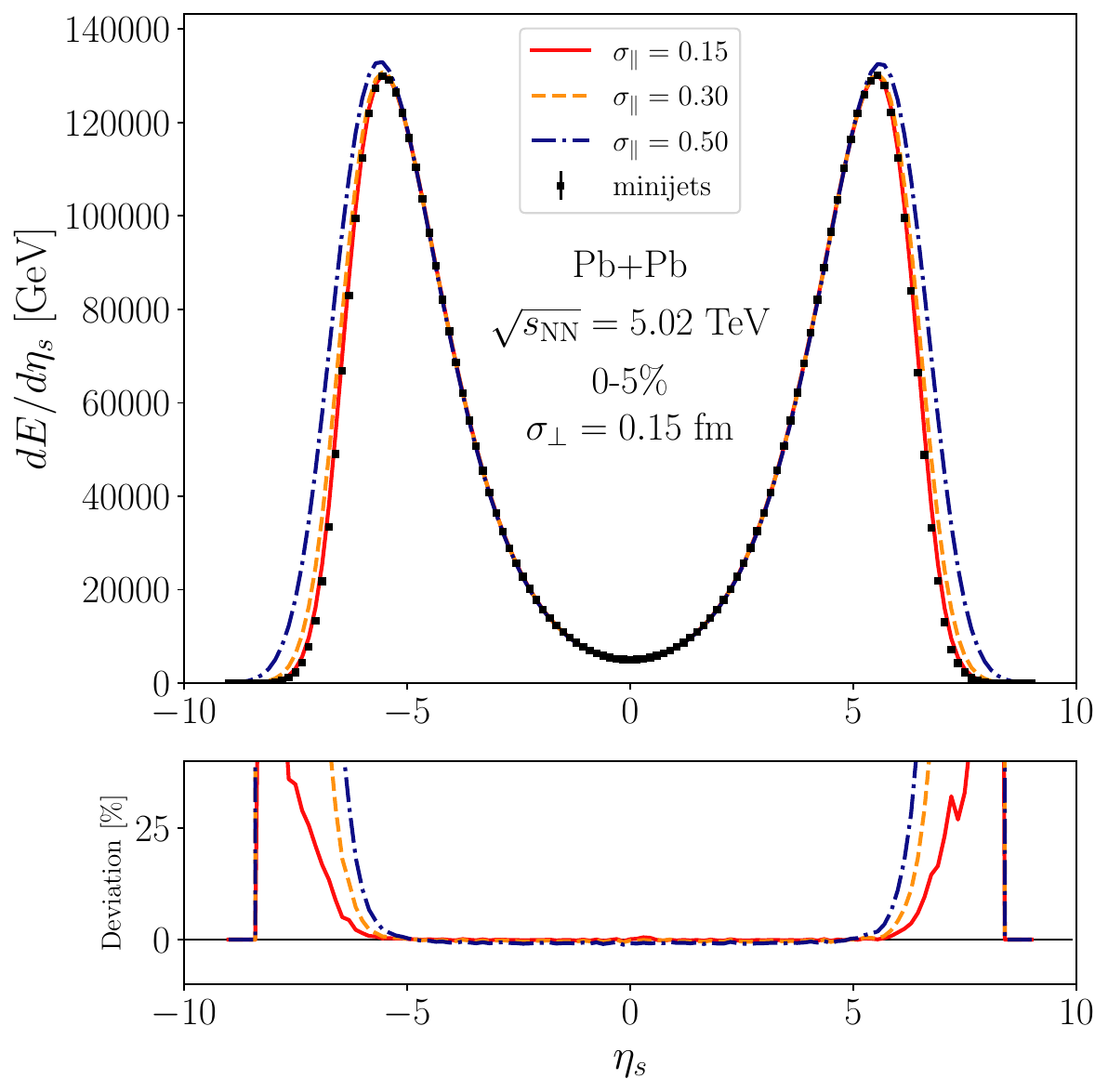}
\includegraphics[width=0.46\textwidth]{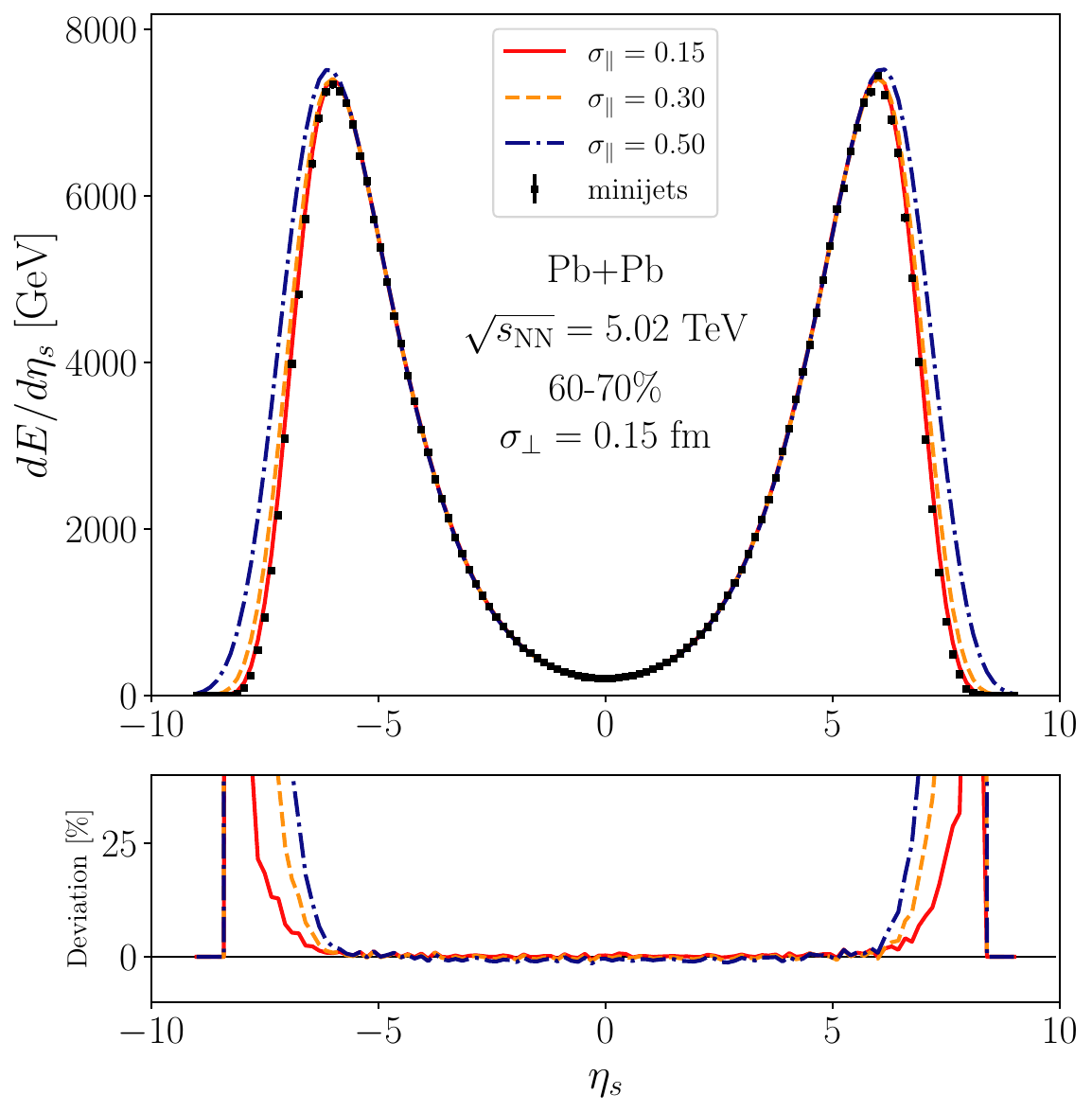}
\caption{Spacetime rapidity distribution of the event-averaged initial energy, $dE/d\eta_s$ as a function of $\eta_s$, at $\tau=\tau_0$ in  0-5 \% central (a) and 60-70 \% central (b) 5.02 TeV Pb+Pb collisions, obtained from the minijets before the smearing (markers) and after the smearing with a fixed transverse width $\sigma_\perp = 0.15$~fm and with different longitudinal widths $\sigma_\parallel$ (solid, dashed and dotted-dashed curves). The smaller panels show the relative difference between the smeared and unsmeared cases.}
\label{fig:smearing_test}
\end{figure}

We note that this way of initializing does not explicitly conserve energy, but with $\sigma_\parallel = 0.15$ the total energy is increased only by $\sim 1$ \%, while with e.g.\ $\sigma_\parallel = 0.5$ already by $\sim 13$ \%. On the other hand, $dE/d\eta_s$ with a rapidity independent distribution of particles would be conserved in the smearing. The MC-EKRT distribution is not rapidity independent, but in practice $dE/d\eta_s$ is almost identical before and after the smearing of parton distribution in the mid-rapidity region. Only at larger rapidities, where experimental data are not available in any case, we start to see the the smeared case $dE/d\eta_s$ deviating from the unsmeared minijet $dE/d\eta_s$. This is shown in Fig.~\ref{fig:smearing_test}, where we compare event-averaged $dE/d\eta_s$ computed from the MC-EKRT partons to those obtained after smearing with different values of $\sigma_\parallel$.

\begin{figure}
    \centering
    \includegraphics[width=0.49\textwidth]{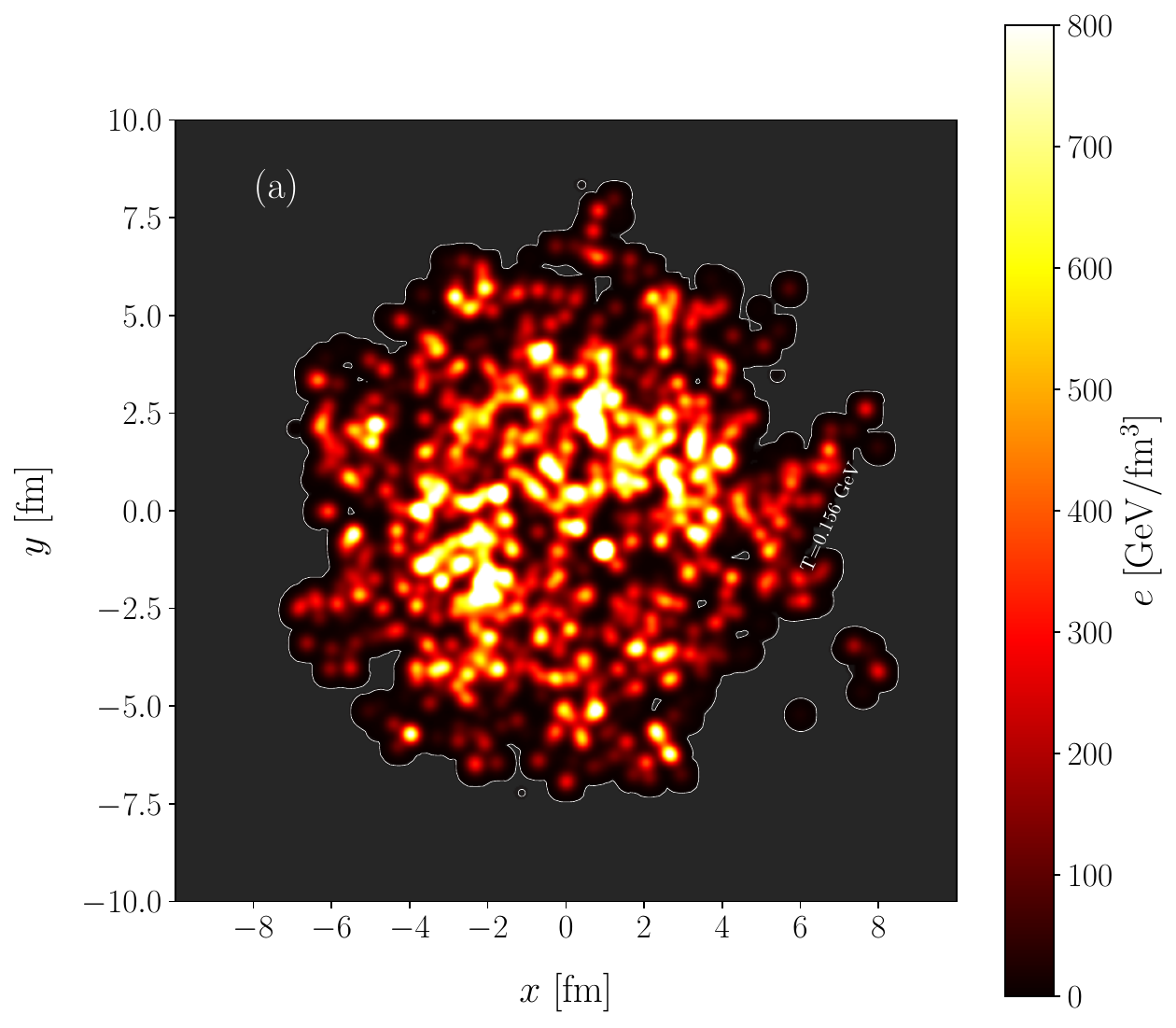}
    \centering
    \includegraphics[width=0.49\textwidth]{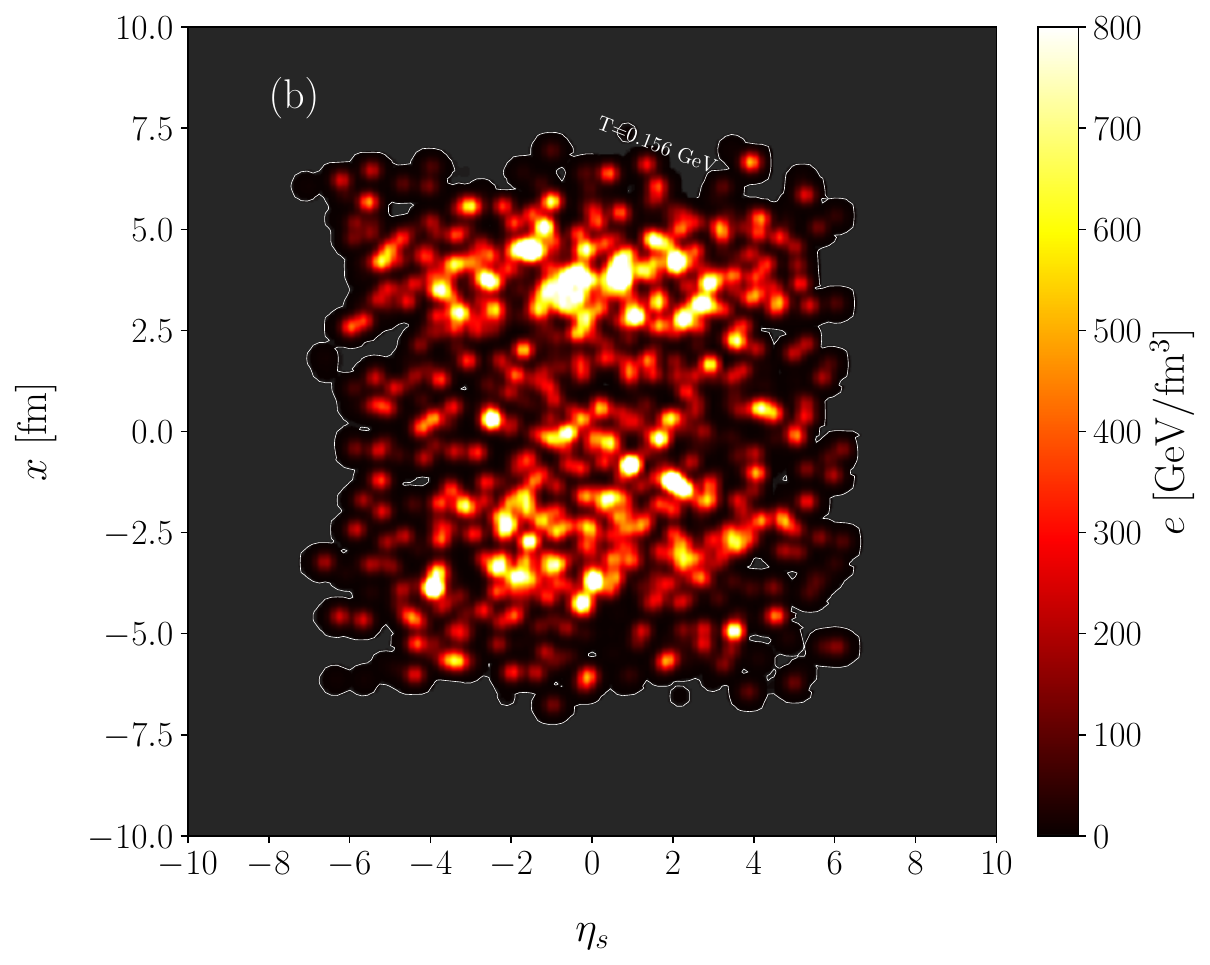}
    \caption{Initial energy density profile above the QCD transition temperature $T_c=0.156$ GeV \cite{HotQCD:2018pds}
    at $\tau=\tau_0$ computed from a single MC-EKRT event in the $x$-$y$ plane (a), and in the $\eta_s$-$x$ plane (b).}
    \label{fig:energyprofile_1ev}
\end{figure}

\subsubsection{Averaging initial conditions}
\label{SS:iniaveragin}
The above construction gives us the initial energy density event-by-event. As an example, the energy density distribution at $\tau=\tau_0$ obtained from a single event is plotted in the $x$-$y$ and $\eta_s$-$x$ planes in panels (a) and (b) of Fig.~\ref{fig:energyprofile_1ev}, respectively. Here we, however, want to avoid computationally very intensive 3+1D event-by-event hydrodynamic simulations, and therefore compute event-averaged initial conditions.
As explained in Sec.~\ref{SS:centrality}, we perform first the centrality selection according to the total initial transverse energy computed from the partons, and average the initial conditions within each centrality class. The hydrodynamic evolution is then computed for each event-averaged initial conditions, i.e.\ one hydrodynamic simulation per centrality class.

We first convert each event-by-event initial energy-density profile to an entropy-density profile using the EoS, and then average the entropy-density profiles and convert the averaged entropy density back to energy density. The reason for this is that the total initial entropy and the final hadron multiplicity have nearly a linear relation, and therefore averaging over the entropy-density profiles rather than over the energy-density profiles is a better approximation for obtaining the event-averaged final multiplicities, and their centrality dependence \cite{Auvinen:2020mpc}. The difference here comes from the non-linear relation between the energy and entropy densities. The linear relation between the multiplicity and the initial entropy is somewhat broken by event-by-event fluctuations in the entropy production due to dissipation, but those fluctuations relative to total entropy production are typically small in central and semi-central collisions~\cite{Niemi:2015qia}.

\section{Results}\label{S:results}

In the following, we have applied MC-EKRT to $5.02$ TeV and $2.76$ TeV Pb+Pb, and $200$ GeV Au+Au collisions. In particular, we explore here how the centrality and pseudorapidity  dependence of charged particle multiplicity at different collision energies is affected by different choices of the Gaussian smearing and shear viscosity. We will also discuss the role of the energy conservation at different collision energies.

For each investigated collision system 100 000 minimum bias events were produced and sorted in centrality classes based on their initial transverse energy $E_T$. The Gaussian smearing widths were chosen to be $\sigma_\perp=0.15$ or $0.4$ fm in the transverse plane and the longitudinal smearing width was fixed to $\sigma_{\parallel}=0.15$. The ratio of shear viscosity to entropy density $\eta/s$ was taken either as constant, tuned to approximately reproduce the elliptic flow measurements at RHIC and LHC, or to follow the temperature dependent $\eta/s=param1$ from Ref.~\cite{Niemi:2015qia} (see Fig.~1 there).

The free parameters in the MC-EKRT model, namely $K$ and $\kappa_{\rm sat}$, were tuned to approximately reproduce the centrality dependence of charged particle multiplicity at midrapidity. The saturation parameter $\kappa_{\rm sat}$ was kept the same for all systems, but the pQCD $K$-factor was tuned for each collision system separately. We note that the parameter values quoted here are specific to these realizations of MC-EKRT computation, and are different for different choices of e.g.\ smoothing and viscosity. Also event-by-event fluctuations would likely change these values.
\subsection{Data comparison with event-averaged initial state}

\subsubsection{Charged particle pseudorapidity distribution}

    \begin{figure*}[!]
        \includegraphics[width = 0.42\textwidth]{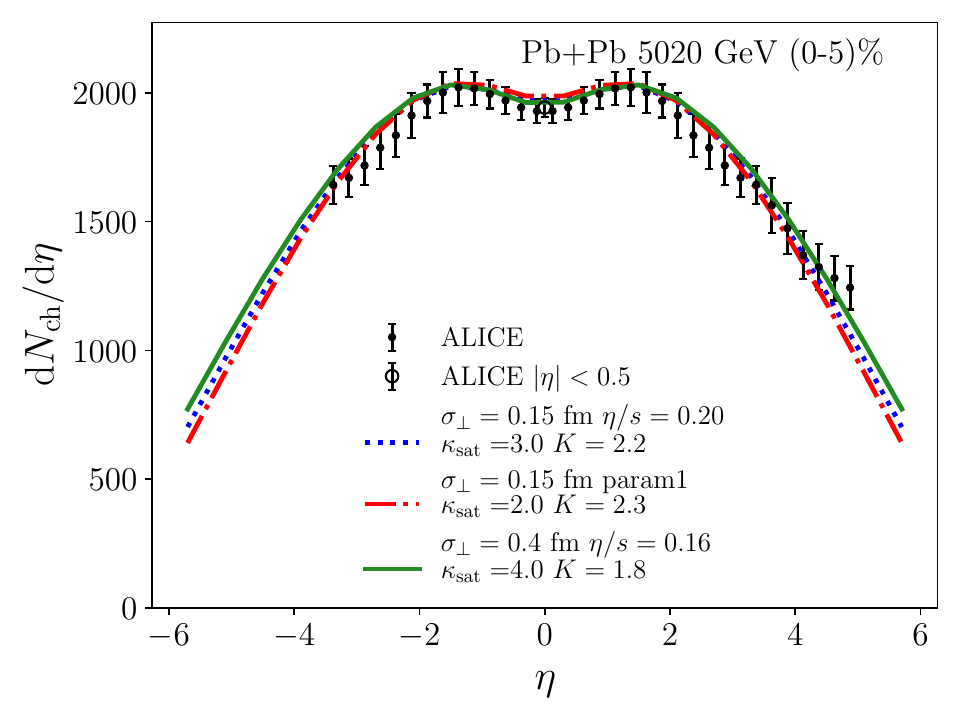}
        \includegraphics[width = 0.42\textwidth]{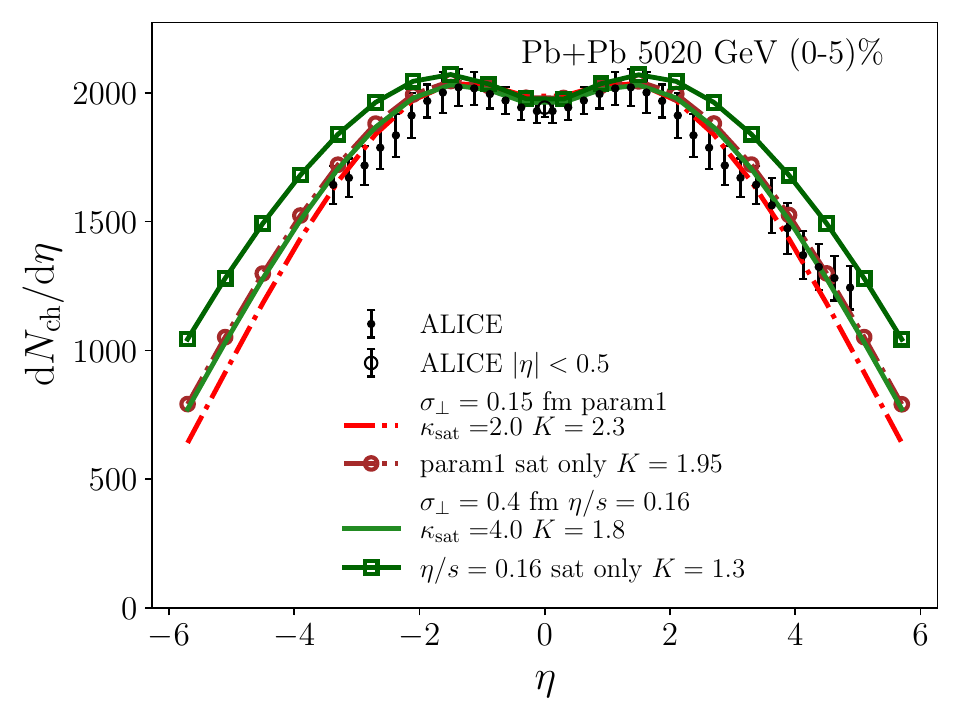}
        \vspace{-0.2cm}
        \includegraphics[width = 0.42\textwidth]{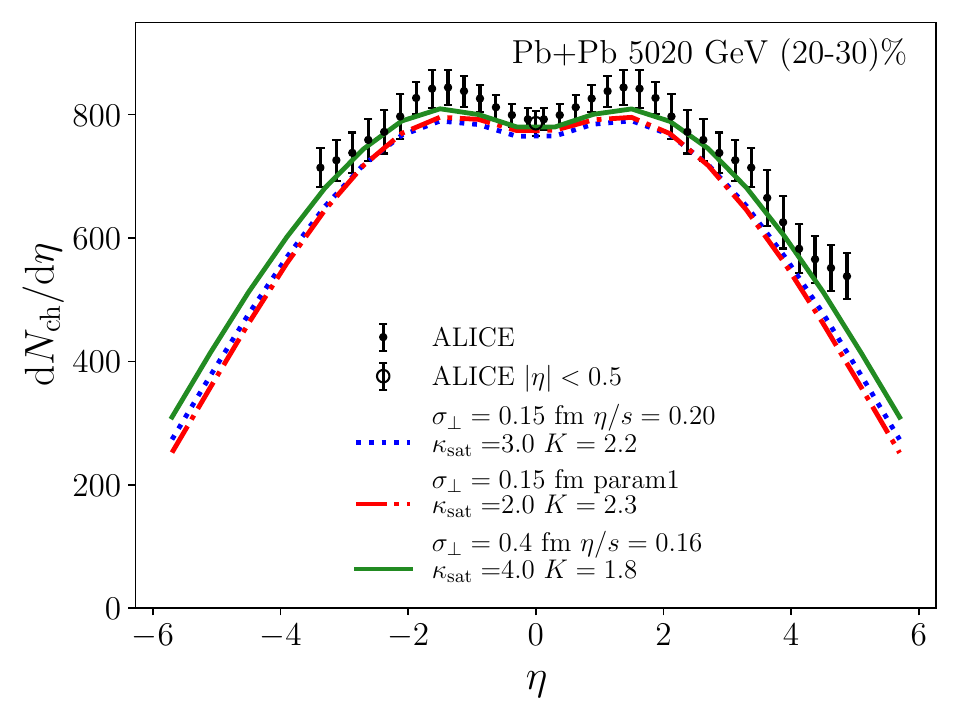}
        \includegraphics[width = 0.42\textwidth]{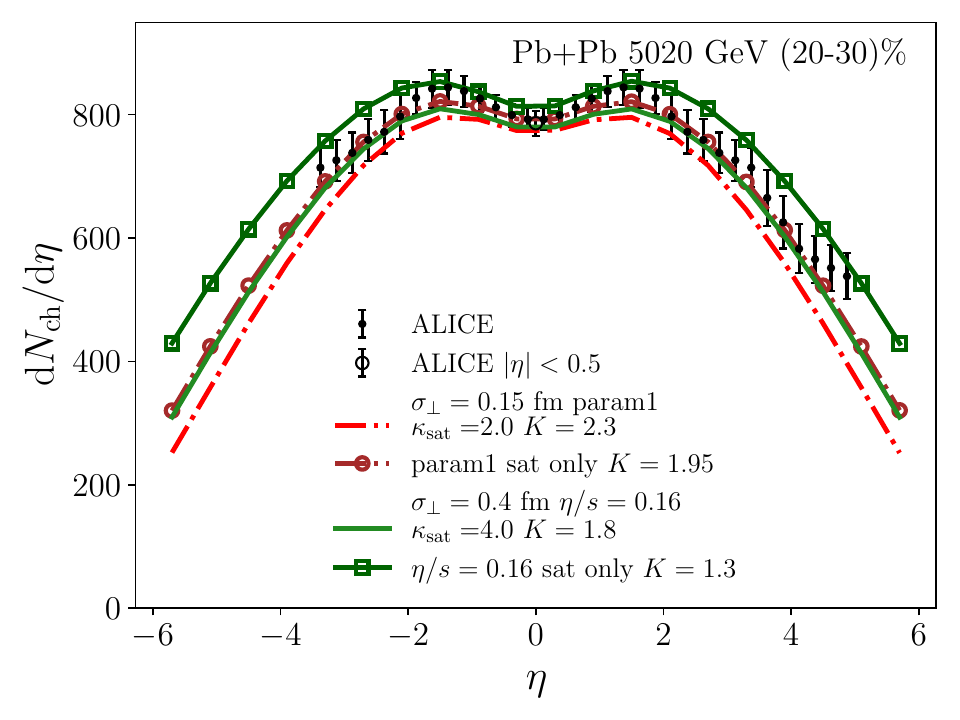}
        \vspace{-0.25cm}
        \includegraphics[width = 0.42\textwidth]{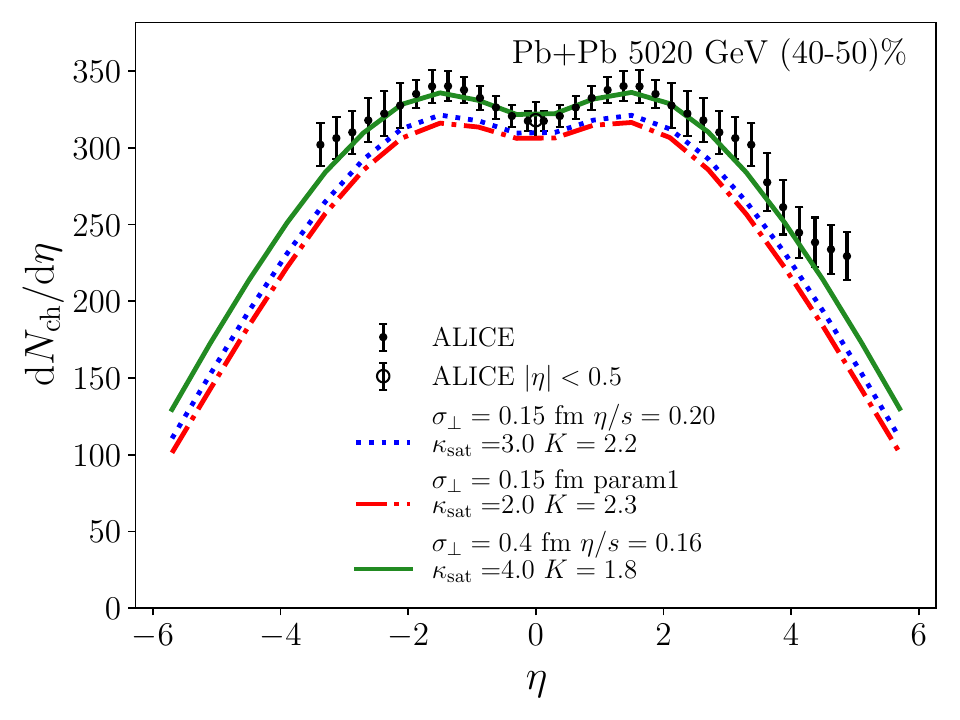}
        \includegraphics[width = 0.42\textwidth]{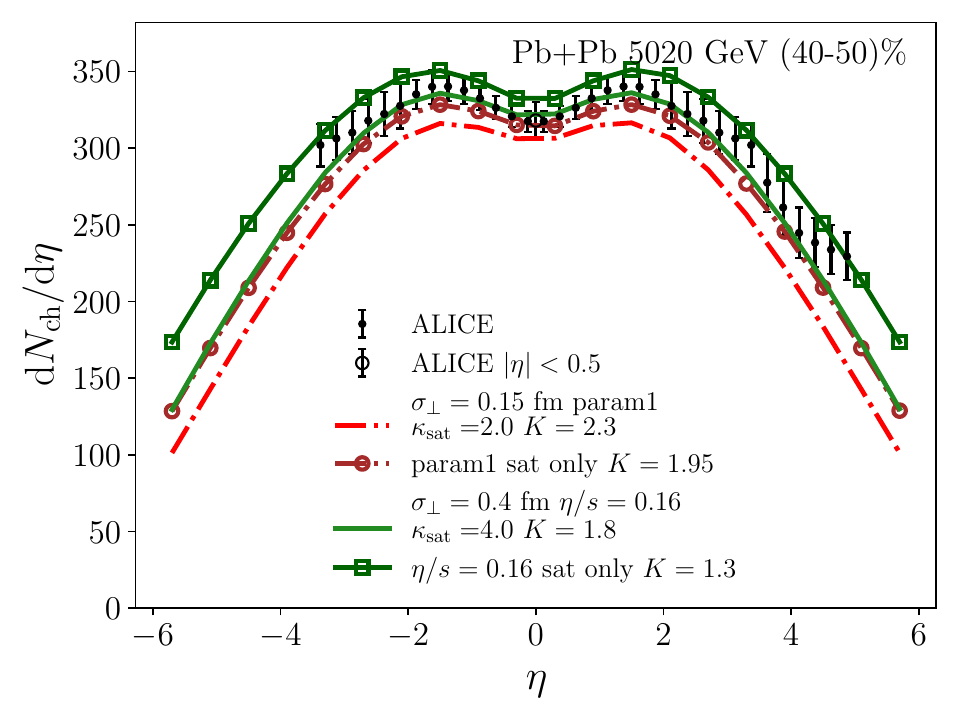}
        \vspace{-0.25cm}
        \includegraphics[width = 0.42\textwidth]{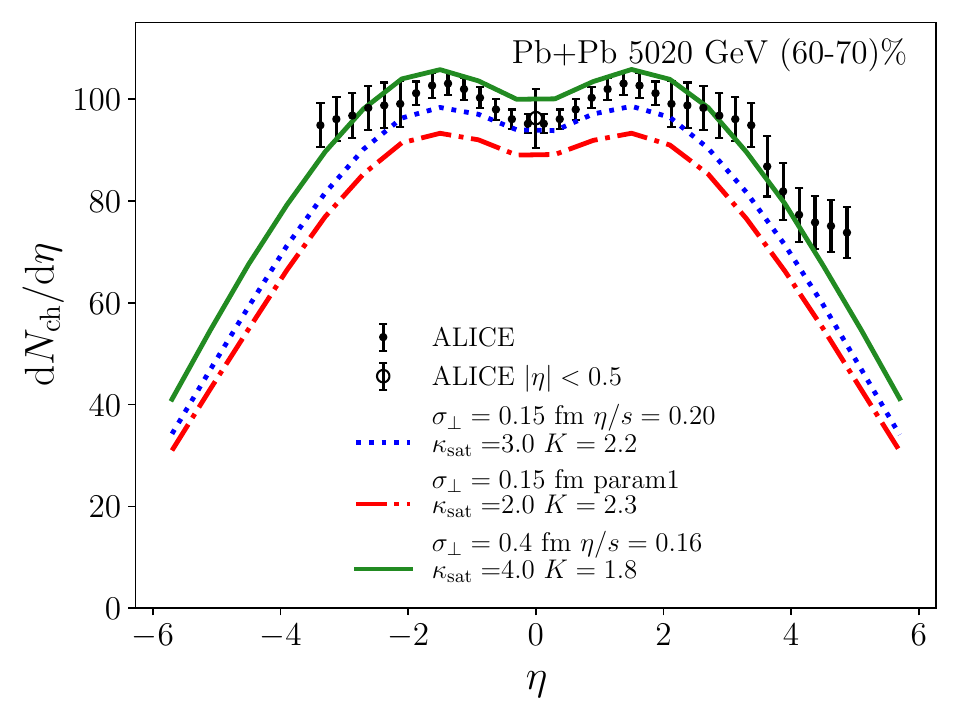}
        \includegraphics[width = 0.42\textwidth]{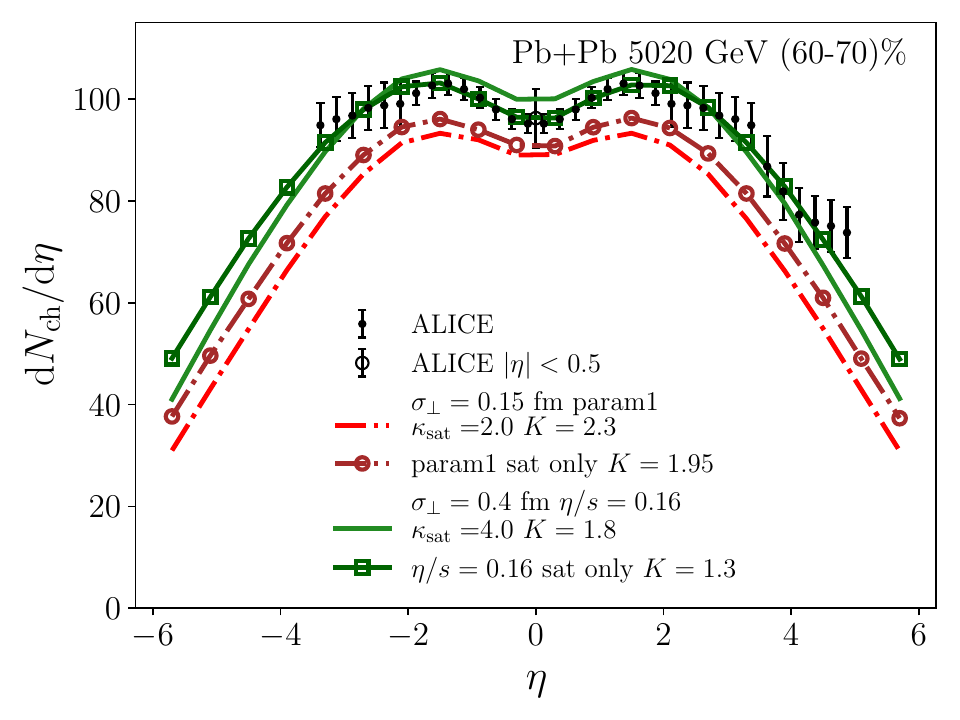}
        \vspace{-0.3cm}
        \caption{Charged particle multiplicity $\mathrm{d}N_{\mathrm{ch}}/\mathrm{d}\eta$ as a function of pseudorapidity in Pb+Pb collisions at $\sqrt{s_{NN}}=5.02$ TeV, compared with ALICE data~\cite{ALICE:2016fbt} (filled markers) and \cite{ALICE:2015juo} (open markers). Left panels show the results with all the filters on, and the curves with markers in the right panels show the results with only the saturation filter on. The solid green and dashed-dotted red curves are the same in the left and right panels.}
        \label{F:dnchdeta5tev}
    \end{figure*}

    \begin{figure*}[!]
        \includegraphics[width = 0.42\textwidth]{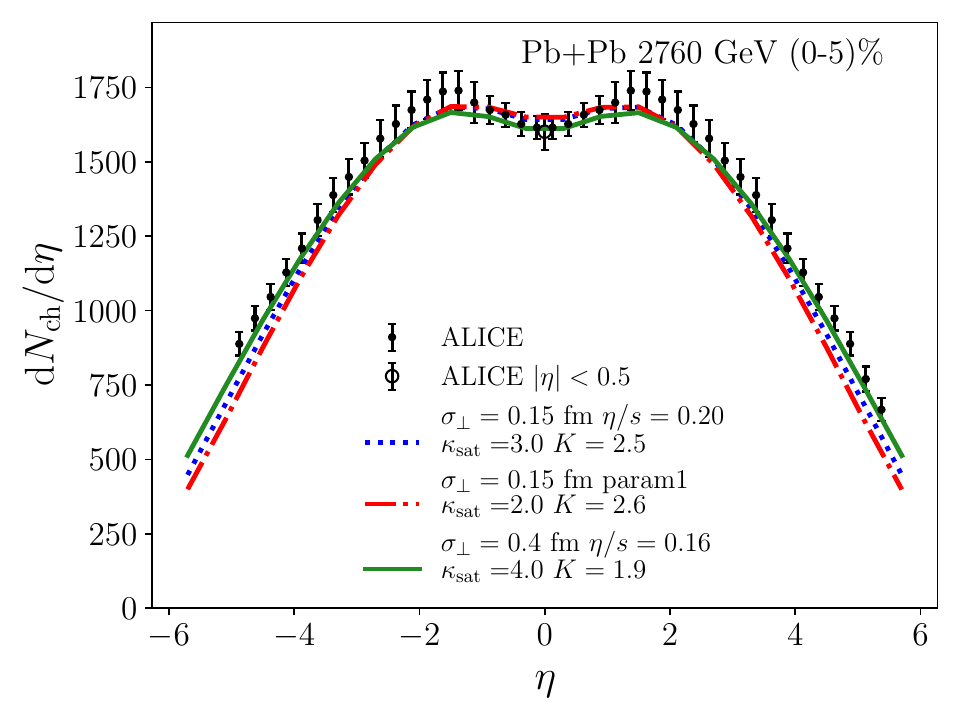}
        \includegraphics[width = 0.42\textwidth]{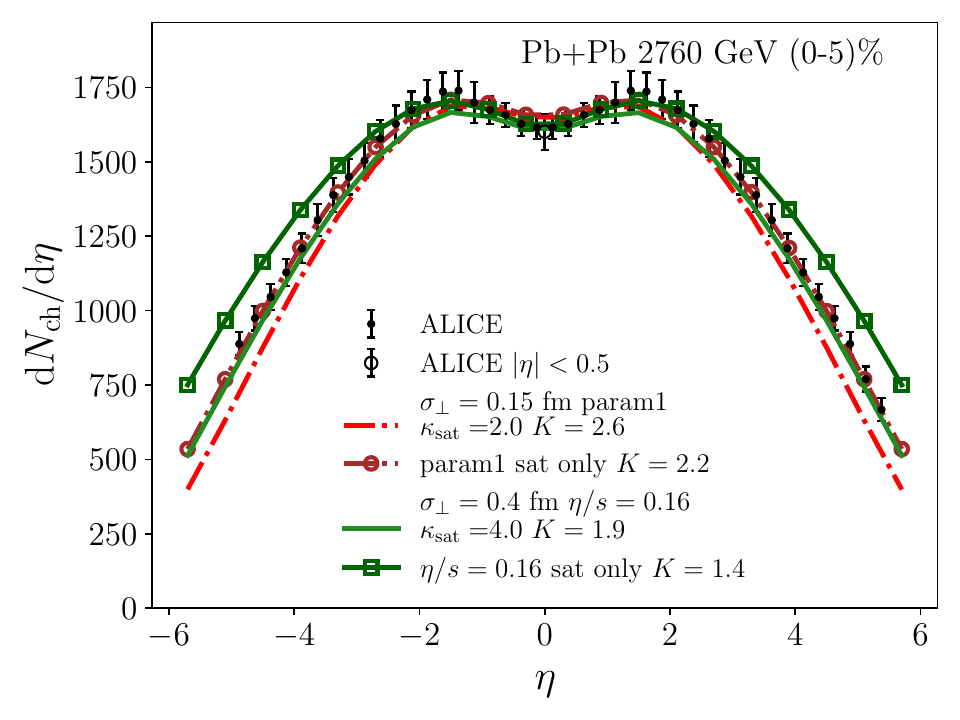}
        \vspace{-0.2cm}
        \includegraphics[width = 0.42\textwidth]{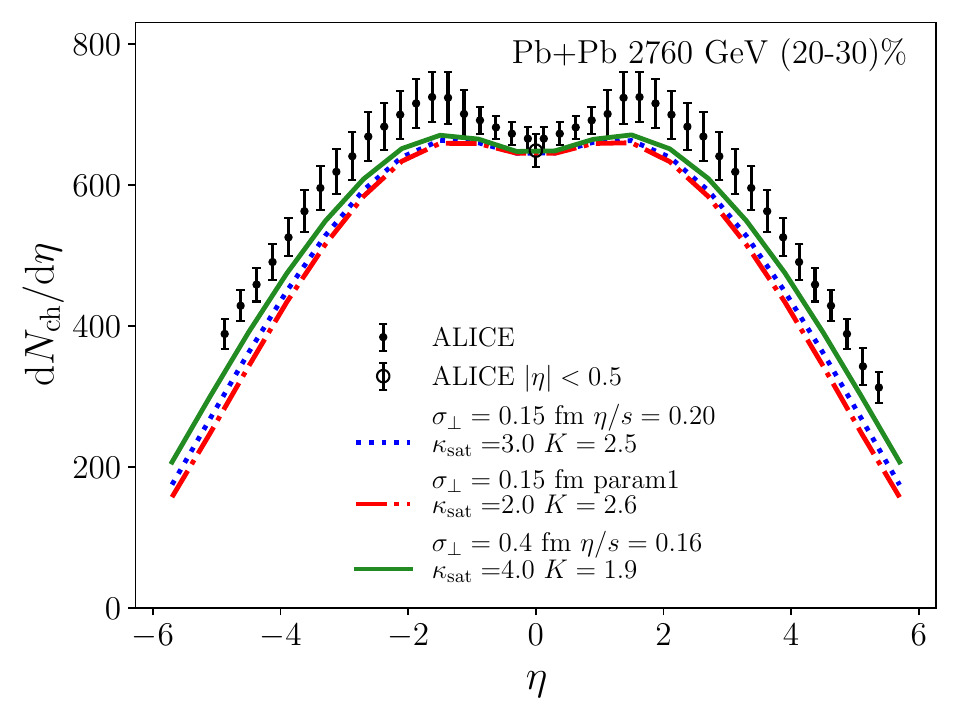}
        \includegraphics[width = 0.42\textwidth]{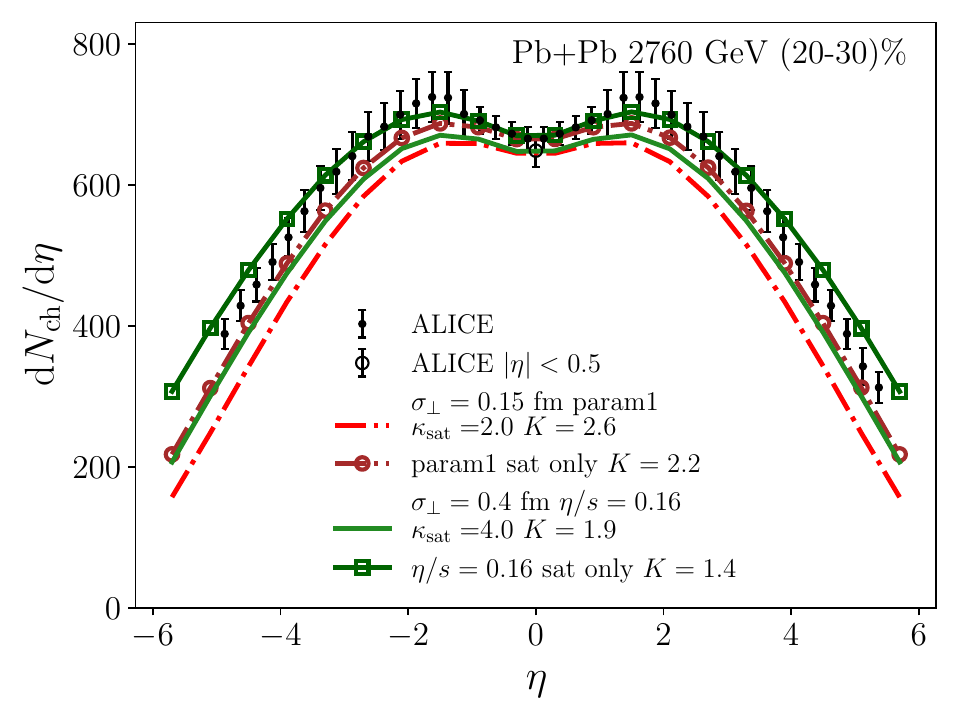}
        \vspace{-0.25cm}
        \includegraphics[width = 0.42\textwidth]{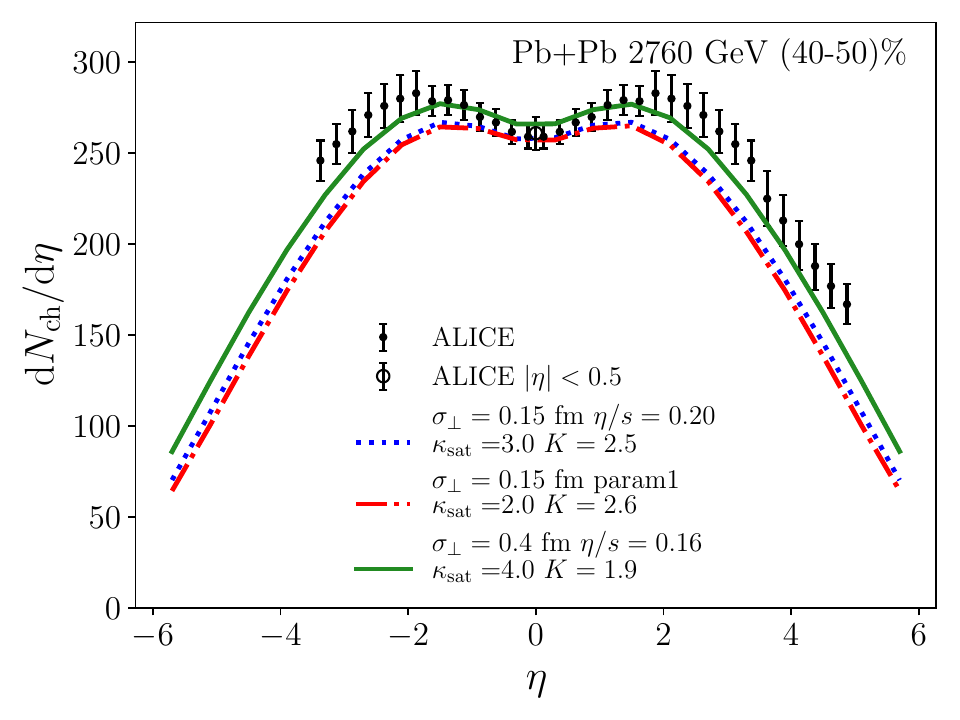}
        \includegraphics[width = 0.42\textwidth]{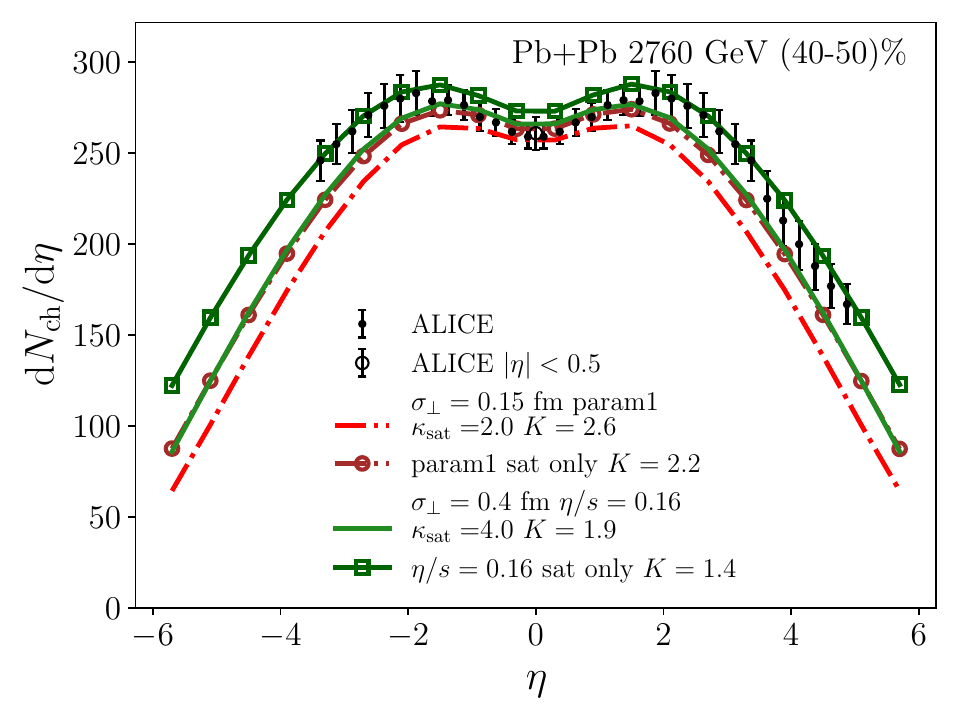}
        \vspace{-0.25cm}
        \includegraphics[width = 0.42\textwidth]{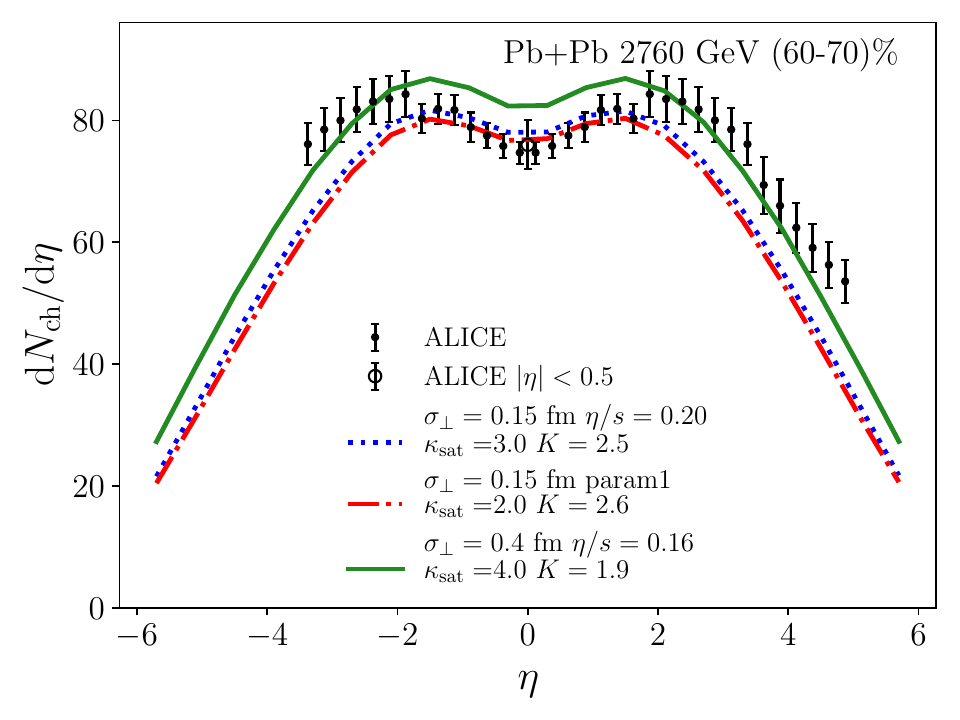}
        \includegraphics[width = 0.42\textwidth]{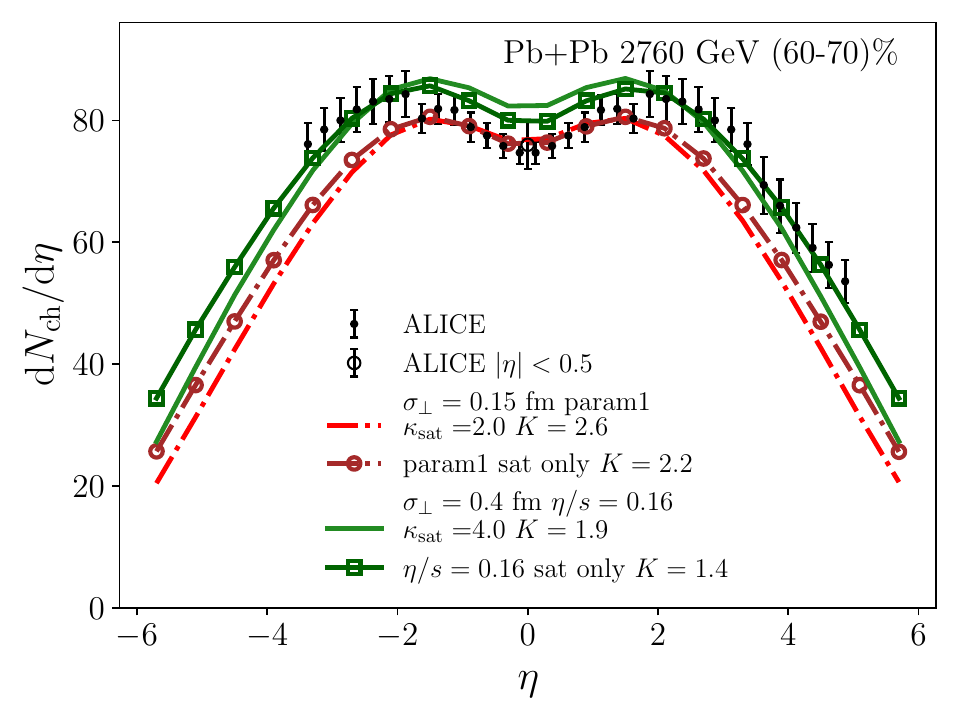}
        \vspace{-0.3cm}
        \caption{Charged particle multiplicity $\mathrm{d}N_{\mathrm{ch}}/\mathrm{d}\eta$ as a function of pseudorapidity in Pb+Pb collisions at $\sqrt{s_{NN}}=2.76$ TeV, compared with ALICE data~\cite{ALICE:2013jfw} (filled markers) and \cite{ALICE:2010mlf} (open markers). Left panels show the results with all the filters on, and the curves with markers in the right panels show the results with only the saturation filter on. The solid green and dashed-dotted red curves are the same in the left and right panels.}
        \label{F:dnchdeta2tev}
    \end{figure*}

    \begin{figure*}[htb!]
        \includegraphics[width = 0.45\textwidth]{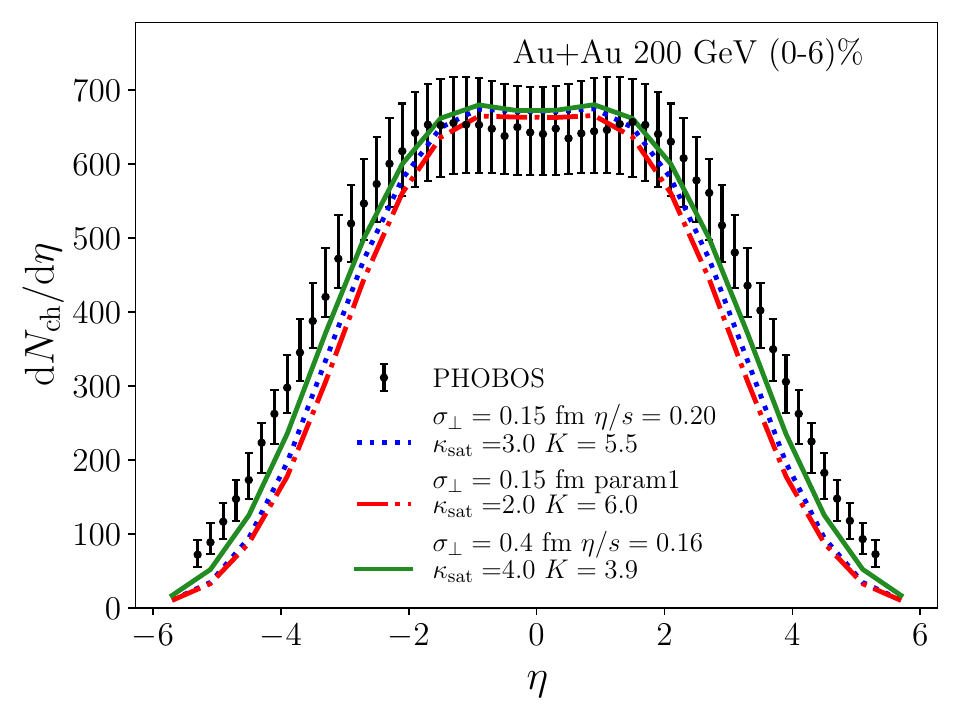}
        \includegraphics[width = 0.45\textwidth]{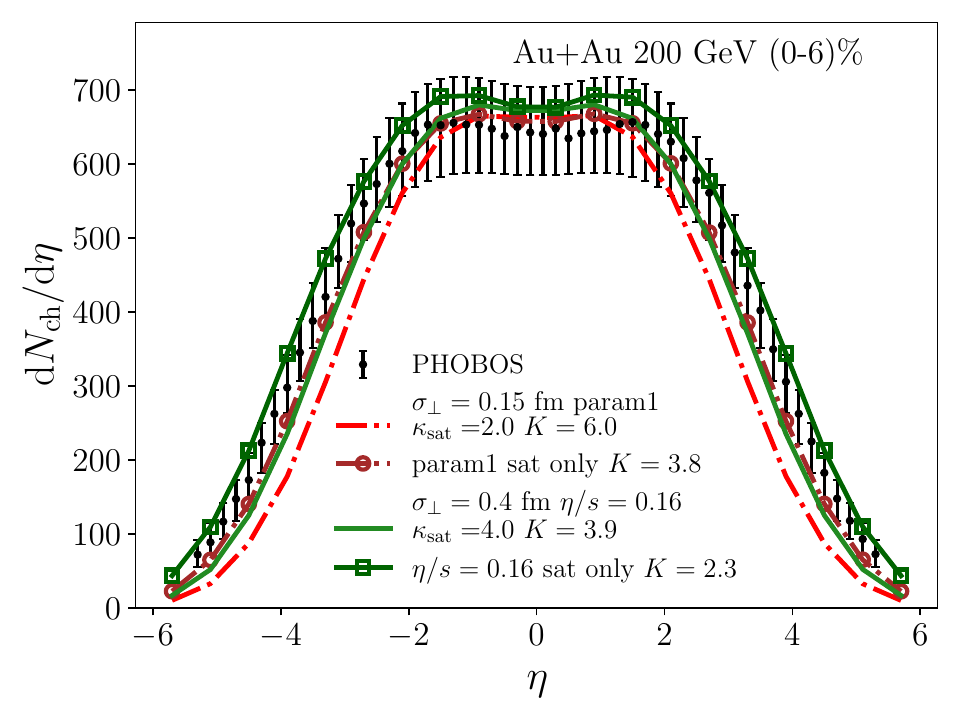}
        \includegraphics[width = 0.45\textwidth]{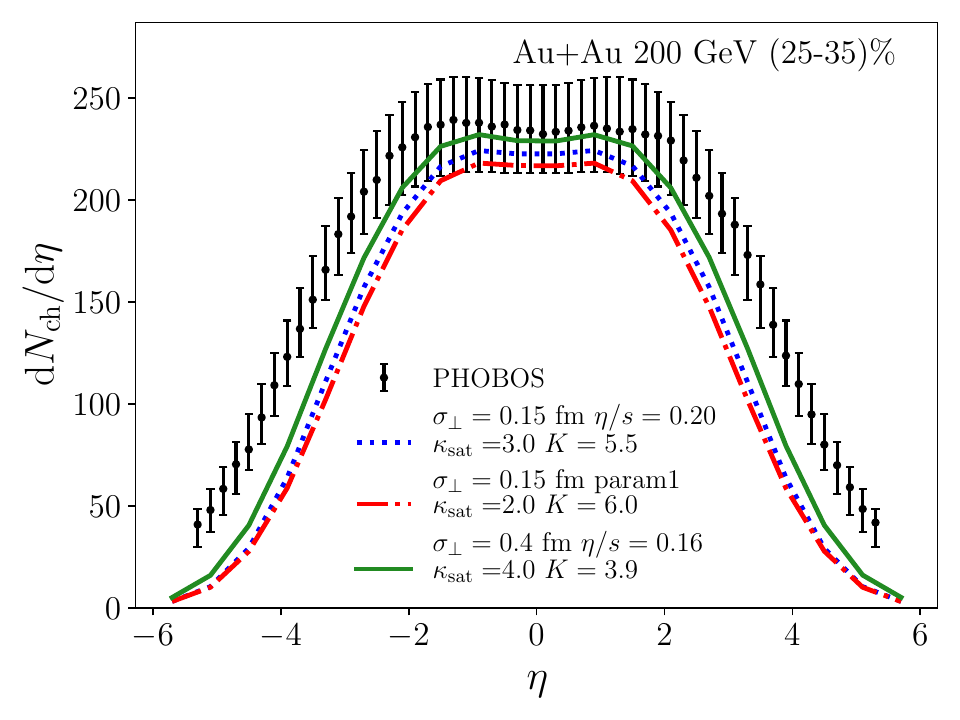}
        \includegraphics[width = 0.45\textwidth]{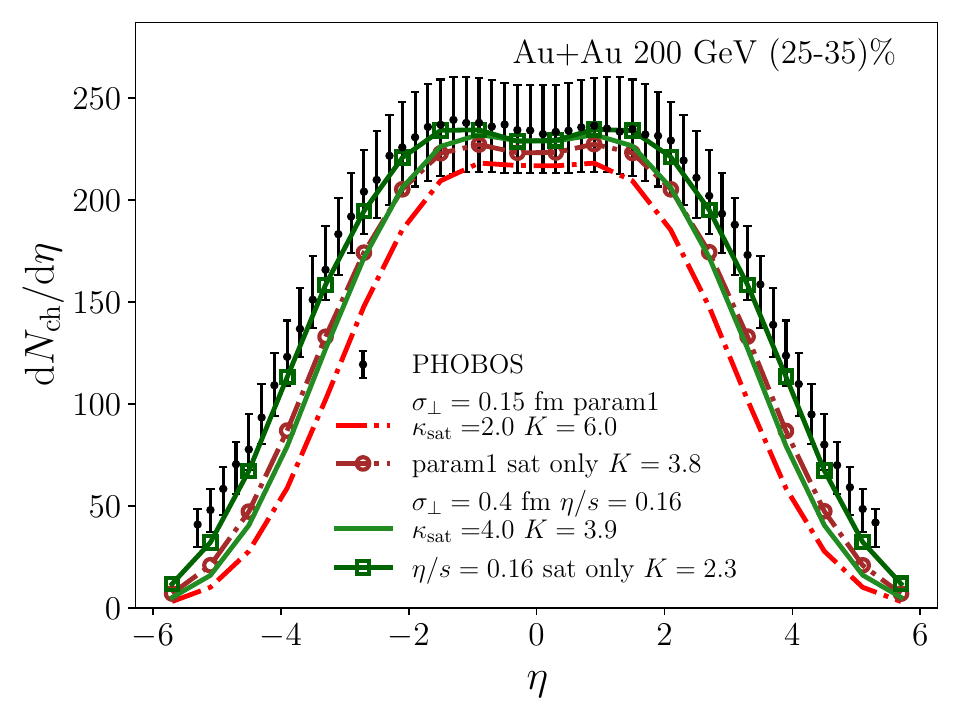}
        \includegraphics[width = 0.45\textwidth]{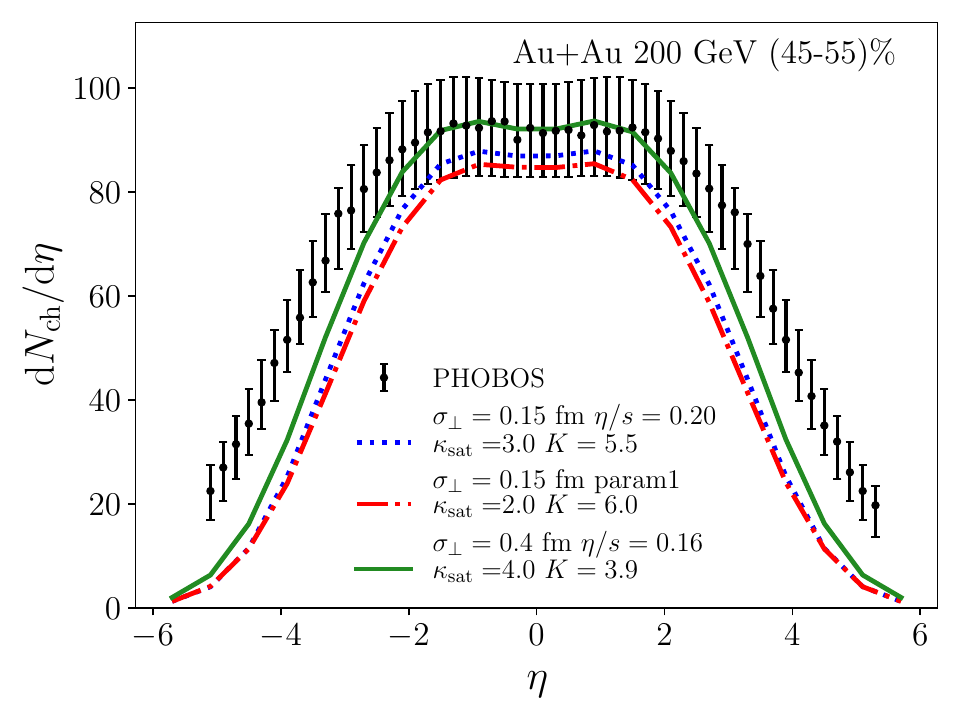}
        \includegraphics[width = 0.45\textwidth]{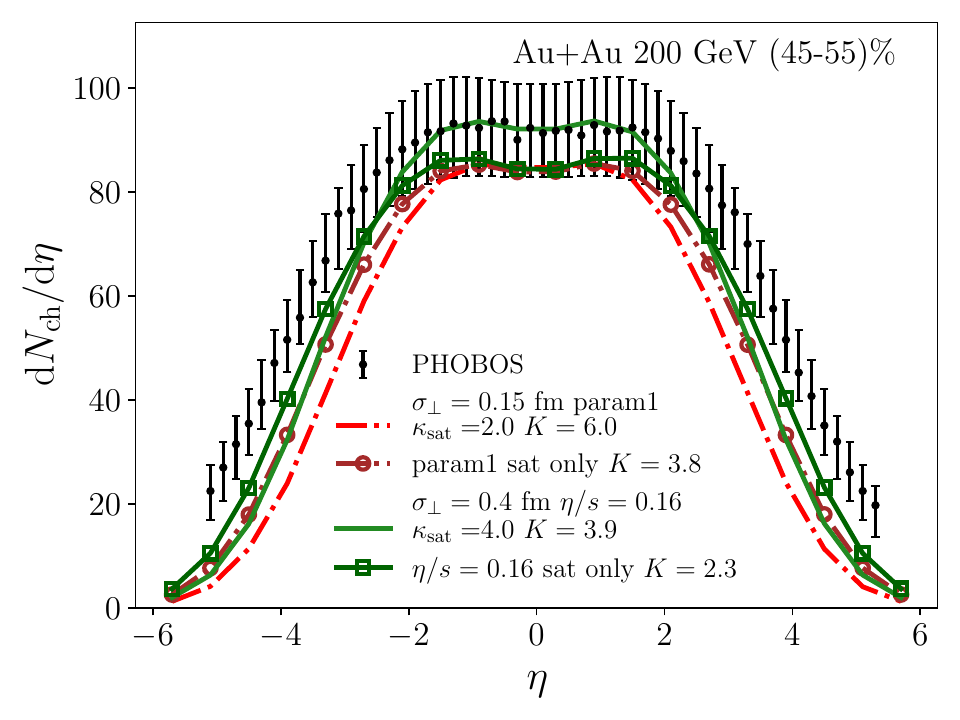}
        \caption{Charged particle multiplicity $\mathrm{d}N_{\mathrm{ch}}/\mathrm{d}\eta$ as a function of pseudorapidity in Au+Au collisions at $\sqrt{s_{NN}}=200$ GeV, compared with PHOBOS data~\cite{Back:2002wb}. Left panels show the results with all the filters on, and the curves with markers in the right panels show the results with only the saturation filter on. The solid green and dashed-dotted red curves are the same in the left and right panels.}
        \label{F:dnchdeta200gev}
    \end{figure*}

Figures~\ref{F:dnchdeta5tev}, \ref{F:dnchdeta2tev}, and \ref{F:dnchdeta200gev} show the charged particle pseudorapidity ($\eta$) distributions for $\sqrt{s_{NN}}=5.02$ TeV Pb+Pb, $2.76$ TeV Pb+Pb, and $200$ GeV Au+Au collisions, respectively. The centrality classes are quoted in the figures. We show all the cases tested here, namely $\sigma_\perp = 0.15$ fm with $\eta/s = 0.20$, $\sigma_\perp = 0.15$ fm with $\eta/s = param1$, and $\sigma_\perp = 0.4$ fm with $\eta/s = 0.16$. The values of $\kappa_{\rm sat}$ and $K$ for each case are indicated in the figures. The left panels show the full results where saturation, energy conservation, and valence-quark number conservation are taken into account. The curves with markers in the right panels show the results with saturation only, demonstrating the role of saturation in the energy conservation, as well as the role of the per-nucleon level energy conservation in narrowing the rapidity distributions.

The values for the $K$ factors that are needed to reproduce the data are increasing with decreasing collision energy. This is in line with the expectation that NLO corrections become increasingly important towards lower collision energy~\cite{Eskola:2000ji,Eskola:2000my}. We can, however, see that the centrality dependence of the multiplicity is well described by collision energy independent $\kappa_{\rm sat}$. This is already a non-trivial result, even if we have some freedom to tune the centrality dependence by changing $\kappa_{\rm sat}$. The range of the centrality dependence with different values of $\kappa_{\rm sat}$ is, as shown in Fig.~\ref{F:parameter_effect}, quite limited. Thus, the centrality dependence of multiplicity is relatively robust prediction of the MC-EKRT model, and the good agreement with the data is similar to the NLO EbyE EKRT model~\cite{Niemi:2015qia}, where 2+1 D fluid dynamics was employed.

A significant new feature in the MC-EKRT model is that we can obtain full 3D initial conditions, and subsequently we can compute the pseudorapidity dependence of the charged particle multiplicity. The overall agreement with the rapidity spectra is encouragingly good. At both LHC energies we can essentially reproduce the measurements in all the centrality classes. Only in the most peripheral collisions with $|\eta| >2$, we can start to see some more significant deviations from the shape of the measured rapidity distribution. In the most central collisions at RHIC the agreement is very similar as at the LHC. In peripheral collisions we start to get too narrow spectrum, but even then the agreement remains good up to $|\eta| \sim 2$.

The transverse smoothing range $\sigma_\perp$ and the $\eta/s$ parametrization slightly affect both the centrality dependence and the width of the rapidity spectra. The energy per unit rapidity is independent of $\sigma_\perp$, but since the conversion from energy density to entropy density is non-linear, the final multiplicity depends on $\sigma_\perp$. As a result, the rapidity spectra get wider with larger smoothing range. Temperature dependence of $\eta/s$ also affects  the width of the rapidity distribution through the entropy production. If $\eta/s$ increases with increasing temperature, the relative entropy production becomes larger at higher temperatures or energy densities, and the rapidity distribution becomes narrower than with a constant $\eta/s$. Even though the main features of the rapidity spectra are here coming from the MC-EKRT model, the finer details of the obtained spectra depend also on the details of the initialization and on the details of the fluid dynamical evolution.

In the right panels of Figs.~\ref{F:dnchdeta5tev}, \ref{F:dnchdeta2tev}, and \ref{F:dnchdeta200gev} we show the charged particle pseudorapidity distributions with saturation only, i.e.\ we do not explicitly impose the nucleon-level energy and valence-quark number conservations. As we can see from the figures, comparing the curves with and without the markers, the rapidity distributions become wider without the per-nucleon energy conservation. This is natural, as dijets with large rapidity carry a lot of energy, and are thus more constrained by the energy conservation. It is interesting to note that the saturation-only results can also reproduce the shape of the rapidity distribution in peripheral Au+Au collisions at RHIC. On the other hand, the saturation-only distributions with $\kappa_{\rm sat} = 4$ at the LHC tend to get too wide in the most central collisions.

We have checked that with the saturation-only $\kappa_{\rm sat} = 4$ central-collision cases, i.e.\ with weaker saturation, the energy conservation of the contributing nucleons is violated on the average already by $\sim 50$ \% at the LHC, and $\sim 20$ \% at RHIC. Interestingly, however, with the saturation-only $\kappa_{\rm sat} = 2$ central-collision cases, i.e.\ with stronger saturation, the average violation is only $\sim 5$ \% at the LHC, and energy is practically conserved at RHIC.

These results suggest that, given strong enough saturation, the total energy budget could be conserved even without a requirement of a tight per-nucleon energy conservation, supporting the view that the high-energy nuclear collisions can be described as collisions of two parton clouds rather than as a collection of sub-collisions of individual nucleons.

    \begin{figure*}[htb!]
        \includegraphics[width = 0.45\textwidth]{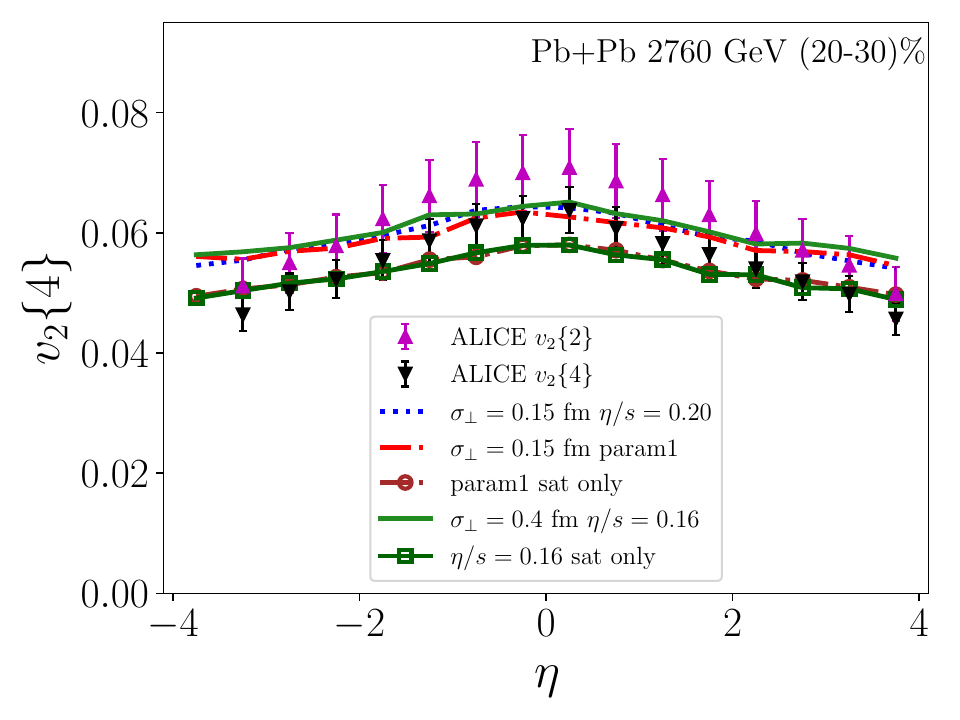}
        \includegraphics[width = 0.45\textwidth]{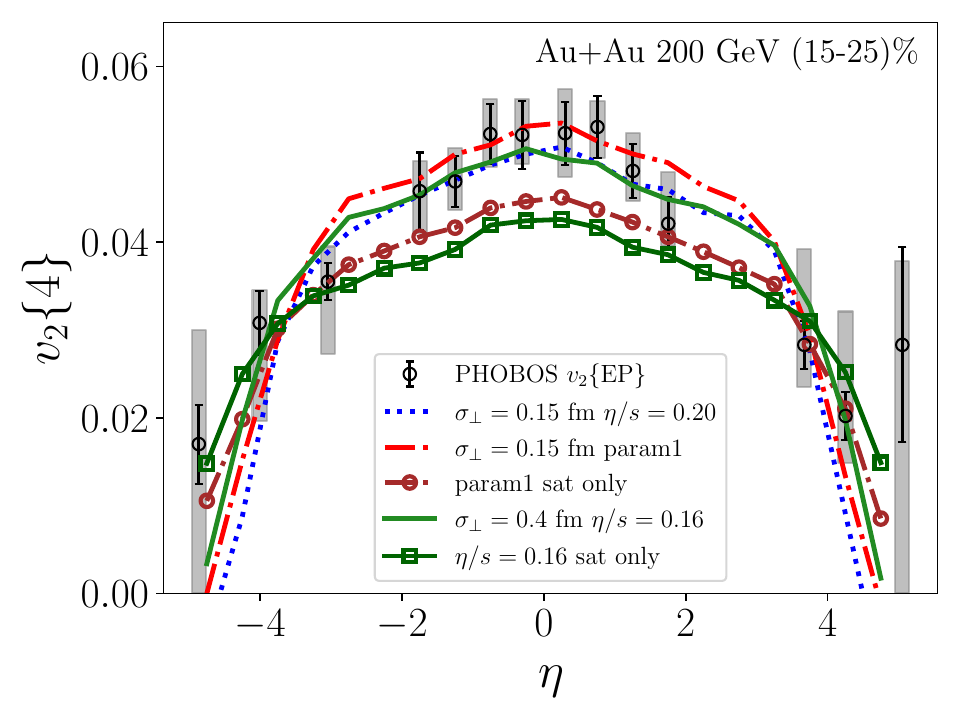}
        \caption{Charged particle $v_2\{4\}$ as a function of pseudorapidity in Pb+Pb collisions at $\sqrt{s_{NN}}=2.76$ TeV, in the 20-30~\% centrality class, compared with ALICE data~\cite{ALICE:2016tlx} (left panel),
        and in Au+Au collisions at $\sqrt{s_{NN}}=200$ GeV, in the 15-25~\% centrality class, compared with PHOBOS hit-based event plane $v_2$ data~\cite{PHOBOS:2004vcu} (right panel). The curves without markers show the results with all filters on, and the ones with markers the saturation-only cases. The parameter setups and the curve labelings are the same as in Figs.~\ref{F:dnchdeta5tev}-\ref{F:dnchdeta200gev}.}
        \label{F:dv2deta}
    \end{figure*}

\nocite{*}

\subsubsection{Charged particle elliptic flow}

Figure~\ref{F:dv2deta} shows the pseudorapidity dependence of elliptic flow, the second-order Fourier coefficient  $v_2\{4\}$ of the azimuthal angle distribution of charged hadrons, in semi-central $2.76$ TeV Pb+Pb and $200$ GeV Au+Au collisions. The model results are calculated using the 4-particle cumulant method~\cite{Bilandzic:2010jr}. Since our initial energy density profiles are averages over multiple events, $v_2\{\mathrm{EP}\} \approx v_2\{2\}\approx v_2\{4\}$ \footnote{PHOBOS states in Ref.~\cite{PHOBOS:2004vcu} that their event plane $v_2\{\mathrm{EP}\}$ results are most consistent with the 4-particle cumulant method, so we consider $v_2\{\mathrm{EP}\}$ and $ v_2\{4\}$ to be comparable in this particular case.}.

The $\eta$-differential flow is determined with respect to a reference flow vector, which is typically constructed from particles in a separate rapidity bin to avoid autocorrelations. For the comparison with the ALICE data \cite{ALICE:2016tlx}, the reference flow vector is calculated using particles in the TPC pseudorapidity acceptance $|\eta_{\mathrm{ref}}|<0.8$ and in addition there is also  a $p_T$ cut $(0.2 <p_T < 5.0)$~GeV. When calculating $v_2(\eta)$ in the rapidity bins with $|\eta|>2.0$, the particles in the $\eta$ bin are correlated with the full reference flow vector. For the rapidity bins with $|\eta|<2.0$, the particles with $\eta<0$ are correlated with the positive-rapidity reference particles $0<\eta_{\mathrm{ref}}<0.8$, while the negative reference $-0.8<\eta_{\mathrm{ref}}<0$ is used for particles with $\eta>0$. In the PHOBOS comparison \cite{PHOBOS:2004vcu}, the reference flow for the $\eta<0$ bins is determined from particles in the pseudorapidity range $0.1 < \eta_{\mathrm{ref}} < 2.0$ and the reference for $\eta > 0$ is determined from particles in the range $-2.0 < \eta_{\mathrm{ref}} < -0.1$.

As our average initial energy density profiles lack event-by-event fluctuations, at present the comparison to $v_2$ data has to be considered more qualitative than quantitative in nature. Nevertheless, the currently observed trends look very promising; the magnitude of $v_2$ is already close to data for both investigated collision systems, and we observe stronger dependence on pseudorapidity at $200$ GeV compared to $2.76$ TeV, as is also suggested by the data. This steeper fall-off of $\mathrm{d}v_2/\mathrm{d}\eta$ at RHIC can be understood as a sign of incomplete conversion of spatial eccentricity into momentum anisotropy due to the shorter lifetime of the hot QCD medium at lower collision energies. This result is rather robust with respect to the implementation details of the MC-EKRT initialization. The largest effect is seen when relaxing the energy conservation requirement, which leads to a visible decrease in $v_2$, but in this case we have not tried to adjust $\eta/s$ to reproduce the data.

\begin{figure*}[!]
\centering
\includegraphics[width = 0.45\textwidth]{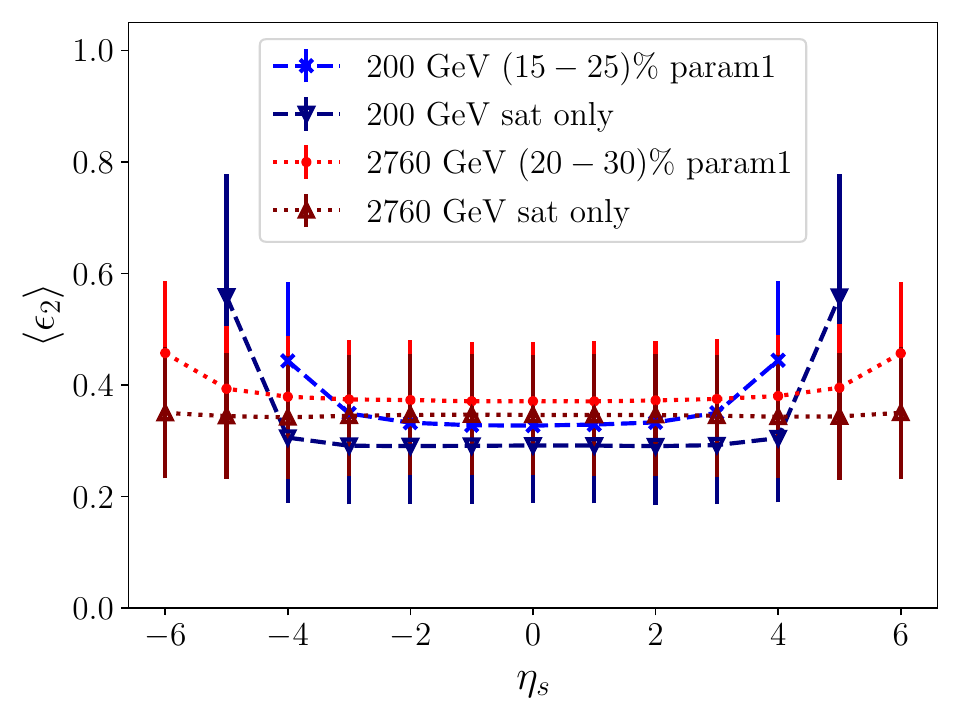}
\includegraphics[width = 0.45\textwidth]{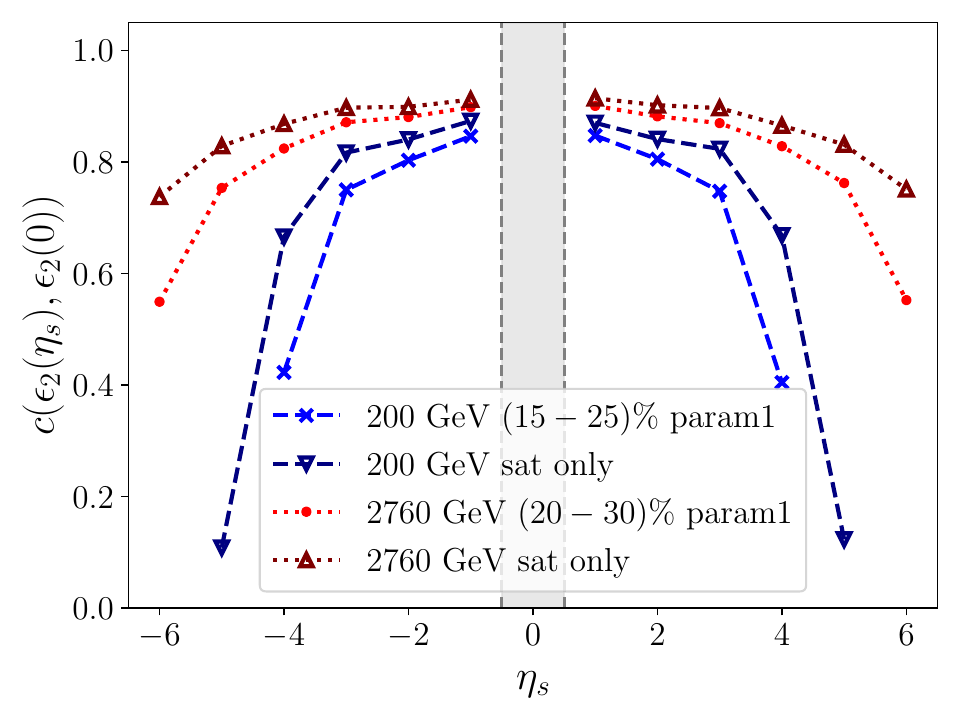}
\caption{Spacetime rapidity dependence of the event-averaged eccentricity (left panel) and the Pearson correlation (right panel) between $\epsilon_2(\eta_s)$ and $\epsilon_2(\eta_s=0)$ (midrapidity bin indicated by the gray band) in 20-30 \% central Pb+Pb collisions at $\sqrt{s_{NN}}=2760$ GeV and in 15-25 \% central Au+Au collisions at $\sqrt{s_{NN}}=200$ GeV. The dashed blue and dotted red curves show the results with all filters on, and the dashed dark blue and dotted brown curves the saturation-only cases. The errorbars show the standard deviation of the calculation.}
 \label{F:e2pearson}
\end{figure*}

\subsection{Event-by-event fluctuations of the initial state eccentrities}

Even though we have not performed here event-by-event fluid dynamical evolution, we can still compute the initial state eccentricities event-by-event, and in particular examine the decorrelation of the eccentrities as a function of spacetime rapidity. The spatial eccentricity vector with the magnitude $\epsilon_2$ pointing at the angle $\Psi_2$ can be defined as a complex number constructed from a weighted average,
\begin{eqnarray}
\epsilon_2 e^{i2\Psi_2}&=&\frac{\sum_k w_k r_k^2 e^{i2\phi_k}}{\sum_k w_k r_k^2} \nonumber \cr
&=& \frac{\sum_k w_k r_k^2 (\cos(2\phi_k) + i \sin(2\phi_k))}{\sum_k w_k r_k^2}.
\label{E:epsilon2}
\end{eqnarray}
Here $r$ and $\phi$ indicate the polar coordinates (radius and angle) in the transverse plane: $r^2 = \hat{x}^2+\hat{y}^2$, $\cos(2\phi)=(\hat{x}^2-\hat{y}^2)/r^2$ and  $\sin(2\phi)=2\hat{x}\hat{y}/r^2$, where we have defined $\hat{x}=x-x_{\mathrm{cm}}$ and $\hat{y}=y-y_{\mathrm{cm}}$ with respect to the center-of-mass point $(x_{\mathrm{cm}}, y_{\mathrm{cm}})=\left(\frac{\sum_k w_k x_k}{\sum_k w_k},  \frac{\sum_k w_k y_k}{\sum_k w_k} \right)$. The weight $w_k$ is the initial energy density at $\tau=\tau_0$ in a hydro cell and the sum is over the cells in a transverse slice of the hydro grid which has the width $\Delta \eta_s$.

Once we have determined the eccentricities for each event, we can compute the Pearson correlation of the eccentricity magnitudes between different rapidity bins $\eta_s$ and $\eta_{s0}$,
\begin{align}
c(\epsilon_2(\eta_s), &\epsilon_2(\eta_{s0})) \notag \\
 = &\frac{\langle (\epsilon_2(\eta_s) - \langle \epsilon_2(\eta_s) \rangle) (\epsilon_2(\eta_{s0}) - \langle \epsilon_2(\eta_{s0}) \rangle) \rangle}{\sigma(\epsilon_2(\eta_s))\sigma(\epsilon_2(\eta_{s0}))},
\end{align}
where $\langle \cdot \rangle$ indicates an average over events and $\sigma$ is the corresponding standard deviation.

In Fig.~\ref{F:e2pearson} we show the event-averaged eccentricities and the Pearson correlations between the eccentricities at finite rapidity $\epsilon_2(\eta_s)$ and midrapidity $\epsilon_2(\eta_{s0}=0)$ in Pb+Pb collisions at $\sqrt{s_{NN}}=2.76$ TeV in the LHC and in Au+Au collisions at $\sqrt{s_{NN}}=200$ GeV at RHIC. The rapidity bin width was chosen to be $\Delta \eta_s=1.0$. The event-averaged eccentricities remain nearly constant close to midrapidity, but both at RHIC and LHC the eccentricity starts to increase at higher rapidities. We also see that if we relax the energy conservation, the mid-rapidity eccentricities decrease by $\sim 10$ \% at the LHC, and $\sim 15$ \% at RHIC, which explains the decrease in $v_2$ in the saturation-only cases in Fig.~\ref{F:dv2deta}.

As seen in the right panel of Fig.~\ref{F:e2pearson}, the Pearson correlation becomes weaker at higher rapidities, and at RHIC the eccentricity  beyond $|\eta_s| \gtrsim 3.5$ is no longer correlated with midrapidity, while at the LHC the correlation spans a considerably larger rapidity range $|\eta_s| \lesssim 5.0$. The decreasing number of particles at RHIC compared to the LHC, and also towards larger rapidities leads to larger fluctuations of eccentricity, and therefore also to a weaker correlation with midrapidity at RHIC and at large rapidities. In the saturation-only cases the correlation is stronger at large rapidities than in the fully filtered cases. To our understanding, also this is a multiplicity effect, originating from the increased multiplicities at larger rapidities.

We also note that the Pearson correlation from the EbyE initial conditions should not be directly compared to the rapidity dependence of elliptic flow in Fig.~\ref{F:dv2deta}. The elliptic flow is computed from event-averaged initial conditions, and all the decorrelation effects disappear in the averaging. We rather expect that the decreasing multiplicity at larger rapidity leads to a shorter lifetime in the fluid evolution, and thus the conversion from eccentricity to elliptic flow is not completed at higher rapidities, and we get smaller $v_2$. In a full EbyE computation both the lifetime effect and the decorrelation effect would be present.

The CMS collaboration has defined the rapidity dependent correlation ratio \cite{CMS:2015xmx} as
\begin{equation}
r_2 = \frac{\langle v_2(-\eta) v_2(\eta_{\mathrm{ref}}) \cos 2\left[ \Psi_2(-\eta) - \Psi_2(\eta_{\mathrm{ref}}) \right] \rangle}{\langle v_2(\eta) v_2(\eta_{\mathrm{ref}}) \cos 2\left[ \Psi_2(\eta) - \Psi_2(\eta_{\mathrm{ref}}) \right] \rangle},
\end{equation}
where the $\eta$ bin is somewhere in the central rapidity region $0<\eta<2.5$, its negative-side counterpart is at $-\eta$, and the reference bin will be somewhere further away in forward rapidity $\eta_{\mathrm{ref}}>3.0$ to reduce nonflow effects. Since we have performed full MC-EKRT + fluid dynamics simulations only for event-averaged initial profiles, we are not able to study event-by-event fluctuations of $v_2$. We can, however, estimate $r_2$ from the pre-averaging eccentricities, assuming $v_2(\eta) \approx k \epsilon_2(\eta_s)$ for some proportionality factor $k$:
\begin{equation}
r_2 \approx \frac{\langle \epsilon_2(-\eta_s) \epsilon_2(\eta_{s,{\mathrm{ref}}}) \cos 2\left[ \Psi_2(-\eta_s) - \Psi_2(\eta_{s,{\mathrm{ref}}}) \right] \rangle}{\langle \epsilon_2(\eta_s) \epsilon_2(\eta_{s,{\mathrm{ref}}}) \cos 2\left[ \Psi_2(\eta_s) - \Psi_2(\eta_{s,{\mathrm{ref}}}) \right] \rangle},
\end{equation}
where $\Psi_2$ is obtained from Eq.~\eqref{E:epsilon2}. Using the same replacement $v_2 \rightarrow \epsilon_2$ we can also investigate the ``twist factor'' $R_{n|n;2} \equiv R_2$, by the ATLAS collaboration \cite{ATLAS:2017rij} where the ratio is controlled by the cosine term:
\begin{widetext}
\begin{equation}
R_2 \approx \frac{\langle \epsilon_2(-\eta_{s,{\mathrm{ref}}}) \epsilon_2(-\eta_s) \epsilon_2(\eta_s) \epsilon_2(\eta_{s,{\mathrm{ref}}}) \cos 2\left[ \Psi_2(-\eta_{s,{\mathrm{ref}}}) - \Psi_2(\eta_{s,{\mathrm{ref}}}) + ( \Psi_2(-\eta_s) - \Psi_2(\eta_s))  \right] \rangle}{\langle \epsilon_2(-\eta_{s,{\mathrm{ref}}}) \epsilon_2(-\eta_s) \epsilon_2(\eta_s) \epsilon_2(\eta_{s,{\mathrm{ref}}}) \cos 2\left[ \Psi_2(-\eta_{s,{\mathrm{ref}}}) - \Psi_2(\eta_{s,{\mathrm{ref}}}) - ( \Psi_2(-\eta_s) - \Psi_2(\eta_s))  \right] \rangle}.
\end{equation}
\end{widetext}

\begin{figure*}[!]
\centering
\includegraphics[width = 0.45\textwidth]{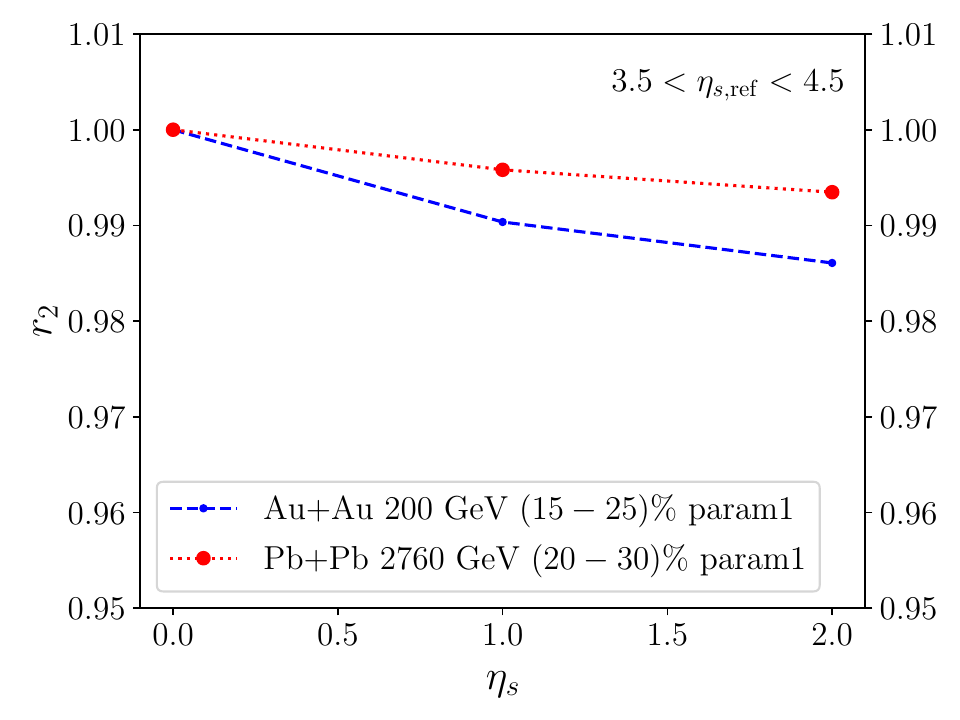}
\includegraphics[width = 0.45\textwidth]{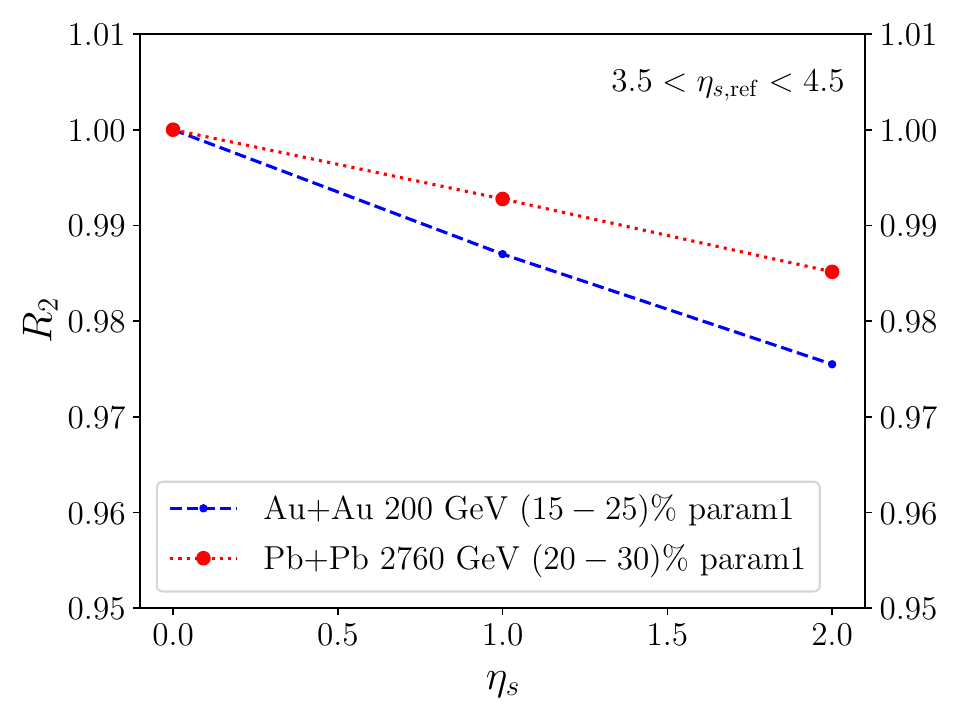}
\caption{Spacetime rapidity dependence of the correlation ratio $r_2$ (left panel) and the twist factor $R_2$ (right panel) in 20-30~\% central Pb+Pb collisions at $\sqrt{s_{NN}}=2760$ GeV and in 15-25~\% central Au+Au collisions at $\sqrt{s_{NN}}=200$ GeV.}
 \label{F:r2R2}
\end{figure*}

We show the eccentricity correlation ratio $r_2$ and twist factor $R_2$ in Fig.~\ref{F:r2R2} in 20-30 \% central Pb+Pb collisions at $\sqrt{s_{NN}}=2.76$ TeV and in 15-25 \% central Au+Au collisions at $\sqrt{s_{NN}}=200$ GeV. Both $r_2$ and $R_2$ show a similar behavior as the Pearson correlator, i.e.\ decorrelation at larger rapidities, and the decorrelation is stronger at RHIC than at the LHC. The decreasing trend of the calculated $r_2$ and $R_2$ is similar as seen in the CMS and ATLAS measurements, but the calculated $r_2$ and $R_2$ show slightly stronger correlations than the measurements do. However, as we do not perform event-by-event fluid dynamical evolution, a direct comparison is not really feasible here \cite{Pang:2015zrq}, but our results should be rather taken as qualitative.

\section{Conclusions}
\label{S:conclusions}

We have developed a new MC-EKRT model for computing initial particle production in ultrarelativistic heavy-ion collisions. This is an extension of the EbyE EKRT model~\cite{Niemi:2015qia}, which has very successfully predicted the midrapidity low-$p_T$ observables from 200 GeV Au+Au collisions at RHIC to the top energy LHC Pb+Pb collisions. The essential new feature in the new Monte-Carlo framework is the implementation of the dynamical minijet-multiplicity-originating fluctuations in the saturation and particle production. Also energy/momentum conservation and  valence-quark number conservation were implemented, together with a new type of spatially dependent nuclear PDFs that cope with the large density fluctuations present in an event-by-event study. As a result, the MC-EKRT model now gives
a full 3-dimensional initial state that can be coupled to 3+1 D fluid dynamics.

We have applied the novel MC-EKRT framework to 5.02 TeV Pb+Pb, 2.76 TeV Pb+Pb, and 200 GeV Au+Au collisions. The 3+1 D spacetime evolution is computed with viscous relativistic hydrodynamics~\cite{Molnar:2014zha}. We have studied the uncertainties related to converting the partonic state, given by the MC-EKRT, to an initial state of fluid dynamics, and also discussed the role of energy conservation in rapidity distributions of charged particle multiplicities and elliptic flow coefficients.

Although the MC-EKRT initial state model gives the full 3--dimensional initial state that include all the EbyE fluctuations, we have here made only an exploratory study of the final observables. The main simplification here is that we have first computed the initial conditions by averaging a large number of EbyE MC-EKRT initial states for each centrality class, and then computed the fluid dynamical evolution only for the averaged initial conditions. Obviously, this limits the number of observables that we can study, but it also decreases the computational cost by a huge amount. The averaging of the initial conditions is performed in such a way that the final multiplicities resemble as closely as possible those that would be obtained by a full EbyE computation.

The comparison with the measured charged particle multiplicities at the LHC and RHIC shows that MC-EKRT can describe the centrality dependence of the multiplicity very well, practically at the same level as the earlier EbyE EKRT implementation. Moreover, the new framework describes the rapidity dependence as well. The overall agreement with the measured shape of the rapidity spectra is very good. This is a non-trivial result, as it is a rather robust outcome from the MC-EKRT model, and essentially dominated by the pQCD minijet production and saturation. Note also that there is no parameter to directly control the rapidity distribution. Only in peripheral collisions at RHIC we start to see larger deviations from the data. Interestingly, we observed that when we give up the detailed nucleon-level energy conservation, the agreement with the data extends all the way to peripheral RHIC collisions. This might indicate that in the view that ultrarelativistic nuclear collision are rather collisions of parton clouds than collisions between individual nucleons, the nucleon-level energy conservation is an unrealistically strict condition.

We have also computed the rapidity dependence of elliptic flow, and the agreement with the LHC and RHIC data is good. The rapidity dependence of the computed $v_2$ is only slightly weaker than that of the data. Even though the computation of the flow coefficients without EbyE fluctuations should be viewed rather as qualitative than quantitative, the fact that the computed rapidity dependence of the elliptic flow is very similar to what is seen in the data is very promising. Moreover, we see that the EbyE initial state eccentricities at different rapidities are slightly decorrelated. As these decorrelations are not accounted for in the averaged initial state, our result suggests that eventually the computation of the true EbyE flow coefficients that include the decorrelations could show a slightly stronger rapidity dependence than the ones now computed from the averaged initial state.

As an outlook, we can see various exciting avenues along which the current MC-EKRT framework can be developed further. First, similarly to Refs.~\cite{Eskola:2000ji,Eskola:2000my}, a well-defined NLO pQCD calculation for the integrated minijet cross section $\sigma_{\text{jet}}^{ab}$, which determines the multiplicity of the candidate dijets here, can and should be done, and also its snPDF and scale dependencies should be charted. Second, pQCD parton showering should be included as a dynamical way to distribute the initially produced parton's energy and momentum into the phase space. Third, also a more detailed spacetime picture of parton production along the lines of Ref.\ \cite{Eskola:1993cz} should be studied, relaxing especially the assumption of all partons being produced at $z=0$ and thus making the initial parton production more isotropic. Fourth, pre-thermal evolution, i.e.\ the effects of the isotropizing and thermalizing secondary collisions of the produced partons \cite{Eskola:1988hp, Xu:2004mz, Kurkela:2014tea, Kurkela:2015qoa, Kurkela:2018vqr, Kurkela:2018oqw, Kurkela:2018wud, Kurkela:2018xxd, Du:2020dvp, Du:2020zqg, Carzon:2023zfp, Zhou:2024ysb} should be considered. After all these developments, an extraction of the full initial energy-momentum tensor $T^{\mu\nu}$ for 3+1 D fluid dynamics could be more realistically done, and effects of e.g.\ initial velocity~\cite{Gyulassy:2001kr, Pang:2012he, Okai:2017ofp} and shear-stress tensor to observables studied. Finally, we note that the MC-EKRT framework provides a promising platform for jet-quenching studies, where both the QCD-matter initial conditions for fluid dynamics and the high-energy partons that are losing energy are consistenly obtained from the same computation, event-by-event.

\acknowledgments
We thank Vadim Guzey, Ilkka Helenius, Pasi Huovinen, Dong Jo Kim, Tuomas Lappi, Aleksas Mazeliauskas, Etele Molnar, Heikki M\"antysaari, Petja Paakkinen, Risto Paatelainen, and Xin-Nian Wang for useful discussions during the preparation of the MC-EKRT framework. In addition, K.J.E.\ is grateful to Keijo Kajantie and Kimmo Tuominen for discussions at the time of writing Ref.~\cite{Eskola:2002qz} which still influenced our ideas here. We acknowledge the financial support from the Wilho, Yrj\"o and Kalle V\"ais\"al\"a Foundation (M.K.) and from the Jenny and Antti Wihuri Foundation (H.H.), and the Academy of Finland Project No. 330448 (K.J.E.). This research was funded as a part of the Center of Excellence in Quark Matter of the Academy of Finland (Projects No. 346325 and 364192). This research is part of the European Research Council Project No. ERC-2018-ADG-835105 YoctoLHC, and the European Union's Horizon 2020 research and innovation program under grant agreement No. 824093 (STRONG-2020). We acknowledge the computation resources from the Finnish IT Center for Science (CSC), project jyy2580, and from the Finnish Computing Competence Infrastructure (FCCI), persistent identifier urn:nbn:fi:research-infras-2016072533.


\begin{thebibliography}{99}
%\cite{Bazavov:2014pvz}
\bibitem{Bazavov:2014pvz}
A.~Bazavov \textit{et al.} [HotQCD],
%``Equation of state in ( 2+1 )-flavor QCD,''
Phys. Rev. D \textbf{90}, 094503 (2014)
doi:10.1103/PhysRevD.90.094503
[arXiv:1407.6387 [hep-lat]].
%1336 citations counted in INSPIRE as of 04 Jun 2024

%\cite{Bazavov:2017dsy}
\bibitem{Bazavov:2017dsy}
A.~Bazavov, P.~Petreczky and J.~H.~Weber,
%``Equation of State in 2+1 Flavor QCD at High Temperatures,''
Phys. Rev. D \textbf{97}, no.1, 014510 (2018)
doi:10.1103/PhysRevD.97.014510
[arXiv:1710.05024 [hep-lat]].
%89 citations counted in INSPIRE as of 02 Apr 2024

%\cite{Borsanyi:2013bia}
\bibitem{Borsanyi:2013bia}
S.~Borsanyi, Z.~Fodor, C.~Hoelbling, S.~D.~Katz, S.~Krieg and K.~K.~Szabo,
%``Full result for the QCD equation of state with 2+1 flavors,''
Phys. Lett. B \textbf{730}, 99-104 (2014)
doi:10.1016/j.physletb.2014.01.007
[arXiv:1309.5258 [hep-lat]].
%1024 citations counted in INSPIRE as of 04 Jun 2024

%\cite{Borsanyi:2010cj}
\bibitem{Borsanyi:2010cj}
S.~Borsanyi, G.~Endrodi, Z.~Fodor, A.~Jakovac, S.~D.~Katz, S.~Krieg, C.~Ratti and K.~K.~Szabo,
%``The QCD equation of state with dynamical quarks,''
JHEP \textbf{11}, 077 (2010)
doi:10.1007/JHEP11(2010)077
[arXiv:1007.2580 [hep-lat]].
%1116 citations counted in INSPIRE as of 13 Jun 2024

%\cite{ALICE:2022wpn}
\bibitem{ALICE:2022wpn}
 [ALICE], ``The ALICE experiment -- A journey through QCD,''
 CERN-EP-2022-227,
 [arXiv:2211.04384 [nucl-ex]].
%135 citations counted in INSPIRE as of 17 Jun 2024

%\cite{Romatschke:2007mq}
\bibitem{Romatschke:2007mq}
P.~Romatschke and U.~Romatschke,
%``Viscosity Information from Relativistic Nuclear Collisions: How Perfect is the Fluid Observed at RHIC?,''
Phys. Rev. Lett. \textbf{99}, 172301 (2007)
doi:10.1103/PhysRevLett.99.172301
[arXiv:0706.1522 [nucl-th]].
%1179 citations counted in INSPIRE as of 13 Jun 2024

%\cite{Luzum:2008cw}
\bibitem{Luzum:2008cw}
M.~Luzum and P.~Romatschke,
%``Conformal Relativistic Viscous Hydrodynamics: Applications to RHIC results at s(NN)**(1/2) = 200-GeV,''
Phys. Rev. C \textbf{78}, 034915 (2008)
[erratum: Phys. Rev. C \textbf{79}, 039903 (2009)]
doi:10.1103/PhysRevC.78.034915
[arXiv:0804.4015 [nucl-th]].
%932 citations counted in INSPIRE as of 13 Jun 2024

%\cite{Bozek:2009dw}
\bibitem{Bozek:2009dw}
P.~Bozek,
%``Bulk and shear viscosities of matter created in relativistic heavy-ion collisions,''
Phys. Rev. C \textbf{81}, 034909 (2010)
doi:10.1103/PhysRevC.81.034909
[arXiv:0911.2397 [nucl-th]].
%220 citations counted in INSPIRE as of 07 Jun 2024

%\cite{Schenke:2010rr}
\bibitem{Schenke:2010rr}
B.~Schenke, S.~Jeon and C.~Gale,
%``Elliptic and triangular flow in event-by-event (3+1)D viscous hydrodynamics,''
Phys. Rev. Lett. \textbf{106}, 042301 (2011)
doi:10.1103/PhysRevLett.106.042301
[arXiv:1009.3244 [hep-ph]].
%760 citations counted in INSPIRE as of 17 Jun 2024

%\cite{Schenke:2010nt}
\bibitem{Schenke:2010nt}
B.~Schenke, S.~Jeon and C.~Gale,
%``(3+1)D hydrodynamic simulation of relativistic heavy-ion collisions,''
Phys. Rev. C \textbf{82}, 014903 (2010)
doi:10.1103/PhysRevC.82.014903
[arXiv:1004.1408 [hep-ph]].
%494 citations counted in INSPIRE as of 12 Jun 2024

%\cite{Schenke:2011bn}
\bibitem{Schenke:2011bn}
B.~Schenke, S.~Jeon and C.~Gale,
%``Higher flow harmonics from (3+1)D event-by-event viscous hydrodynamics,''
Phys. Rev. C \textbf{85}, 024901 (2012)
doi:10.1103/PhysRevC.85.024901
[arXiv:1109.6289 [hep-ph]].
%314 citations counted in INSPIRE as of 13 Jun 2024

%\cite{Song:2011qa}
\bibitem{Song:2011qa}
H.~Song, S.~A.~Bass and U.~Heinz,
%``Elliptic flow in 200 A GeV Au+Au collisions and 2.76 A TeV Pb+Pb collisions: insights from viscous hydrodynamics + hadron cascade hybrid model,''
Phys. Rev. C \textbf{83}, 054912 (2011)
[erratum: Phys. Rev. C \textbf{87}, no.1, 019902 (2013)]
doi:10.1103/PhysRevC.83.054912
[arXiv:1103.2380 [nucl-th]].
%174 citations counted in INSPIRE as of 07 May 2024

%\cite{Niemi:2011ix}
\bibitem{Niemi:2011ix}
H.~Niemi, G.~S.~Denicol, P.~Huovinen, E.~Molnar and D.~H.~Rischke,
%``Influence of the shear viscosity of the quark-gluon plasma on elliptic flow in ultrarelativistic heavy-ion collisions,''
Phys. Rev. Lett. \textbf{106}, 212302 (2011)
doi:10.1103/PhysRevLett.106.212302
[arXiv:1101.2442 [nucl-th]].
%258 citations counted in INSPIRE as of 07 May 2024

%\cite{Gale:2012rq}
\bibitem{Gale:2012rq}
C.~Gale, S.~Jeon, B.~Schenke, P.~Tribedy and R.~Venugopalan,
%``Event-by-event anisotropic flow in heavy-ion collisions from combined Yang-Mills and viscous fluid dynamics,''
Phys. Rev. Lett. \textbf{110}, no.1, 012302 (2013)
doi:10.1103/PhysRevLett.110.012302
[arXiv:1209.6330 [nucl-th]].
%634 citations counted in INSPIRE as of 17 Jun 2024

%\cite{Niemi:2012ry}
\bibitem{Niemi:2012ry}
H.~Niemi, G.~S.~Denicol, P.~Huovinen, E.~Molnar and D.~H.~Rischke,
%``Influence of a temperature-dependent shear viscosity on the azimuthal asymmetries of transverse momentum spectra in ultrarelativistic heavy-ion collisions,''
Phys. Rev. C \textbf{86}, 014909 (2012)
doi:10.1103/PhysRevC.86.014909
[arXiv:1203.2452 [nucl-th]].
%159 citations counted in INSPIRE as of 07 May 2024

%\cite{Noronha-Hostler:2013gga}
\bibitem{Noronha-Hostler:2013gga}
J.~Noronha-Hostler, G.~S.~Denicol, J.~Noronha, R.~P.~G.~Andrade and F.~Grassi,
%``Bulk Viscosity Effects in Event-by-Event Relativistic Hydrodynamics,''
Phys. Rev. C \textbf{88}, no.4, 044916 (2013)
doi:10.1103/PhysRevC.88.044916
[arXiv:1305.1981 [nucl-th]].
%189 citations counted in INSPIRE as of 07 Jun 2024

%\cite{Niemi:2015qia}
\bibitem{Niemi:2015qia}
H.~Niemi, K.~J.~Eskola and R.~Paatelainen,
%``Event-by-event fluctuations in a perturbative QCD + saturation + hydrodynamics model: Determining QCD matter shear viscosity in ultrarelativistic heavy-ion collisions,''
Phys. Rev. C \textbf{93}, no.2, 024907 (2016)
doi:10.1103/PhysRevC.93.024907
[arXiv:1505.02677 [hep-ph]].
%321 citations counted in INSPIRE as of 03 Jun 2024

%\cite{Ryu:2015vwa}
\bibitem{Ryu:2015vwa}
S.~Ryu, J.~F.~Paquet, C.~Shen, G.~S.~Denicol, B.~Schenke, S.~Jeon and C.~Gale,
%``Importance of the Bulk Viscosity of QCD in Ultrarelativistic Heavy-Ion Collisions,''
Phys. Rev. Lett. \textbf{115}, no.13, 132301 (2015)
doi:10.1103/PhysRevLett.115.132301
[arXiv:1502.01675 [nucl-th]].
%367 citations counted in INSPIRE as of 29 May 2024

%\cite{Karpenko:2015xea}
\bibitem{Karpenko:2015xea}
I.~A.~Karpenko, P.~Huovinen, H.~Petersen and M.~Bleicher,
%``Estimation of the shear viscosity at finite net-baryon density from $A+A$ collision data at $\sqrt{s_\mathrm{NN}} = 7.7-200$ GeV,''
Phys. Rev. C \textbf{91}, no.6, 064901 (2015)
doi:10.1103/PhysRevC.91.064901
[arXiv:1502.01978 [nucl-th]].
%171 citations counted in INSPIRE as of 24 May 2024

%\cite{Giacalone:2017dud}
\bibitem{Giacalone:2017dud}
G.~Giacalone, J.~Noronha-Hostler, M.~Luzum and J.~Y.~Ollitrault,
%``Hydrodynamic predictions for 5.44 TeV Xe+Xe collisions,''
Phys. Rev. C \textbf{97}, no.3, 034904 (2018)
doi:10.1103/PhysRevC.97.034904
[arXiv:1711.08499 [nucl-th]].
%118 citations counted in INSPIRE as of 07 Jun 2024

%\cite{Nijs:2020roc}
\bibitem{Nijs:2020roc}
G.~Nijs, W.~van der Schee, U.~G\"ursoy and R.~Snellings,
%``Bayesian analysis of heavy ion collisions with the heavy ion computational framework Trajectum,''
Phys. Rev. C \textbf{103}, no.5, 054909 (2021)
doi:10.1103/PhysRevC.103.054909
[arXiv:2010.15134 [nucl-th]].
%154 citations counted in INSPIRE as of 14 Jun 2024

%\cite{Hirvonen:2022xfv}
\bibitem{Hirvonen:2022xfv}
H.~Hirvonen, K.~J.~Eskola and H.~Niemi,
%``Flow correlations from a hydrodynamics model with dynamical freeze-out and initial conditions based on perturbative QCD and saturation,''
Phys. Rev. C \textbf{106}, no.4, 044913 (2022)
doi:10.1103/PhysRevC.106.044913
[arXiv:2206.15207 [hep-ph]].
%7 citations counted in INSPIRE as of 14 Jun 2024

%\cite{Novak:2013bqa}
\bibitem{Novak:2013bqa}
J.~Novak, K.~Novak, S.~Pratt, J.~Vredevoogd, C.~Coleman-Smith and R.~Wolpert,
%``Determining Fundamental Properties of Matter Created in Ultrarelativistic Heavy-Ion Collisions,''
Phys. Rev. C \textbf{89}, no.3, 034917 (2014)
doi:10.1103/PhysRevC.89.034917
[arXiv:1303.5769 [nucl-th]].
%131 citations counted in INSPIRE as of 12 Jun 2024

%\cite{Bernhard:2016tnd}
\bibitem{Bernhard:2016tnd}
J.~E.~Bernhard, J.~S.~Moreland, S.~A.~Bass, J.~Liu and U.~Heinz,
%``Applying Bayesian parameter estimation to relativistic heavy-ion collisions: simultaneous characterization of the initial state and quark-gluon plasma medium,''
Phys. Rev. C \textbf{94}, no.2, 024907 (2016)
doi:10.1103/PhysRevC.94.024907
[arXiv:1605.03954 [nucl-th]].
%474 citations counted in INSPIRE as of 17 Jun 2024

%\cite{Bass:2017zyn}
\bibitem{Bass:2017zyn}
S.~A.~Bass, J.~E.~Bernhard and J.~S.~Moreland,
%``Determination of Quark-Gluon-Plasma Parameters from a Global Bayesian Analysis,''
Nucl. Phys. A \textbf{967}, 67-73 (2017)
doi:10.1016/j.nuclphysa.2017.05.052
[arXiv:1704.07671 [nucl-th]].
%38 citations counted in INSPIRE as of 07 May 2024

%\cite{Bernhard:2019bmu}
\bibitem{Bernhard:2019bmu}
J.~E.~Bernhard, J.~S.~Moreland and S.~A.~Bass,
%``Bayesian estimation of the specific shear and bulk viscosity of quark\textendash{}gluon plasma,''
Nature Phys. \textbf{15}, no.11, 1113-1117 (2019)
doi:10.1038/s41567-019-0611-8
%331 citations counted in INSPIRE as of 17 Jun 2024

%\cite{JETSCAPE:2020mzn}
\bibitem{JETSCAPE:2020mzn}
D.~Everett \textit{et al.} [JETSCAPE],
%``Multisystem Bayesian constraints on the transport coefficients of QCD matter,''
Phys. Rev. C \textbf{103}, no.5, 054904 (2021)
doi:10.1103/PhysRevC.103.054904
[arXiv:2011.01430 [hep-ph]].
%228 citations counted in INSPIRE as of 13 Jun 2024

%\cite{Auvinen:2020mpc}
\bibitem{Auvinen:2020mpc}
J.~Auvinen, K.~J.~Eskola, P.~Huovinen, H.~Niemi, R.~Paatelainen and P.~Petreczky,
%``Temperature dependence of $\eta/s$ of strongly interacting matter: Effects of the equation of state and the parametric form of $(\eta/s)(T)$,''
Phys. Rev. C \textbf{102}, no.4, 044911 (2020)
doi:10.1103/PhysRevC.102.044911
[arXiv:2006.12499 [nucl-th]].
%43 citations counted in INSPIRE as of 30 May 2024

%\cite{Parkkila:2021tqq}
\bibitem{Parkkila:2021tqq}
J.~E.~Parkkila, A.~Onnerstad and D.~J.~Kim,
%``Bayesian estimation of the specific shear and bulk viscosity of the quark-gluon plasma with additional flow harmonic observables,''
Phys. Rev. C \textbf{104}, no.5, 054904 (2021)
doi:10.1103/PhysRevC.104.054904
[arXiv:2106.05019 [hep-ph]].
%51 citations counted in INSPIRE as of 10 Jun 2024

%\cite{Parkkila:2021yha}
\bibitem{Parkkila:2021yha}
J.~E.~Parkkila, A.~Onnerstad, S.~F.~Taghavi, C.~Mordasini, A.~Bilandzic, M.~Virta and D.~J.~Kim,
%``New constraints for QCD matter from improved Bayesian parameter estimation in heavy-ion collisions at LHC,''
Phys. Lett. B \textbf{835}, 137485 (2022)
doi:10.1016/j.physletb.2022.137485
[arXiv:2111.08145 [hep-ph]].
%47 citations counted in INSPIRE as of 07 Jun 2024

%\cite{Nijs:2023yab}
\bibitem{Nijs:2023yab}
G.~Nijs and W.~van der Schee,
%``A generalized hydrodynamizing initial stage for Heavy Ion Collisions,''
[arXiv:2304.06191 [nucl-th]].
%10 citations counted in INSPIRE as of 14 Jun 2024

%\cite{Shen:2022oyg}
\bibitem{Shen:2022oyg}
C.~Shen and B.~Schenke,
%``Longitudinal dynamics and particle production in relativistic nuclear collisions,''
Phys. Rev. C \textbf{105}, no.6, 064905 (2022)
doi:10.1103/PhysRevC.105.064905
[arXiv:2203.04685 [nucl-th]].
%44 citations counted in INSPIRE as of 07 Jun 2024

%\cite{Bozek:2017qir}
\bibitem{Bozek:2017qir}
P.~Bozek and W.~Broniowski,
%``Longitudinal decorrelation measures of flow magnitude and event-plane angles in ultrarelativistic nuclear collisions,''
Phys. Rev. C \textbf{97}, no.3, 034913 (2018)
doi:10.1103/PhysRevC.97.034913
[arXiv:1711.03325 [nucl-th]].
%49 citations counted in INSPIRE as of 24 May 2024

%\cite{Bozek:2012qs}
\bibitem{Bozek:2012qs}
P.~Bozek and I.~Wyskiel-Piekarska,
%``Particle spectra in Pb-Pb collisions at $\sqrt{S_{NN}}$ = 2.76 TeV,''
Phys. Rev. C \textbf{85}, 064915 (2012)
doi:10.1103/PhysRevC.85.064915
[arXiv:1203.6513 [nucl-th]].
%157 citations counted in INSPIRE as of 07 May 2024

%\cite{Pang:2018zzo}
\bibitem{Pang:2018zzo}
L.~G.~Pang, H.~Petersen and X.~N.~Wang,
%``Pseudorapidity distribution and decorrelation of anisotropic flow within the open-computing-language implementation CLVisc hydrodynamics,''
Phys. Rev. C \textbf{97}, no.6, 064918 (2018)
doi:10.1103/PhysRevC.97.064918
[arXiv:1802.04449 [nucl-th]].
%134 citations counted in INSPIRE as of 07 Jun 2024

%\cite{Pang:2012he}
\bibitem{Pang:2012he}
L.~Pang, Q.~Wang and X.~N.~Wang,
%``Effects of initial flow velocity fluctuation in event-by-event (3+1)D hydrodynamics,''
Phys. Rev. C \textbf{86}, 024911 (2012)
doi:10.1103/PhysRevC.86.024911
[arXiv:1205.5019 [nucl-th]].
%216 citations counted in INSPIRE as of 04 Jun 2024

%\cite{Denicol:2015nhu}
\bibitem{Denicol:2015nhu}
G.~Denicol, A.~Monnai and B.~Schenke,
%``Moving forward to constrain the shear viscosity of QCD matter,''
Phys. Rev. Lett. \textbf{116}, no.21, 212301 (2016)
doi:10.1103/PhysRevLett.116.212301
[arXiv:1512.01538 [nucl-th]].
%89 citations counted in INSPIRE as of 26 May 2024

%\cite{Pang:2015zrq}
\bibitem{Pang:2015zrq}
L.~G.~Pang, H.~Petersen, G.~Y.~Qin, V.~Roy and X.~N.~Wang,
%``Decorrelation of anisotropic flow along the longitudinal direction,''
Eur. Phys. J. A \textbf{52}, no.4, 97 (2016)
doi:10.1140/epja/i2016-16097-x
[arXiv:1511.04131 [nucl-th]].
%86 citations counted in INSPIRE as of 04 Jun 2024


%\cite{Auvinen:2017fjw}
\bibitem{Auvinen:2017fjw}
J.~Auvinen, J.~E.~Bernhard, S.~A.~Bass and I.~Karpenko,
%``Investigating the collision energy dependence of $\eta$/s in the beam energy scan at the BNL Relativistic Heavy Ion Collider using Bayesian statistics,''
Phys. Rev. C \textbf{97}, no.4, 044905 (2018)
doi:10.1103/PhysRevC.97.044905
[arXiv:1706.03666 [hep-ph]].
%55 citations counted in INSPIRE as of 10 Jun 2024

%\cite{JETSCAPE:2023nuf}
\bibitem{JETSCAPE:2023nuf}
A.~Mankolli \textit{et al.} [JETSCAPE],
%``3D Multi-system Bayesian Calibration with Energy Conservation to Study Rapidity-dependent Dynamics of Nuclear Collisions,''
[arXiv:2401.00402 [nucl-th]].
%0 citations counted in INSPIRE as of 04 Jun 2024

%\cite{Soeder:2023vdn}
\bibitem{Soeder:2023vdn}
D.~Soeder, W.~Ke, J.~F.~Paquet and S.~A.~Bass,
%``Bayesian parameter estimation with a new three-dimensional initial-conditions model for ultrarelativistic heavy-ion collisions,''
[arXiv:2306.08665 [nucl-th]].
%6 citations counted in INSPIRE as of 31 May 2024

%\cite{Hirvonen:2023lqy}
\bibitem{Hirvonen:2023lqy}
H.~Hirvonen, K.~J.~Eskola and H.~Niemi,
%``Deep learning for flow observables in ultrarelativistic heavy-ion collisions,''
Phys. Rev. C \textbf{108}, no.3, 034905 (2023)
doi:10.1103/PhysRevC.108.034905
[arXiv:2303.04517 [hep-ph]].
%4 citations counted in INSPIRE as of 14 Jun 2024

%\cite{Hirvonen:2024ycx}
\bibitem{Hirvonen:2024ycx}
H.~Hirvonen, K.~J.~Eskola and H.~Niemi,
%``Deep learning for flow observables in high energy heavy-ion collisions,''
[arXiv:2404.02602 [hep-ph]].
%0 citations counted in INSPIRE as of 07 Jun 2024

%\cite{Schenke:2012wb}
\bibitem{Schenke:2012wb}
B.~Schenke, P.~Tribedy and R.~Venugopalan,
%``Fluctuating Glasma initial conditions and flow in heavy ion collisions,''
Phys. Rev. Lett. \textbf{108}, 252301 (2012)
doi:10.1103/PhysRevLett.108.252301
[arXiv:1202.6646 [nucl-th]].
%631 citations counted in INSPIRE as of 13 Jun 2024

%\cite{Eskola:1999fc}
\bibitem{Eskola:1999fc}
K.~J.~Eskola, K.~Kajantie, P.~V.~Ruuskanen and K.~Tuominen,
%``Scaling of transverse energies and multiplicities with atomic number and energy in ultrarelativistic nuclear collisions,''
Nucl. Phys. B \textbf{570}, 379-389 (2000)
doi:10.1016/S0550-3213(99)00720-8
[arXiv:hep-ph/9909456 [hep-ph]].
%394 citations counted in INSPIRE as of 04 Apr 2024

%\cite{Eskola:2000xq}
\bibitem{Eskola:2000xq}
K.~J.~Eskola, K.~Kajantie and K.~Tuominen,
%``Centrality dependence of multiplicities in ultrarelativistic nuclear collisions,''
Phys. Lett. B \textbf{497}, 39-43 (2001)
doi:10.1016/S0370-2693(00)01341-1
[arXiv:hep-ph/0009246 [hep-ph]].
%105 citations counted in INSPIRE as of 07 May 2024

%\cite{Paatelainen:2013eea}
\bibitem{Paatelainen:2013eea}
R.~Paatelainen, K.~J.~Eskola, H.~Niemi and K.~Tuominen,
%``Fluid dynamics with saturated minijet initial conditions in ultrarelativistic heavy-ion collisions,''
Phys. Lett. B \textbf{731}, 126-130 (2014)
doi:10.1016/j.physletb.2014.02.018
[arXiv:1310.3105 [hep-ph]].
%53 citations counted in INSPIRE as of 22 Apr 2024

%\cite{Pierog:2009zt}
\bibitem{Pierog:2009zt}
T.~Pierog and K.~Werner,
%``EPOS Model and Ultra High Energy Cosmic Rays,''
Nucl. Phys. B Proc. Suppl. \textbf{196}, 102-105 (2009)
doi:10.1016/j.nuclphysbps.2009.09.017
[arXiv:0905.1198 [hep-ph]].
%259 citations counted in INSPIRE as of 14 May 2024

%\cite{Pierog:2013ria}
\bibitem{Pierog:2013ria}
T.~Pierog, I.~Karpenko, J.~M.~Katzy, E.~Yatsenko and K.~Werner,
%``EPOS LHC: Test of collective hadronization with data measured at the CERN Large Hadron Collider,''
Phys. Rev. C \textbf{92}, no.3, 034906 (2015)
doi:10.1103/PhysRevC.92.034906
[arXiv:1306.0121 [hep-ph]].
%1489 citations counted in INSPIRE as of 17 Jun 2024

%\cite{Werner:2013tya}
\bibitem{Werner:2013tya}
K.~Werner, B.~Guiot, I.~Karpenko and T.~Pierog,
%``Analysing radial flow features in p-Pb and p-p collisions at several TeV by studying identified particle production in EPOS3,''
Phys. Rev. C \textbf{89}, no.6, 064903 (2014)
doi:10.1103/PhysRevC.89.064903
[arXiv:1312.1233 [nucl-th]].
%272 citations counted in INSPIRE as of 03 Jun 2024

%\cite{Werner:2023zvo}
\bibitem{Werner:2023zvo}
K.~Werner,
%``Revealing a deep connection between factorization and saturation: New insight into modeling high-energy proton-proton and nucleus-nucleus scattering in the EPOS4 framework,''
Phys. Rev. C \textbf{108}, no.6, 064903 (2023)
doi:10.1103/PhysRevC.108.064903
[arXiv:2301.12517 [hep-ph]].
%25 citations counted in INSPIRE as of 17 Jun 2024

%\cite{Werner:2023jps}
\bibitem{Werner:2023jps}
K.~Werner,
%``Core-corona procedure and microcanonical hadronization to understand strangeness enhancement in proton-proton and heavy ion collisions in the EPOS4 framework,''
Phys. Rev. C \textbf{109}, no.1, 014910 (2024)
doi:10.1103/PhysRevC.109.014910
[arXiv:2306.10277 [hep-ph]].
%18 citations counted in INSPIRE as of 17 Jun 2024

%\cite{Werner:2023fne}
\bibitem{Werner:2023fne}
K.~Werner and B.~Guiot,
%``Perturbative QCD concerning light and heavy flavor in the EPOS4 framework,''
Phys. Rev. C \textbf{108}, no.3, 034904 (2023)
doi:10.1103/PhysRevC.108.034904
[arXiv:2306.02396 [hep-ph]].
%18 citations counted in INSPIRE as of 17 Jun 2024

%\cite{Pang:2014pxa}
\bibitem{Pang:2014pxa}
L.~G.~Pang, G.~Y.~Qin, V.~Roy, X.~N.~Wang and G.~L.~Ma,
%``Longitudinal decorrelation of anisotropic flows in heavy-ion collisions at the CERN Large Hadron Collider,''
Phys. Rev. C \textbf{91}, no.4, 044904 (2015)
doi:10.1103/PhysRevC.91.044904
[arXiv:1410.8690 [nucl-th]].
%65 citations counted in INSPIRE as of 14 Jun 2024


%\cite{Kanakubo:2019ogh}
\bibitem{Kanakubo:2019ogh}
Y.~Kanakubo, Y.~Tachibana and T.~Hirano,
%``Unified description of hadron yield ratios from dynamical core-corona initialization,''
Phys. Rev. C \textbf{101}, no.2, 024912 (2020)
doi:10.1103/PhysRevC.101.024912
[arXiv:1910.10556 [nucl-th]].
%56 citations counted in INSPIRE as of 07 Jun 2024

%\cite{Kanakubo:2021qcw}
\bibitem{Kanakubo:2021qcw}
Y.~Kanakubo, Y.~Tachibana and T.~Hirano,
%``Interplay between core and corona components in high-energy nuclear collisions,''
Phys. Rev. C \textbf{105}, no.2, 024905 (2022)
doi:10.1103/PhysRevC.105.024905
[arXiv:2108.07943 [nucl-th]].
%23 citations counted in INSPIRE as of 07 Jun 2024

%\cite{Bierlich:2018xfw}
\bibitem{Bierlich:2018xfw}
C.~Bierlich, G.~Gustafson, L.~L\"onnblad and H.~Shah,
%``The Angantyr model for Heavy-Ion Collisions in PYTHIA8,''
JHEP \textbf{10}, 134 (2018)
doi:10.1007/JHEP10(2018)134
[arXiv:1806.10820 [hep-ph]].
%169 citations counted in INSPIRE as of 13 Jun 2024

%\cite{Carzon:2019qja}
\bibitem{Carzon:2019qja}
P.~Carzon, M.~Martinez, M.~D.~Sievert, D.~E.~Wertepny and J.~Noronha-Hostler,
%``Monte Carlo event generator for initial conditions of conserved charges in nuclear geometry,''
Phys. Rev. C \textbf{105}, no.3, 034908 (2022)
doi:10.1103/PhysRevC.105.034908
[arXiv:1911.12454 [nucl-th]].
%30 citations counted in INSPIRE as of 10 Jun 2024

%\cite{Garcia-Montero:2023gex}
\bibitem{Garcia-Montero:2023gex}
O.~Garcia-Montero, H.~Elfner and S.~Schlichting,
%``McDIPPER: A novel saturation-based 3+1D initial-state model for heavy ion collisions,''
Phys. Rev. C \textbf{109}, no.4, 044916 (2024)
doi:10.1103/PhysRevC.109.044916
[arXiv:2308.11713 [hep-ph]].
%10 citations counted in INSPIRE as of 07 Jun 2024

%\cite{Kajantie:1987pd}
\bibitem{Kajantie:1987pd}
K.~Kajantie, P.~V.~Landshoff and J.~Lindfors,
%``Minijet Production in High-Energy Nucleus-Nucleus Collisions,''
Phys. Rev. Lett. \textbf{59}, 2527 (1987)
doi:10.1103/PhysRevLett.59.2527
%172 citations counted in INSPIRE as of 29 Feb 2024

%\cite{Eskola:1988yh}
\bibitem{Eskola:1988yh}
K.~J.~Eskola, K.~Kajantie and J.~Lindfors,
%``Quark and Gluon Production in High-Energy Nucleus-Nucleus Collisions,''
Nucl. Phys. B \textbf{323}, 37-52 (1989)
doi:10.1016/0550-3213(89)90586-5
%329 citations counted in INSPIRE as of 08 Apr 2024

%\cite{Eskola:1996ce}
\bibitem{Eskola:1996ce}
K.~J.~Eskola and K.~Kajantie,
%``Baryon to entropy ratio in very high-energy nuclear collisions,''
Z. Phys. C \textbf{75}, 515-522 (1997)
doi:10.1007/s002880050495
[arXiv:nucl-th/9610015 [nucl-th]].
%76 citations counted in INSPIRE as of 07 May 2024

%\cite{Eskola:2005ue}
\bibitem{Eskola:2005ue}
K.~J.~Eskola, H.~Honkanen, H.~Niemi, P.~V.~Ruuskanen and S.~S.~Rasanen,
%``RHIC-tested predictions for low-p($T$) and high-p($T$) hadron spectra in nearly central Pb + Pb collisions at the CERN LHC,''
Phys. Rev. C \textbf{72}, 044904 (2005)
doi:10.1103/PhysRevC.72.044904
[arXiv:hep-ph/0506049 [hep-ph]].
%118 citations counted in INSPIRE as of 07 May 2024

%\cite{Kolb:2001qz}
\bibitem{Kolb:2001qz}
P.~F.~Kolb, U.~W.~Heinz, P.~Huovinen, K.~J.~Eskola and K.~Tuominen,
%``Centrality dependence of multiplicity, transverse energy, and elliptic flow from hydrodynamics,''
Nucl. Phys. A \textbf{696}, 197-215 (2001)
doi:10.1016/S0375-9474(01)01114-9
[arXiv:hep-ph/0103234 [hep-ph]].
%351 citations counted in INSPIRE as of 17 May 2024

%\cite{Niemi:2008ta}
\bibitem{Niemi:2008ta}
H.~Niemi, K.~J.~Eskola and P.~V.~Ruuskanen,
%``Elliptic flow in nuclear collisions at the Large Hadron Collider,''
Phys. Rev. C \textbf{79}, 024903 (2009)
doi:10.1103/PhysRevC.79.024903
[arXiv:0806.1116 [hep-ph]].
%45 citations counted in INSPIRE as of 08 May 2024

%\cite{Eskola:2000my}
\bibitem{Eskola:2000my}
K.~J.~Eskola and K.~Tuominen,
%``Transverse energy from minijets in ultrarelativistic nuclear collisions: A Next-to-leading order analysis,''
Phys. Rev. D \textbf{63}, 114006 (2001)
doi:10.1103/PhysRevD.63.114006
[arXiv:hep-ph/0010319 [hep-ph]].
%30 citations counted in INSPIRE as of 07 May 2024

%\cite{Eskola:2000ji}
\bibitem{Eskola:2000ji}
K.~J.~Eskola and K.~Tuominen,
%``Production of transverse energy from minijets in next-to-leading order perturbative QCD,''
Phys. Lett. B \textbf{489}, 329-336 (2000)
doi:10.1016/S0370-2693(00)00946-1
[arXiv:hep-ph/0002008 [hep-ph]].
%32 citations counted in INSPIRE as of 07 May 2024

%\cite{Niemi:2015voa}
\bibitem{Niemi:2015voa}
H.~Niemi, K.~J.~Eskola, R.~Paatelainen and K.~Tuominen,
%``Predictions for 5.023 TeV Pb + Pb collisions at the CERN Large Hadron Collider,''
Phys. Rev. C \textbf{93}, no.1, 014912 (2016)
doi:10.1103/PhysRevC.93.014912
[arXiv:1511.04296 [hep-ph]].
%96 citations counted in INSPIRE as of 11 Jun 2024

%\cite{Eskola:2017bup}
\bibitem{Eskola:2017bup}
K.~J.~Eskola, H.~Niemi, R.~Paatelainen and K.~Tuominen,
%``Predictions for multiplicities and flow harmonics in 5.44 TeV Xe+Xe collisions at the CERN Large Hadron Collider,''
Phys. Rev. C \textbf{97}, no.3, 034911 (2018)
doi:10.1103/PhysRevC.97.034911
[arXiv:1711.09803 [hep-ph]].
%48 citations counted in INSPIRE as of 17 May 2024

%\cite{Niemi:2018ijm}
\bibitem{Niemi:2018ijm}
H.~Niemi, K.~J.~Eskola, R.~Paatelainen and K.~Tuominen,
%``Latest predictions from the EbyE NLO EKRT model,''
Nucl. Phys. A \textbf{982}, 443-446 (2019)
doi:10.1016/j.nuclphysa.2018.10.013
[arXiv:1807.02378 [nucl-th]].
%3 citations counted in INSPIRE as of 14 May 2024

%\cite{Helenius:2012wd}
\bibitem{Helenius:2012wd}
I.~Helenius, K.~J.~Eskola, H.~Honkanen and C.~A.~Salgado,
%``Impact-Parameter Dependent Nuclear Parton Distribution Functions: EPS09s and EKS98s and Their Applications in Nuclear Hard Processes,''
JHEP \textbf{07}, 073 (2012)
doi:10.1007/JHEP07(2012)073
[arXiv:1205.5359 [hep-ph]].
%179 citations counted in INSPIRE as of 07 May 2024

\bibitem{MCEKRT}
The MC-EKRT code will be available at url:github.com/mialkuha/MC-EKRT

%\cite{Molnar:2014zha}
\bibitem{Molnar:2014zha}
E.~Molnar, H.~Holopainen, P.~Huovinen and H.~Niemi,
%``Influence of temperature-dependent shear viscosity on elliptic flow at backward and forward rapidities in ultrarelativistic heavy-ion collisions,''
Phys. Rev. C \textbf{90}, no.4, 044904 (2014)
doi:10.1103/PhysRevC.90.044904
[arXiv:1407.8152 [nucl-th]].
%48 citations counted in INSPIRE as of 07 Jun 2024

%\cite{Paatelainen:2012at}
\bibitem{Paatelainen:2012at}
R.~Paatelainen, K.~J.~Eskola, H.~Holopainen and K.~Tuominen,
%``Multiplicities and $p_T$ spectra in ultrarelativistic heavy ion collisions from a next-to-leading order improved perturbative QCD + saturation + hydrodynamics model,''
Phys. Rev. C \textbf{87}, no.4, 044904 (2013)
doi:10.1103/PhysRevC.87.044904
[arXiv:1211.0461 [hep-ph]].
%42 citations counted in INSPIRE as of 21 May 2024

%\cite{Loizides:2017ack}
\bibitem{Loizides:2017ack}
C.~Loizides, J.~Kamin and D.~d'Enterria,
%``Improved Monte Carlo Glauber predictions at present and future nuclear colliders,''
Phys. Rev. C \textbf{97}, no.5, 054910 (2018)
[erratum: Phys. Rev. C \textbf{99}, no.1, 019901 (2019)]
doi:10.1103/PhysRevC.97.054910
[arXiv:1710.07098 [nucl-ex]].
%272 citations counted in INSPIRE as of 12 Jun 2024

%\cite{DeJager:1974liz}
\bibitem{DeJager:1974liz}
C.~W.~De Jager, H.~De Vries and C.~De Vries,
%``Nuclear charge and magnetization density distribution parameters from elastic electron scattering,''
Atom. Data Nucl. Data Tabl. \textbf{14}, 479-508 (1974)
[erratum: Atom. Data Nucl. Data Tabl. \textbf{16}, 580-580 (1975)]
doi:10.1016/S0092-640X(74)80002-1
%982 citations counted in INSPIRE as of 17 Jun 2024

%\cite{DeVries:1987atn}
\bibitem{DeVries:1987atn}
H.~De Vries, C.~W.~De Jager and C.~De Vries,
%``Nuclear charge and magnetization density distribution parameters from elastic electron scattering,''
Atom. Data Nucl. Data Tabl. \textbf{36}, 495-536 (1987)
doi:10.1016/0092-640X(87)90013-1
%1900 citations counted in INSPIRE as of 17 Jun 2024

%\cite{Woods:1954zz}
\bibitem{Woods:1954zz}
R.~D.~Woods and D.~S.~Saxon,
%``Diffuse Surface Optical Model for Nucleon-Nuclei Scattering,''
Phys. Rev. \textbf{95}, 577-578 (1954)
doi:10.1103/PhysRev.95.577
%494 citations counted in INSPIRE as of 10 Jun 2024

%\cite{Loizides:2014vua}
\bibitem{Loizides:2014vua}
C.~Loizides, J.~Nagle and P.~Steinberg,
%``Improved version of the PHOBOS Glauber Monte Carlo,''
SoftwareX \textbf{1-2}, 13-18 (2015)
doi:10.1016/j.softx.2015.05.001
[arXiv:1408.2549 [nucl-ex]].
%199 citations counted in INSPIRE as of 28 May 2024

%\cite{COMPETE:2002jcr}
\bibitem{COMPETE:2002jcr}
J.~R.~Cudell \textit{et al.} [COMPETE],
%``Benchmarks for the forward observables at RHIC, the Tevatron Run II and the LHC,''
Phys. Rev. Lett. \textbf{89}, 201801 (2002)
doi:10.1103/PhysRevLett.89.201801
[arXiv:hep-ph/0206172 [hep-ph]].
%304 citations counted in INSPIRE as of 11 Jun 2024

%\cite{TOTEM:2017asr}
\bibitem{TOTEM:2017asr}
G.~Antchev \textit{et al.} [TOTEM],
%``First measurement of elastic, inelastic and total cross-section at $\sqrt{s}=13$ TeV by TOTEM and overview of cross-section data at LHC energies,''
Eur. Phys. J. C \textbf{79}, no.2, 103 (2019)
doi:10.1140/epjc/s10052-019-6567-0
[arXiv:1712.06153 [hep-ex]].
%231 citations counted in INSPIRE as of 05 Jun 2024

%\cite{Wang:1991hta}
\bibitem{Wang:1991hta}
X.~N.~Wang and M.~Gyulassy,
%``HIJING: A Monte Carlo model for multiple jet production in p p, p A and A A collisions,''
Phys. Rev. D \textbf{44}, 3501-3516 (1991)
doi:10.1103/PhysRevD.44.3501
%2025 citations counted in INSPIRE as of 14 Jun 2024

%\cite{ZEUS:2002wfj}
\bibitem{ZEUS:2002wfj}
S.~Chekanov \textit{et al.} [ZEUS],
%``Exclusive photoproduction of J / psi mesons at HERA,''
Eur. Phys. J. C \textbf{24}, 345-360 (2002)
doi:10.1007/s10052-002-0953-7
[arXiv:hep-ex/0201043 [hep-ex]].
%403 citations counted in INSPIRE as of 11 Jun 2024

%\cite{Eskola:2022vpi}
\bibitem{Eskola:2022vpi}
K.~J.~Eskola, C.~A.~Flett, V.~Guzey, T.~L\"oyt\"ainen and H.~Paukkunen,
%``Exclusive J/\ensuremath{\psi} photoproduction in ultraperipheral Pb+Pb collisions at the CERN Large Hadron Collider calculated at next-to-leading order perturbative QCD,''
Phys. Rev. C \textbf{106}, no.3, 035202 (2022)
doi:10.1103/PhysRevC.106.035202
[arXiv:2203.11613 [hep-ph]].
%46 citations counted in INSPIRE as of 14 Jun 2024

%\cite{Flett:2021xsl}
\bibitem{Flett:2021xsl}
C.~A.~Flett, ``Exclusive Observables to NLO and Low x PDF Phenomenology at the LHC,''
Ph.D. thesis, University of Liverpool, 2021
%3 citations counted in INSPIRE as of 17 May 2024

%\cite{Wang:1990qp}
\bibitem{Wang:1990qp}
X.~N.~Wang,
%``Role of multiple mini - jets in high-energy hadronic reactions,''
Phys. Rev. D \textbf{43}, 104-112 (1991)
doi:10.1103/PhysRevD.43.104
%155 citations counted in INSPIRE as of 04 Jun 2024

%\cite{Durand:1987yv}
\bibitem{Durand:1987yv}
L.~Durand and P.~Hong,
%``QCD and Rising Total Cross-Sections,''
Phys. Rev. Lett. \textbf{58}, 303-306 (1987)
doi:10.1103/PhysRevLett.58.303
%315 citations counted in INSPIRE as of 14 May 2024

%\cite{Eichten:1984eu}
\bibitem{Eichten:1984eu}
E.~Eichten, I.~Hinchliffe, K.~D.~Lane and C.~Quigg,
%``Super Collider Physics,''
Rev. Mod. Phys. \textbf{56}, 579-707 (1984)
doi:10.1103/RevModPhys.56.579
%2574 citations counted in INSPIRE as of 17 Jun 2024

%\cite{Sarcevic:1988tu}
\bibitem{Sarcevic:1988tu}
I.~Sarcevic, S.~D.~Ellis and P.~Carruthers,
%``QCD MINIJET CROSS-SECTIONS,''
Phys. Rev. D \textbf{40}, 1446 (1989)
doi:10.1103/PhysRevD.40.1446
%53 citations counted in INSPIRE as of 17 Jun 2024

%\cite{Eskola:2009uj}
\bibitem{Eskola:2009uj}
K.~J.~Eskola, H.~Paukkunen and C.~A.~Salgado,
%``EPS09: A New Generation of NLO and LO Nuclear Parton Distribution Functions,''
JHEP \textbf{04}, 065 (2009)
doi:10.1088/1126-6708/2009/04/065
[arXiv:0902.4154 [hep-ph]].
%1278 citations counted in INSPIRE as of 31 May 2024

%\cite{Eskola:2016oht}
\bibitem{Eskola:2016oht}
K.~J.~Eskola, P.~Paakkinen, H.~Paukkunen and C.~A.~Salgado,
%``EPPS16: Nuclear parton distributions with LHC data,''
Eur. Phys. J. C \textbf{77}, no.3, 163 (2017)
doi:10.1140/epjc/s10052-017-4725-9
[arXiv:1612.05741 [hep-ph]].
%624 citations counted in INSPIRE as of 13 Jun 2024

%\cite{Eskola:2021nhw}
\bibitem{Eskola:2021nhw}
K.~J.~Eskola, P.~Paakkinen, H.~Paukkunen and C.~A.~Salgado,
%``EPPS21: a global QCD analysis of nuclear PDFs,''
Eur. Phys. J. C \textbf{82}, no.5, 413 (2022)
doi:10.1140/epjc/s10052-022-10359-0
[arXiv:2112.12462 [hep-ph]].
%103 citations counted in INSPIRE as of 07 Jun 2024

%\cite{Helenius:2021tof}
\bibitem{Helenius:2021tof}
I.~Helenius, M.~Walt and W.~Vogelsang,
%``NNLO nuclear parton distribution functions with electroweak-boson production data from the LHC,''
Phys. Rev. D \textbf{105}, no.9, 094031 (2022)
doi:10.1103/PhysRevD.105.094031
[arXiv:2112.11904 [hep-ph]].
%27 citations counted in INSPIRE as of 07 Jun 2024

%\cite{AbdulKhalek:2022fyi}
\bibitem{AbdulKhalek:2022fyi}
R.~Abdul Khalek, R.~Gauld, T.~Giani, E.~R.~Nocera, T.~R.~Rabemananjara and J.~Rojo,
%``nNNPDF3.0: evidence for a modified partonic structure in heavy nuclei,''
Eur. Phys. J. C \textbf{82}, no.6, 507 (2022)
doi:10.1140/epjc/s10052-022-10417-7
[arXiv:2201.12363 [hep-ph]].
%87 citations counted in INSPIRE as of 07 Jun 2024

%\cite{Duwentaster:2022kpv}
\bibitem{Duwentaster:2022kpv}
P.~Duwent\"aster, T.~Je\v{z}o, M.~Klasen, K.~Kova\v{r}\'\i{}k, A.~Kusina, K.~F.~Muzakka, F.~I.~Olness, R.~Ruiz, I.~Schienbein and J.~Y.~Yu,
%``Impact of heavy quark and quarkonium data on nuclear gluon PDFs,''
Phys. Rev. D \textbf{105}, no.11, 114043 (2022)
doi:10.1103/PhysRevD.105.114043
[arXiv:2204.09982 [hep-ph]].
%24 citations counted in INSPIRE as of 13 Jun 2024

%\cite{Eskola:1991ec}
\bibitem{Eskola:1991ec}
K.~J.~Eskola,
%``Shadowing effects on quark and gluon production in ultrarelativistic heavy ion collisions,''
Z. Phys. C \textbf{51}, 633-642 (1991)
doi:10.1007/BF01565590
%24 citations counted in INSPIRE as of 29 Feb 2024

%\cite{Klein:2003dj}
\bibitem{Klein:2003dj}
S.~R.~Klein and R.~Vogt,
%``Inhomogeneous shadowing effects on J / psi production in dA collisions,''
Phys. Rev. Lett. \textbf{91}, 142301 (2003)
doi:10.1103/PhysRevLett.91.142301
[arXiv:nucl-th/0305046 [nucl-th]].
%77 citations counted in INSPIRE as of 07 May 2024

%\cite{Vogt:2004hf}
\bibitem{Vogt:2004hf}
R.~Vogt,
%``Shadowing effects on the nuclear suppression factor, R (dAu) , in d+Au interactions,''
Phys. Rev. C \textbf{70}, 064902 (2004)
doi:10.1103/PhysRevC.70.064902
%33 citations counted in INSPIRE as of 07 May 2024

%\cite{Frankfurt:2011cs}
\bibitem{Frankfurt:2011cs}
L.~Frankfurt, V.~Guzey and M.~Strikman,
%``Leading Twist Nuclear Shadowing Phenomena in Hard Processes with Nuclei,''
Phys. Rept. \textbf{512}, 255-393 (2012)
doi:10.1016/j.physrep.2011.12.002
[arXiv:1106.2091 [hep-ph]].
%215 citations counted in INSPIRE as of 10 Jun 2024

%\cite{Dulat:2015mca}
\bibitem{Dulat:2015mca}
S.~Dulat, T.~J.~Hou, J.~Gao, M.~Guzzi, J.~Huston, P.~Nadolsky, J.~Pumplin, C.~Schmidt, D.~Stump and C.~P.~Yuan,
%``New parton distribution functions from a global analysis of quantum chromodynamics,''
Phys. Rev. D \textbf{93}, no.3, 033006 (2016)
doi:10.1103/PhysRevD.93.033006
[arXiv:1506.07443 [hep-ph]].
%1880 citations counted in INSPIRE as of 18 Jun 2024

%\cite{Buckley:2014ana}
\bibitem{Buckley:2014ana}
A.~Buckley, J.~Ferrando, S.~Lloyd, K.~Nordstr\"om, B.~Page, M.~R\"ufenacht, M.~Sch\"onherr and G.~Watt,
%``LHAPDF6: parton density access in the LHC precision era,''
Eur. Phys. J. C \textbf{75}, 132 (2015)
doi:10.1140/epjc/s10052-015-3318-8
[arXiv:1412.7420 [hep-ph]].
%1621 citations counted in INSPIRE as of 23 Jun 2024

% \cite{Hirvonen:workinprogress}
\bibitem{Hirvonen:workinprogress}
H.~Hirvonen, M.~Kuha, J.~Auvinen, K.~J.~Eskola, Y.~Kanakubo, H.~Niemi,
work in progress

%\cite{CMS:2014qvs}
\bibitem{CMS:2014qvs}
S.~Chatrchyan \textit{et al.} [CMS],
%``Studies of dijet transverse momentum balance and pseudorapidity distributions in pPb collisions at $\sqrt{s_{\mathrm{NN}}} = 5.02$ $\,\text {TeV}$,''
Eur. Phys. J. C \textbf{74}, no.7, 2951 (2014)
doi:10.1140/epjc/s10052-014-2951-y
[arXiv:1401.4433 [nucl-ex]].
%153 citations counted in INSPIRE as of 13 Jun 2024

%\cite{Denicol:2012cn}
\bibitem{Denicol:2012cn}
G.~S.~Denicol, H.~Niemi, E.~Molnar and D.~H.~Rischke,
%``Derivation of transient relativistic fluid dynamics from the Boltzmann equation,''
Phys. Rev. D \textbf{85}, 114047 (2012)
[erratum: Phys. Rev. D \textbf{91}, no.3, 039902 (2015)]
doi:10.1103/PhysRevD.85.114047
[arXiv:1202.4551 [nucl-th]].
%584 citations counted in INSPIRE as of 31 May 2024

%\cite{Israel:1979wp}
\bibitem{Israel:1979wp}
W.~Israel and J.~M.~Stewart,
%``Transient relativistic thermodynamics and kinetic theory,''
Annals Phys. \textbf{118}, 341-372 (1979)
doi:10.1016/0003-4916(79)90130-1
%1464 citations counted in INSPIRE as of 16 Jun 2024

%\cite{Molnar:2013lta}
\bibitem{Molnar:2013lta}
E.~Moln\'ar, H.~Niemi, G.~S.~Denicol and D.~H.~Rischke,
%``Relative importance of second-order terms in relativistic dissipative fluid dynamics,''
Phys. Rev. D \textbf{89}, no.7, 074010 (2014)
doi:10.1103/PhysRevD.89.074010
[arXiv:1308.0785 [nucl-th]].
%88 citations counted in INSPIRE as of 22 Apr 2024

%\cite{Denicol:2010xn}
\bibitem{Denicol:2010xn}
G.~S.~Denicol, T.~Koide and D.~H.~Rischke,
%``Dissipative relativistic fluid dynamics: a new way to derive the equations of motion from kinetic theory,''
Phys. Rev. Lett. \textbf{105}, 162501 (2010)
doi:10.1103/PhysRevLett.105.162501
[arXiv:1004.5013 [nucl-th]].
%272 citations counted in INSPIRE as of 26 May 2024

%\cite{Huovinen:2009yb}
\bibitem{Huovinen:2009yb}
P.~Huovinen and P.~Petreczky,
%``QCD Equation of State and Hadron Resonance Gas,''
Nucl. Phys. A \textbf{837}, 26-53 (2010)
doi:10.1016/j.nuclphysa.2010.02.015
[arXiv:0912.2541 [hep-ph]].
%602 citations counted in INSPIRE as of 16 May 2024

%\cite{Huovinen:2007xh}
\bibitem{Huovinen:2007xh}
P.~Huovinen,
%``Chemical freeze-out temperature in hydrodynamical description of Au+Au collisions at s(NN)**(1/2) = 200-GeV,''
Eur. Phys. J. A \textbf{37}, 121-128 (2008)
doi:10.1140/epja/i2007-10611-3
[arXiv:0710.4379 [nucl-th]].
%79 citations counted in INSPIRE as of 22 Apr 2024

%\cite{borisbook}
\bibitem{borisbook}
J.~P.~Boris and D.~L.~Book, J.~Comput.~Phys. \textbf{11}, 38 (1973).

%\cite{Molnar:2009tx}
\bibitem{Molnar:2009tx}
E.~Molnar, H.~Niemi and D.~H.~Rischke,
%``Numerical tests of causal relativistic dissipative fluid dynamics,''
Eur. Phys. J. C \textbf{65}, 615-635 (2010)
doi:10.1140/epjc/s10052-009-1194-9
[arXiv:0907.2583 [nucl-th]].
%65 citations counted in INSPIRE as of 29 Feb 2024

%\cite{Cooper:1974mv}
\bibitem{Cooper:1974mv}
F.~Cooper and G.~Frye,
%``Comment on the Single Particle Distribution in the Hydrodynamic and Statistical Thermodynamic Models of Multiparticle Production,''
Phys. Rev. D \textbf{10}, 186 (1974)
doi:10.1103/PhysRevD.10.186
%1143 citations counted in INSPIRE as of 12 Jun 2024

\bibitem{vanLeeuwen}
F.~Debbasch, W.~A.~van Leeuwen, Physica A \textbf{388}, 1079-1104 (2009).

%\cite{Eskola:1993cz}
\bibitem{Eskola:1993cz}
K.~J.~Eskola and X.~N.~Wang,
%``Space-time structure of initial parton production in ultrarelativistic heavy ion collisions,''
Phys. Rev. D \textbf{49}, 1284-1292 (1994)
doi:10.1103/PhysRevD.49.1284
[arXiv:nucl-th/9307011 [nucl-th]].
%81 citations counted in INSPIRE as of 04 Jun 2024

%\cite{HotQCD:2018pds}
\bibitem{HotQCD:2018pds}
A.~Bazavov \textit{et al.} [HotQCD],
%``Chiral crossover in QCD at zero and non-zero chemical potentials,''
Phys. Lett. B \textbf{795}, 15-21 (2019)
doi:10.1016/j.physletb.2019.05.013
[arXiv:1812.08235 [hep-lat]].
%548 citations counted in INSPIRE as of 17 Jun 2024

%\cite{ALICE:2016fbt}
\bibitem{ALICE:2016fbt}
J.~Adam \textit{et al.} [ALICE],
%``Centrality dependence of the pseudorapidity density distribution for charged particles in Pb-Pb collisions at $\sqrt{s_{\rm NN}}=5.02$ TeV,''
Phys. Lett. B \textbf{772}, 567-577 (2017)
doi:10.1016/j.physletb.2017.07.017
[arXiv:1612.08966 [nucl-ex]].
%133 citations counted in INSPIRE as of 24 May 2024

%\cite{ALICE:2015juo}
\bibitem{ALICE:2015juo}
J.~Adam \textit{et al.} [ALICE],
%``Centrality Dependence of the Charged-Particle Multiplicity Density at Midrapidity in Pb-Pb Collisions at $\sqrt{s_{\rm NN}}$ = 5.02 TeV,''
Phys. Rev. Lett. \textbf{116}, no.22, 222302 (2016)
doi:10.1103/PhysRevLett.116.222302
[arXiv:1512.06104 [nucl-ex]].
%350 citations counted in INSPIRE as of 17 Jun 2024

%\cite{ALICE:2013jfw}
\bibitem{ALICE:2013jfw}
E.~Abbas \textit{et al.} [ALICE],
%``Centrality dependence of the pseudorapidity density distribution for charged particles in Pb-Pb collisions at $\sqrt{s_{\rm NN}}$ = 2.76 TeV,''
Phys. Lett. B \textbf{726}, 610-622 (2013)
doi:10.1016/j.physletb.2013.09.022
[arXiv:1304.0347 [nucl-ex]].
%230 citations counted in INSPIRE as of 16 May 2024

%\cite{ALICE:2010mlf}
\bibitem{ALICE:2010mlf}
K.~Aamodt \textit{et al.} [ALICE],
%``Centrality dependence of the charged-particle multiplicity density at mid-rapidity in Pb-Pb collisions at $\sqrt{s_{NN}}=2.76$ TeV,''
Phys. Rev. Lett. \textbf{106}, 032301 (2011)
doi:10.1103/PhysRevLett.106.032301
[arXiv:1012.1657 [nucl-ex]].
%814 citations counted in INSPIRE as of 26 May 2024

%\cite{Back:2002wb}
\bibitem{Back:2002wb}
B.~B.~Back, M.~D.~Baker, D.~S.~Barton, R.~R.~Betts, M.~Ballintijn, A.~A.~Bickley, R.~Bindel, A.~Budzanowski, W.~Busza and A.~Carroll, \textit{et al.}
%``The Significance of the fragmentation region in ultrarelativistic heavy ion collisions,''
Phys. Rev. Lett. \textbf{91}, 052303 (2003)
doi:10.1103/PhysRevLett.91.052303
[arXiv:nucl-ex/0210015 [nucl-ex]].
%400 citations counted in INSPIRE as of 20 May 2024

%\cite{ALICE:2016tlx}
\bibitem{ALICE:2016tlx}
J.~Adam \textit{et al.} [ALICE],
%``Pseudorapidity dependence of the anisotropic flow of charged particles in Pb-Pb collisions at $\sqrt{s_{\rm NN}}=2.76$ TeV,''
Phys. Lett. B \textbf{762}, 376-388 (2016)
doi:10.1016/j.physletb.2016.07.017
[arXiv:1605.02035 [nucl-ex]].
%70 citations counted in INSPIRE as of 03 Jun 2024

%\cite{PHOBOS:2004vcu}
\bibitem{PHOBOS:2004vcu}
B.~B.~Back \textit{et al.} [PHOBOS],
%``Centrality and pseudorapidity dependence of elliptic flow for charged hadrons in Au+Au collisions at s(NN)**(1/2) = 200-GeV,''
Phys. Rev. C \textbf{72}, 051901 (2005)
doi:10.1103/PhysRevC.72.051901
[arXiv:nucl-ex/0407012 [nucl-ex]].
%319 citations counted in INSPIRE as of 10 May 2024

%\cite{Bilandzic:2010jr}
\bibitem{Bilandzic:2010jr}
A.~Bilandzic, R.~Snellings and S.~Voloshin,
%``Flow analysis with cumulants: Direct calculations,''
Phys. Rev. C \textbf{83}, 044913 (2011)
doi:10.1103/PhysRevC.83.044913
[arXiv:1010.0233 [nucl-ex]].
%473 citations counted in INSPIRE as of 17 Jun 2024

%\cite{CMS:2015xmx}
\bibitem{CMS:2015xmx}
V.~Khachatryan \textit{et al.} [CMS],
%``Evidence for transverse momentum and pseudorapidity dependent event plane fluctuations in PbPb and pPb collisions,''
Phys. Rev. C \textbf{92}, no.3, 034911 (2015)
doi:10.1103/PhysRevC.92.034911
[arXiv:1503.01692 [nucl-ex]].
%231 citations counted in INSPIRE as of 17 Jun 2024

%\cite{ATLAS:2017rij}
\bibitem{ATLAS:2017rij}
M.~Aaboud \textit{et al.} [ATLAS],
%``Measurement of longitudinal flow decorrelations in Pb+Pb collisions at $\sqrt{s_{\text {NN}}}=2.76$ and 5.02 TeV with the ATLAS detector,''
Eur. Phys. J. C \textbf{78}, no.2, 142 (2018)
doi:10.1140/epjc/s10052-018-5605-7
[arXiv:1709.02301 [nucl-ex]].
%108 citations counted in INSPIRE as of 24 May 2024

%\cite{Eskola:1988hp}
\bibitem{Eskola:1988hp}
K.~J.~Eskola, K.~Kajantie and J.~Lindfors,
%``Perturbative Thermalization in Nuclear Collisions at Very High-energies?,''
Phys. Lett. B \textbf{214}, 613-616 (1988)
doi:10.1016/0370-2693(88)90130-X
%31 citations counted in INSPIRE as of 29 Feb 2024

%\cite{Xu:2004mz}
\bibitem{Xu:2004mz}
Z.~Xu and C.~Greiner,
%``Thermalization of gluons in ultrarelativistic heavy ion collisions by including three-body interactions in a parton cascade,''
Phys. Rev. C \textbf{71}, 064901 (2005)
doi:10.1103/PhysRevC.71.064901
[arXiv:hep-ph/0406278 [hep-ph]].
%525 citations counted in INSPIRE as of 26 May 2024

%\cite{Kurkela:2014tea}
\bibitem{Kurkela:2014tea}
A.~Kurkela and E.~Lu,
%``Approach to Equilibrium in Weakly Coupled Non-Abelian Plasmas,''
Phys. Rev. Lett. \textbf{113}, no.18, 182301 (2014)
doi:10.1103/PhysRevLett.113.182301
[arXiv:1405.6318 [hep-ph]].
%117 citations counted in INSPIRE as of 29 May 2024

%\cite{Kurkela:2015qoa}
\bibitem{Kurkela:2015qoa}
A.~Kurkela and Y.~Zhu,
%``Isotropization and hydrodynamization in weakly coupled heavy-ion collisions,''
Phys. Rev. Lett. \textbf{115}, no.18, 182301 (2015)
doi:10.1103/PhysRevLett.115.182301
[arXiv:1506.06647 [hep-ph]].
%254 citations counted in INSPIRE as of 13 Jun 2024

%\cite{Kurkela:2018vqr}
\bibitem{Kurkela:2018vqr}
A.~Kurkela, A.~Mazeliauskas, J.~F.~Paquet, S.~Schlichting and D.~Teaney,
%``Effective kinetic description of event-by-event pre-equilibrium dynamics in high-energy heavy-ion collisions,''
Phys. Rev. C \textbf{99}, no.3, 034910 (2019)
doi:10.1103/PhysRevC.99.034910
[arXiv:1805.00961 [hep-ph]].
%167 citations counted in INSPIRE as of 17 Jun 2024

%\cite{Kurkela:2018oqw}
\bibitem{Kurkela:2018oqw}
A.~Kurkela and A.~Mazeliauskas,
%``Chemical equilibration in weakly coupled QCD,''
Phys. Rev. D \textbf{99}, no.5, 054018 (2019)
doi:10.1103/PhysRevD.99.054018
[arXiv:1811.03068 [hep-ph]].
%73 citations counted in INSPIRE as of 09 May 2024

%\cite{Kurkela:2018wud}
\bibitem{Kurkela:2018wud}
A.~Kurkela, A.~Mazeliauskas, J.~F.~Paquet, S.~Schlichting and D.~Teaney,
%``Matching the Nonequilibrium Initial Stage of Heavy Ion Collisions to Hydrodynamics with QCD Kinetic Theory,''
Phys. Rev. Lett. \textbf{122}, no.12, 122302 (2019)
doi:10.1103/PhysRevLett.122.122302
[arXiv:1805.01604 [hep-ph]].
%178 citations counted in INSPIRE as of 04 Jun 2024

%\cite{Kurkela:2018xxd}
\bibitem{Kurkela:2018xxd}
A.~Kurkela and A.~Mazeliauskas,
%``Chemical Equilibration in Hadronic Collisions,''
Phys. Rev. Lett. \textbf{122}, 142301 (2019)
doi:10.1103/PhysRevLett.122.142301
[arXiv:1811.03040 [hep-ph]].
%67 citations counted in INSPIRE as of 05 Jun 2024

%\cite{Du:2020dvp}
\bibitem{Du:2020dvp}
X.~Du and S.~Schlichting,
%``Equilibration of weakly coupled QCD plasmas,''
Phys. Rev. D \textbf{104}, no.5, 054011 (2021)
doi:10.1103/PhysRevD.104.054011
[arXiv:2012.09079 [hep-ph]].
%46 citations counted in INSPIRE as of 30 May 2024

%\cite{Du:2020zqg}
\bibitem{Du:2020zqg}
X.~Du and S.~Schlichting,
%``Equilibration of the Quark-Gluon Plasma at Finite Net-Baryon Density in QCD Kinetic Theory,''
Phys. Rev. Lett. \textbf{127}, no.12, 122301 (2021)
doi:10.1103/PhysRevLett.127.122301
[arXiv:2012.09068 [hep-ph]].
%41 citations counted in INSPIRE as of 17 May 2024

%\cite{Carzon:2023zfp}
\bibitem{Carzon:2023zfp}
P.~Carzon, M.~Martinez, J.~Noronha-Hostler, P.~Plaschke, S.~Schlichting and M.~Sievert,
%``Pre-equilibrium evolution of conserved charges with initial conditions in the ICCING Monte Carlo event generator,''
Phys. Rev. C \textbf{108}, no.6, 064905 (2023)
doi:10.1103/PhysRevC.108.064905
[arXiv:2301.04572 [nucl-th]].
%8 citations counted in INSPIRE as of 30 May 2024

%\cite{Zhou:2024ysb}
\bibitem{Zhou:2024ysb}
F.~Zhou, J.~Brewer and A.~Mazeliauskas,
%``Minijet quenching in non-equilibrium quark-gluon plasma,''
[arXiv:2402.09298 [hep-ph]].
%1 citations counted in INSPIRE as of 13 Jun 2024

%\cite{Gyulassy:2001kr}
\bibitem{Gyulassy:2001kr}
M.~Gyulassy, I.~Vitev, X.~N.~Wang and P.~Huovinen,
%``Transverse expansion and high p(T) azimuthal asymmetry at RHIC,''
Phys. Lett. B \textbf{526}, 301-308 (2002)
doi:10.1016/S0370-2693(02)01157-7
[arXiv:nucl-th/0109063 [nucl-th]].
%132 citations counted in INSPIRE as of 04 Jun 2024

%\cite{Okai:2017ofp}
\bibitem{Okai:2017ofp}
M.~Okai, K.~Kawaguchi, Y.~Tachibana and T.~Hirano,
%``New approach to initializing hydrodynamic fields and mini-jet propagation in quark-gluon fluids,''
Phys. Rev. C \textbf{95}, no.5, 054914 (2017)
doi:10.1103/PhysRevC.95.054914
[arXiv:1702.07541 [nucl-th]].
%53 citations counted in INSPIRE as of 07 Jun 2024

%\cite{Eskola:2002qz}
\bibitem{Eskola:2002qz}
K.~J.~Eskola, K.~Kajantie, P.~V.~Ruuskanen and K.~Tuominen,
%``Rapidity dependence of particle production in ultrarelativistic nuclear collisions,''
Phys. Lett. B \textbf{543}, 208-216 (2002)
doi:10.1016/S0370-2693(02)02457-7
[arXiv:hep-ph/0204034 [hep-ph]].
%33 citations counted in INSPIRE as of 07 May 2024

\end{thebibliography}
\end{document}